\documentclass[journal,twocolumn,romanappendices]{IEEEtran-v17}

\pdfoutput=1

\newlength{\figwidth}

\usepackage[nosort]{cite}
\usepackage{url}
\usepackage[intlimits]{amsmath}
\usepackage{amsfonts}
\usepackage{footnote}

\usepackage{bbm}
\usepackage{bbold}
\usepackage{graphicx}
\usepackage{paralist}
\usepackage{fancyref}
\usepackage[stretch=16,shrink=16,step=4]{microtype}
\usepackage{pdfsync}

\usepackage{color}
\definecolor{links}{rgb}{0.7,0,0}   
\definecolor{urls}{rgb}{0,0,0.8}    
\definecolor{cites}{rgb}{0,0,0.8}   

\usepackage{flushend}

\makeatletter
\def\@IEEEinterspaceratioM{0.265}
\def\@IEEEinterspaceMINratioM{0.1651}
\def\@IEEEinterspaceMAXratioM{0.38}

\def\@IEEEinterspaceratioB{0.31}
\def\@IEEEinterspaceMINratioB{0.19}
\def\@IEEEinterspaceMAXratioB{0.38}
\@IEEEtunefonts
\makeatother
\usepackage{bm}
\usepackage{upgreek}
\usepackage{amssymb}
\usepackage{amsfonts}
\usepackage{mathrsfs}
\usepackage{xspace}

\makeatletter
\def\amsbb{\use@mathgroup \M@U \symAMSb}
\makeatother

\newcommand{\safemath}[2]{\newcommand{#1}{\ensuremath{#2}\xspace}}
\safemath{\opE}{\amsbb{E}}
\newcommand{\Ex}[2]{\ensuremath{\amsbb{E}_{#1}\mathopen{}\left[#2\right]}} 	
\safemath{\prob}{\amsbb{P}}
\safemath{\bigO}{\mathcal{O}}
\safemath{\littleo}{\mathit{o}}

\newcommand{\tp}[1]{\ensuremath{#1^{\mathrm{T}}}} 		
\newcommand{\herm}[1]{\ensuremath{#1^{\mathrm{H}}}} 	
\newcommand{\abs}[1]{\left\lvert#1\right\rvert}		
\newcommand{\tr}{\mathrm{tr}}
\newcommand{\rank}{\mathrm{rank}}
\newcommand{\vol}{\mathrm{vol}}
\newcommand{\fnorm}[1]{\ensuremath{\left\|#1\right\|_{\mathsf{F}}}}
\newcommand{\Int}{\mathrm{Int}}

\newtheorem{thm}{Theorem}
\newtheorem{lemma}[thm]{Lemma}
\newtheorem{dfn}{Definition}
\newtheorem{rem}{Remark}
\newtheorem{prop}[thm]{Proposition}
\newcommand{\indfun}[1]{\mathbbmss{1}\{#1\}}

\safemath{\matA}{\mathsf{A}}
\safemath{\matB}{\mathsf{B}}
\safemath{\matC}{\mathsf{C}}
\safemath{\matD}{\mathsf{D}}
\safemath{\matE}{\mathsf{E}}
\safemath{\matF}{\mathsf{F}}
\safemath{\matG}{\mathsf{G}}
\safemath{\matH}{\mathsf{H}}
\safemath{\matI}{\mathsf{I}}
\safemath{\matJ}{\mathsf{J}}
\safemath{\matK}{\mathsf{K}}
\safemath{\matL}{\mathsf{L}}
\safemath{\matM}{\mathsf{M}}
\safemath{\matN}{\mathsf{N}}
\safemath{\matO}{\mathsf{O}}
\safemath{\matP}{\mathsf{P}}
\safemath{\matQ}{\mathsf{Q}}
\safemath{\matR}{\mathsf{R}}
\safemath{\matS}{\mathsf{S}}
\safemath{\matT}{\mathsf{T}}
\safemath{\matU}{\mathsf{U}}
\safemath{\matV}{\mathsf{V}}
\safemath{\matW}{\mathsf{W}}
\safemath{\matX}{\mathsf{X}}
\safemath{\matY}{\mathsf{Y}}
\safemath{\matZ}{\mathsf{Z}}
\safemath{\matSigma}{\mathsf{\Sigma}}
\safemath{\matPhi}{\mathsf{\Phi}}
\safemath{\matLambda}{\mathsf{\Lambda}}

\safemath{\randveca}{\bm{A}}
\safemath{\randvecb}{\bm{B}}
\safemath{\randvecc}{\bm{C}}
\safemath{\randvecd}{\bm{D}}
\safemath{\randvece}{\bm{E}}
\safemath{\randvecf}{\bm{F}}
\safemath{\randvecg}{\bm{G}}
\safemath{\randvech}{\bm{H}}
\safemath{\randveci}{\bm{I}}
\safemath{\randvecj}{\bm{J}}
\safemath{\randveck}{\bm{K}}
\safemath{\randvecl}{\bm{L}}
\safemath{\randvecm}{\bm{M}}
\safemath{\randvecn}{\bm{N}}
\safemath{\randveco}{\bm{O}}
\safemath{\randvecp}{\bm{P}}
\safemath{\randvecq}{\bm{Q}}
\safemath{\randvecr}{\bm{R}}
\safemath{\randvecs}{\bm{S}}
\safemath{\randvect}{\bm{T}}
\safemath{\randvecu}{\bm{U}}
\safemath{\randvecv}{\bm{V}}
\safemath{\randvecw}{\bm{W}}
\safemath{\randvecx}{\bm{X}}
\safemath{\randvecy}{\bm{Y}}
\safemath{\randvecz}{\bm{Z}}

\safemath{\randmatA}{\amsbb{A}}
\safemath{\randmatB}{\amsbb{B}}
\safemath{\randmatC}{\amsbb{C}}
\safemath{\randmatD}{\amsbb{D}}
\safemath{\randmatE}{\amsbb{E}}
\safemath{\randmatF}{\amsbb{F}}
\safemath{\randmatG}{\amsbb{G}}
\safemath{\randmatH}{\amsbb{H}}
\safemath{\randmatI}{\amsbb{I}}
\safemath{\randmatJ}{\amsbb{J}}
\safemath{\randmatK}{\amsbb{K}}
\safemath{\randmatL}{\amsbb{L}}
\safemath{\randmatM}{\amsbb{M}}
\safemath{\randmatN}{\amsbb{N}}
\safemath{\randmatO}{\amsbb{O}}
\safemath{\randmatP}{\amsbb{P}}
\safemath{\randmatQ}{\amsbb{Q}}
\safemath{\randmatR}{\amsbb{R}}
\safemath{\randmatS}{\amsbb{S}}
\safemath{\randmatT}{\amsbb{T}}
\safemath{\randmatU}{\amsbb{U}}
\safemath{\randmatV}{\amsbb{V}}
\safemath{\randmatW}{\amsbb{W}}
\safemath{\randmatX}{\amsbb{X}}
\safemath{\randmatY}{\amsbb{Y}}
\safemath{\randmatZ}{\amsbb{Z}}
\safemath{\randmatSigma}{\mathbb{\Sigma}}

\safemath{\pdff}{f}
\safemath{\pdfp}{p}
\safemath{\pdfq}{q}

\safemath{\cdfF}{F}
\safemath{\cdfP}{P}
\safemath{\cdfQ}{Q}

\safemath{\veca}{\bm{a}}
\safemath{\vecb}{\bm{b}}
\safemath{\vecc}{\bm{c}}
\safemath{\vecd}{\bm{d}}
\safemath{\vece}{\bm{e}}
\safemath{\vecf}{\bm{f}}
\safemath{\vecg}{\bm{g}}
\safemath{\vech}{\bm{h}}
\safemath{\veci}{\bm{i}}
\safemath{\vecj}{\bm{j}}
\safemath{\veck}{\bm{k}}
\safemath{\vecl}{\bm{l}}
\safemath{\vecm}{\bm{m}}
\safemath{\vecn}{\bm{n}}
\safemath{\veco}{\bm{o}}
\safemath{\vecp}{\bm{p}}
\safemath{\vecq}{\bm{q}}
\safemath{\vecr}{\bm{r}}
\safemath{\vecs}{\bm{s}}
\safemath{\vect}{\bm{t}}
\safemath{\vecu}{\bm{u}}
\safemath{\vecv}{\bm{v}}
\safemath{\vecw}{\bm{w}}
\safemath{\vecx}{\bm{x}}
\safemath{\vecy}{\bm{y}}
\safemath{\vecz}{\bm{z}}

\safemath{\veclambda}{\bm{\lambda}}

\safemath{\setA}{\mathcal{A}}
\safemath{\setB}{\mathcal{B}}
\safemath{\setC}{\mathcal{C}}
\safemath{\setD}{\mathcal{D}}
\safemath{\setE}{\mathcal{E}}
\safemath{\setF}{\mathcal{F}}
\safemath{\setG}{\mathcal{G}}
\safemath{\setH}{\mathcal{H}}
\safemath{\setI}{\mathcal{I}}
\safemath{\setJ}{\mathcal{J}}
\safemath{\setK}{\mathcal{K}}
\safemath{\setL}{\mathcal{L}}
\safemath{\setM}{\mathcal{M}}
\safemath{\setN}{\mathcal{N}}
\safemath{\setO}{\mathcal{O}}
\safemath{\setP}{\mathcal{P}}
\safemath{\setQ}{\mathcal{Q}}
\safemath{\setR}{\mathcal{R}}
\safemath{\setS}{\mathcal{S}}
\safemath{\setT}{\mathcal{T}}
\safemath{\setU}{\mathcal{U}}
\safemath{\setV}{\mathcal{V}}
\safemath{\setW}{\mathcal{W}}
\safemath{\setX}{\mathcal{X}}
\safemath{\setY}{\mathcal{Y}}
\safemath{\setZ}{\mathcal{Z}}
\safemath{\emptySet}{\varnothing}

\safemath{\veczero}{\mathbf{0}} 

\safemath{\diag}{\mathrm{diag}}

\safemath{\jpg}{\mathcal{CN}}			
\safemath{\complexset}{\amsbb{C}}
\safemath{\realset}{\amsbb{R}}
\safemath{\posrealset}{\realset_{+}}
\safemath{\integerset}{\amsbb{N}}

\usepackage{bm}
\usepackage{upgreek}
\usepackage{amssymb}
\usepackage{amsfonts}
\usepackage{mathrsfs}
\usepackage{xspace}

\safemath{\mi}{I}
\safemath{\difent}{\mathrm{h}}		
\safemath{\NonnegReal}{\mathbb{R}^{+}}
\safemath{\re}{\mathrm{re}}
\safemath{\Real}{\mathrm{Re}} 
\safemath{\gradient}{\nabla}

\safemath{\indist}{\cdfP} 
\safemath{\outdist}{\cdfQ} 
\safemath{\inpdf}{\pdfp} 
\safemath{\outpdf}{\pdfq} 
\safemath{\testdist}{\cdfP} 

\safemath{\powallocvec}{\vecv}
\safemath{\powalloc}{v}

\safemath{\encoder}{f} 
\safemath{\decoder}{g} 

\safemath{\msg}{J} 
\safemath{\csir}{\mathrm{rx}}
\safemath{\csit}{\mathrm{tx}}
\safemath{\csi}{\mathrm{csi}}
\safemath{\nocsi}{\mathrm{no}}
\safemath{\csirt}{\mathrm{rt}}
\safemath{\iso}{\mathrm{iso}}

\safemath{\Rcsirt}{R_{\csirt}} 
\safemath{\Rcsir}{R_{\csir}}
\safemath{\Rcsit}{R_{\csit}}
\safemath{\Rnocsit}{R_{\nocsi}} 
\safemath{\Rnocsi}{R_{\nocsi}} 
\safemath{\Rnormal}{R_{\setN}}
\safemath{\Rnormalcsirt}{\Rcsirt^{\setN}}
\safemath{\Rnormaliso}{\Rcriso^{\setN}}

\safemath{\Rnoiso}{R_{\nocsi,\mathrm{iso}}}
\safemath{\Rcriso}{R_{\csir,\mathrm{iso}}}
\safemath{\Rrtiso}{R_{\csirt,\mathrm{iso}}}
\safemath{\Cnocsit}{C_{\epsilon}^{\nocsi}}
\safemath{\Ccsit}{C_{\epsilon}^{\csit}}
\safemath{\Ciso}{C_{\epsilon}^{\iso}}
\safemath{\Csimo}{C_{\epsilon}}

\safemath{\cdistno}{F_{\nocsi}}
\safemath{\cdistcsit}{F_{\csit}}
\safemath{\cdistiso}{F_{\iso}}
\safemath{\cdistsimo}{F}

\safemath{\matQiso}{\matQ_{\iso}}

\safemath{\optcov}{\matQ^\ast}
\safemath{\halfcov}{\matU}

\safemath{\sigmart}{\sigma}
\safemath{\murt}{\mu}

\safemath{\inset}{\setF} 
\safemath{\insetiso}{\inset_{\iso}}
\safemath{\insetMIMO}{\setF}
\safemath{\insetcov}{\setU_{\txant}}
\safemath{\insetcove}{\setU_{\txant}^{\mathrm{e}}}

\safemath{\equalR}{R_{\mathrm{e}}}

\safemath{\randtemp}{K}

\safemath{\Vnocsit}{V_{\error}^{\mathrm{no}}}
\safemath{\Vcsit}{V_{\error}^{\mathrm{rt}}}

\safemath{\cadist}{F_C} 

\safemath{\deF}{d_0} 

\safemath{\bl}{n} 
\safemath{\blt}{\tilde{\bl}}
\safemath{\error}{\epsilon} 
\safemath{\NumCode}{M}
\safemath{\RXant}{r}
\safemath{\rxant}{r}
\safemath{\txant}{t}
\safemath{\txantop}{t^{\ast}}
\safemath{\minant}{m}
\safemath{\minantop}{\minant^\ast}
\safemath{\snr}{\rho}
\safemath{\const}{k}

\safemath{\spanm}{\mathrm{span}}
\safemath{\ubb}{b_{0}}

\safemath{\covmat}{\matU}
\safemath{\covmatY}{\matV}
\safemath{\randcmY}{\randmatV}

\safemath{\condcov}{\matQ\in\insetcov}

\safemath{\BoundFU}{k_{\delta}}

\safemath{\argpn}{\xi} 

\safemath{\angletest}{Z} 

\safemath{\Lcsirt}{L^{\csirt}}
\safemath{\Lcsir}{L^{\csir}}
\safemath{\Scsirt}{S^{\csirt}}
\safemath{\Scsir}{S^{\csir}}
\safemath{\Lsimo}{L}
\safemath{\Ssimo}{S}

\safemath{\errorach}{P_\mathrm{e}} 

\newcommand{\given}{\,\vert\,}				
\safemath{\define}{\triangleq}			

\safemath{\altbl}{\tilde{\bl}}
\safemath{\constL}{k_{\mathrm{L}}}
\safemath{\constU}{k_{\mathrm{U}}}
\safemath{\funcL}{\tilde{q}}
\safemath{\funcU}{q}
\safemath{\ConstThm}{k_0}
\safemath{\randrevec}{\randvecy}
\safemath{\revec}{\vecy}
\safemath{\trcwd}{\vecx}
\safemath{\randtrcwd}{\randvecx}
\safemath{\randnoisevec}{\randvecw}
\safemath{\transmitcwd}{\vecx_1} 

\safemath{\pickcwdnoch}{\vecx_0} 
\safemath{\pickcwd}{\vecx_0} 

\safemath{\inseqrand}{\randvecx}
\safemath{\inseq}{\vecx}

\safemath{\outseq}{\matY}
\safemath{\outseqrand}{\randmatY}

\safemath{\altT}{\widetilde{T}}
\safemath{\altU}{\widetilde{U}}

\safemath{\altG}{\widetilde{G}}
\safemath{\altD}{\widetilde{D}}
\safemath{\altu}{\tilde{u}}

\safemath{\altmean}{\tilde{\mu}}
\safemath{\altvar}{\tilde{\sigma}}
\safemath{\altf}{\funcL}
\safemath{\altg}{\tilde{g}}
\safemath{\altgamma}{\tilde{\gamma}}
\safemath{\altdelta}{\tilde{\delta}}
\safemath{\altk}{\tilde{k}}
\safemath{\altconst}{\altk}
\safemath{\altbmLambda}{\widetilde{\bm{\Lambda}}}
\safemath{\altbmlambda}{\tilde{\bm{\lambda}}}
\safemath{\altLambda}{\widetilde{\Lambda}}

\safemath{\constant}{\tilde{k}}
\safemath{\constthree}{\const_3}
\safemath{\constfour}{\const_{4}}
\safemath{\constfive}{\const_{5}}
\safemath{\constup}{\const_{\mathrm{c}}}
\safemath{\constlow}{\const_{\mathrm{a}}}
\safemath{\constrt}{\const}

\safemath{\constrm}{\mathrm{const}}

\safemath{\blup}{\bl_{\mathrm{c}}}
\safemath{\bllow}{\bl_{\mathrm{a}}}

\safemath{\deltaup}{\altdelta}
\safemath{\deltalow}{\delta_{\mathrm{a}}}

\safemath{\CF}{\varphi}
\safemath{\gth}{g_{\mathrm{th}}} 
\safemath{\randratio}{Z}
\safemath{\randratiotwo}{Z_2}
\safemath{\ratiotwo}{z_2}

\safemath{\volform}{\mathrm{d}V}
\safemath{\surform}{\mathrm{d}S}
\safemath{\surformp}{\mathrm{d}\widetilde{S}}
\safemath{\pdfH}{f_{\randmatH}}
\safemath{\funct}{\varphi}
\safemath{\functQ}{\funct_{\gamma,\matQ}}
\safemath{\functt}{\funct_{\gamma}}
\safemath{\functb}{\bar{\funct}_{\gamma}}
\safemath{\functh}{\hat{\funct}_\gamma}
\safemath{\pdfgeneral}{f}
\safemath{\ulower}{\hat{u}}
\safemath{\randulower}{\hat{U}}
\safemath{\UgammaQ}{U(\gamma,\matQ)}
\safemath{\barU}{\overline{U}}

\newcommand{\mnorm}[1]{\ensuremath{\left\|#1\right\|_{\mathsf{F}}}}

\safemath{\condpdfbarG}{\tilde{f}_{\overline{G}_1}} 
\safemath{\condpdfG}{\tilde{f}_{G_1}} 

\safemath{\auxconstone}{k_1}
\safemath{\auxconsttwo}{k_2}
\safemath{\auxconstthree}{k_3}
\safemath{\auxconstfour}{k_4}

\begin{document}

\IEEEoverridecommandlockouts

\title{Quasi-Static Multiple-Antenna Fading Channels at Finite Blocklength}

\author{Wei Yang,~\IEEEmembership{Student Member,~IEEE}, Giuseppe Durisi,~\IEEEmembership{Senior Member,~IEEE}, \\Tobias Koch,~\IEEEmembership{Member,~IEEE}, and Yury Polyanskiy,~\IEEEmembership{Member,~IEEE}
\thanks{This work was supported in part by the Swedish Research Council under grant 2012-4571, by the Ericsson Research Foundation under grant~FOSTIFT-12:022, by a Marie Curie FP7 Integration Grant within the 7th European Union Framework Programme under Grant 333680, by the Spanish government (TEC2009-14504-C02-01, CSD2008-00010, and TEC2012-38800-C03-01), and by the National Science Foundation under Grant CCF-1253205.
 The material of this paper was presented in part at the 2013 and 2014 IEEE International Symposium on Information Theory.}
\thanks{W. Yang and G. Durisi are with the Department of Signals and Systems, Chalmers University of Technology, 41296,  Gothenburg, Sweden (e-mail: \{ywei, durisi\}@chalmers.se).}
 \thanks{T. Koch is with the Signal Theory and Communications Department, Universidad Carlos III de Madrid, 28911, Legan\'{e}s, Spain (e-mail: koch@tsc.uc3m.es).}
  \thanks{Y. Polyanskiy is with the Department of Electrical Engineering and Computer Science, MIT, Cambridge, MA, 02139 USA (e-mail: yp@mit.edu).}
}
	
\maketitle 		


\begin{abstract}
This paper investigates the maximal achievable rate for a given blocklength and error probability over quasi-static multiple-input multiple-output (MIMO) fading channels, with and without
channel state information (CSI) at the transmitter and/or the receiver.
The principal finding is that outage capacity, despite being an asymptotic quantity, is a sharp proxy for the finite-blocklength fundamental limits of slow-fading channels.
Specifically, the channel dispersion is shown to be zero regardless of whether the fading realizations are available at both transmitter and receiver, at only one of them, or at neither of them.
These results follow from analytically tractable converse and achievability bounds.
Numerical evaluation of these bounds verifies that zero dispersion may indeed imply fast convergence to the outage capacity as the blocklength increases.
In the example of a particular $1 \times 2$ single-input multiple-output (SIMO) Rician fading channel, the blocklength required to achieve $90\%$ of capacity is about an order of magnitude smaller compared to the blocklength required for an AWGN channel with the same capacity.
For this specific scenario, the coding/decoding schemes adopted in the LTE-Advanced standard are benchmarked against the finite-blocklength achievability and converse bounds.
\end{abstract}


\section{Introduction}\label{sec:introduction}
%

Consider a delay-constrained communication system operating over a slowly-varying fading channel.
In such a scenario, it is plausible to assume that the duration of each of the transmitted codewords is smaller than the coherence time of the channel, so the random fading coefficients stay constant over the duration of each codeword~\cite[p.~2631]{biglieri98-10a},\cite[Sec.~5.4.1]{tse05a}.
We shall refer to this channel model as \emph{quasi-static fading channel}.\footnote{\label{footnote:terms}The term ``quasi-static'' is widely used in the communication literature~(see, e.g., \cite[Sec.~5.4.1]{tse05a},\cite{foschini98}). The quasi-static channel model belongs to the general class of \emph{composite channels}~\cite[p.~2631]{biglieri98-10a},\cite{effros98-08} (also known as \emph{mixed channels}~\cite[Sec.~3.3]{han03a}).}

When communicating over quasi-static fading channels at a given rate~$R$, the realization of the random fading coefficient may be very small, in which case the block (frame) error probability~$\error$ is bounded away from zero even if the blocklength~$n$ tends to infinity. In this case, the channel is said to be in \emph{outage}. For fading distributions for which the fading coefficient can be arbitrarily small (such as for Rayleigh, Rician, or Nakagami fading), the probability of an outage is positive.
Hence, the overall block error probability $\epsilon$ is bounded away from zero for every positive rate $R>0$, in which case the Shannon capacity is zero.
More generally, the Shannon capacity depends on the fading probability density function (pdf) only through its support~\cite{effros10-07,verdu94-07a}.

For applications in which a positive block error probability $\epsilon>0$ is acceptable, the maximal achievable rate as a function of the outage probability (also known as \emph{capacity versus outage})~\cite[p.~2631]{biglieri98-10a},~\cite{ozarow94}, may be a more relevant performance metric than Shannon capacity.
The capacity versus outage coincides with the $\epsilon$-capacity $C_{\epsilon}$ (which is the largest achievable rate under the assumption that the block error probability is less than $\epsilon>0$) at the points where $C_{\epsilon}$ is a continuous function of $\epsilon$~\cite[Sec.~IV]{verdu94-07a}.

For the sake of simplicity, let us consider for a moment a single-antenna communication system operating over a  quasi-static flat-fading channel.
The outage probability as a function of the rate $R$ is defined by
\begin{equation}\label{eq:outage_prob}
F(R) = \prob\mathopen{}\left[\log(1+|H|^2 \rho) < R\right].
\end{equation}
Here, $H$ denotes the random channel gain and $\rho$ is the signal-to-noise ratio (SNR).
For a given $\epsilon>0$, the outage capacity (or $\epsilon$-capacity) $C_{\epsilon}$  is the supremum of all rates $R$ satisfying $F(R)\leq\epsilon$. The rationale behind this definition is that, for every realization of the fading coefficient $H=h$, the quasi-static fading channel can be viewed as an AWGN channel with channel gain $|h|^2$, for which communication with arbitrarily small block error probability is feasible if and only if $R<\log(1+|h|^2\rho)$, provided that the blocklength $n$ is sufficiently large. Thus, the outage probability can be interpreted as the probability that the channel gain~$H$ is too small to allow for communication with arbitrarily small block error probability.

A major criticism of this definition is that it is somewhat contradictory to the underlying motivation of the channel model. Indeed, while $\log(1+|h|^2\rho)$ is meaningful only for codewords of sufficiently large blocklength, the assumption that the fading coefficient is constant during the transmission of the codeword is only reasonable if the blocklength is smaller than the coherence time of the channel. In other words, it is \emph{prima facie} unclear whether for those blocklengths for which the quasi-static channel model is reasonable, the outage capacity  is a meaningful performance metric.
%

In order to shed light on this issue, we study the maximal achievable rate $R^{*}(\bl, \error)$ for a given blocklength~$\bl$ and block error probability~$\error$ over a quasi-static multiple-input multiple-output (MIMO) fading channel, subject to a per-codeword power constraint.

\paragraph*{Previous results} 
\label{par:previous_results}

Building upon Dobrushin's and Strassen's asymptotic results, Polyanskiy, Poor, and Verd\'u recently showed that for various channels with positive Shannon capacity $C$, 
the maximal achievable rate can be tightly approximated by~\cite{polyanskiy10-05}
\begin{IEEEeqnarray}{rCl}
    \label{eq:approx_R_introduction}
    R^{*}(\bl,\error)  = C -\sqrt{\frac{V}{\bl}}Q^{-1}(\error) + \bigO\mathopen{}\left(\frac{\log \bl }{\bl}\right).
\end{IEEEeqnarray}
 Here, $Q^{-1}(\cdot)$ denotes the inverse of the Gaussian $Q$-function
 \begin{IEEEeqnarray}{rCL}
   Q(x)&\define&\int\nolimits_{x}^{\infty}\!\!\frac{1}{\sqrt{2\pi}}e^{-t^2/2} dt
 \end{IEEEeqnarray}
 and $V$ is the \emph{channel dispersion}~\cite[Def.~1]{polyanskiy10-05}.
 The approximation~(\ref{eq:approx_R_introduction}) implies that to sustain the desired error probability~$\error$ at a finite blocklength~$\bl$, one pays a penalty on the rate (compared to the channel capacity) that is proportional to $1/\sqrt{\bl}$.

  Recent works have extended~\eqref{eq:approx_R_introduction} to some ergodic fading channels.
  Specifically, the dispersion of single-input single-output (SISO) stationary fading channels for the case when channel state information (CSI) is available at the receiver was derived in \cite{polyanskiy11-08a}.
  This result was extended to block-memoryless fading channels in~\cite{yang12-09}.
Upper and lower bounds on the second-order coding rate of quasi-static MIMO Rayleigh-fading channels have been reported in~\cite{hoydis13} for the asymptotically ergodic setup when the number of antennas grows linearly with the blocklength. A lower bound on $R^*(\bl,\error)$ for the imperfect CSI case has been developed in~\cite{potter13-07}. The second-order coding rate of single-antenna quasi-static fading channels for the case of perfect CSI and long-term power constraint has been derived in~\cite{yang14-07b}.

\paragraph*{Contributions} We provide achievability and converse bounds on $R^\ast(\bl,\error)$ for quasi-static MIMO fading channels.
We consider both the case when the transmitter has full transmit CSI (CSIT) and, hence, can perform spatial water-filling, and the case when no CSIT is available.
Our converse results are obtained under the assumption of perfect receive CSI (CSIR), whereas the achievability results are derived under the assumption of no CSIR.

By analyzing the asymptotic behavior of our achievability and converse bounds, we show that under mild conditions on the fading distribution,\footnote{These conditions are satisfied by the fading distributions commonly used in the wireless communication literature (e.g., Rayleigh, Rician, Nakagami).}
\begin{IEEEeqnarray}{rCl}
\label{eq:zero-dispersion}
R^\ast(\bl,\error) =C_{\epsilon} +\bigO\mathopen{}\left(\frac{\log\bl}{\bl}\right).
\end{IEEEeqnarray}
This results holds both for the case of perfect CSIT and for the case of no CSIT, and independently on whether CSIR is available at the receiver or not.
By comparing~\eqref{eq:approx_R_introduction} with~\eqref{eq:zero-dispersion}, we observe that  for the quasi-static fading case, the $1/\sqrt{\bl}$ rate penalty is absent. In other words, the \error-dispersion (see~\cite[Def.~2]{polyanskiy10-05} or \eqref{eq:def-dispersion-epsilon} below) of quasi-static fading channels is \emph{zero}.
This suggests that the maximal achievable rate $R^*(n,\epsilon)$ converges quickly to $C_{\epsilon}$ as $n$ tends to infinity, thereby indicating that the outage capacity is indeed a meaningful performance metric for delay-constrained communication over slowly-varying fading channels. Fast convergence to the outage capacity provides mathematical support to the observation reported by several researchers in the past that the outage probability describes accurately the performance over quasi-static fading channels of actual codes (see~\cite{caire99-05} and references therein).

The following example supports our claims: for a $1\times 2$ single-input multiple-output (SIMO) Rician-fading channel with $C_{\epsilon}=1$ bit$/$channel use and $\epsilon=10^{-3}$, the blocklength required to achieve $90\%$ of $C_{\epsilon}$ for the perfect CSIR case is between $120$ and $320$ (see Fig.~\ref{fig:bounds-simo} on p.~\pageref{fig:bounds-simo}), which is about an order of magnitude smaller compared to the blocklength required for an AWGN channel with the same capacity (see~\cite[Fig.~12]{polyanskiy10-05}).
%
%
%
%

Fast convergence to the outage capacity further suggests that communication strategies that are optimal with respect to outage capacity may perform also well at finite blocklength.
 Note, however, that this need not be true for very small blocklengths, where the $\mathcal{O}(n^{-1}\log n)$ term in~\eqref{eq:zero-dispersion} may dominate. Thus, for small $n$ the derived achievability and converse bounds on $R^*(n,\epsilon)$ may behave differently than the outage capacity.
Table~\ref{tab:wisdom} summarizes how the outage capacity and the achievability/converse bounds on $R^*(n,\epsilon)$ derived in this paper depend on system parameters such as the availability of CSI and the number of antennas at the transmitter/receiver. These observations may be relevant for delay-constrained communication over slowly-varying fading channels.

\paragraph*{Proof techniques} Our converse bounds on $R^{*}(\bl,\error)$ are based on the meta-converse theorem~\cite[Th.~30]{polyanskiy10-05}.
Our achievability bounds on $R^{*}(\bl,\error)$ are based on the $\kappa\beta$ bound~\cite[Th.~25]{polyanskiy10-05} applied to a stochastically degraded channel, whose choice is motivated by geometric considerations.
The main tools used to establish~\eqref{eq:zero-dispersion}  are a Cramer-Esseen-type central-limit theorem~\cite[Th.~VI.1]{petrov75} and a result on the speed of convergence of $\prob[B> A/\sqrt{\bl} ]$ to $\prob[B>0]$ for $\bl\to\infty$, where $A$ and $B$ are independent random variables.

\begin{savenotes}
 \begin{table*}[ht]\label{tab:wisdom}
   \caption{Outage capacity vs. finite blocklength wisdom; $\txant$ is the number of transmit antennas.}
 \begin{center}
  \renewcommand{\arraystretch}{1.3}
 \begin{tabular}{|p{5.5cm}||c|c|}
 \hline
Wisdom & $C_{\epsilon}$ & Bounds on $R^*(n,\epsilon)$\\\hline\hline
 CSIT is beneficial & only if $\txant>1$ & only if $\txant>1$ \\ \hline
 CSIR is beneficial & no~{\cite[p.~2632]{biglieri98-10a}} & yes \\\hline
 With CSIT, waterfilling is optimal & yes~{\cite{telatar99-11a}} & no \\\hline
 With CSIT, the channel is reciprocal\footnotemark & yes~{\cite{telatar99-11a}} & only with CSIR \\\hline
\end{tabular}
\end{center}
\end{table*}
\end{savenotes}

\paragraph*{Notation}
Upper case letters such as $X$ denote scalar random variables and their realizations are written in lower case, e.g.,~$x$.
We use boldface upper case letters to denote random vectors, e.g., $\randvecx$, and boldface lower case letters for their realizations, e.g., $\vecx$.
Upper case letters of two special fonts are used to denote deterministic matrices (e.g., $\matY$) and random matrices (e.g., $\randmatY$).
The superscripts~$\tp{}$ and $\herm{}$ stand for transposition and Hermitian transposition, respectively.
%
%
We use $\tr(\matA)$ and $\det(\matA)$ to denote the trace and determinant of the matrix $\matA$, respectively, and use $\spanm(\matA)$ to designate the subspace spanned by the column vectors of $\matA$.
The Frobenius norm of a matrix $\matA$ is denoted by $\fnorm{\matA} \define \sqrt{\tr(\matA\herm{\matA})}$.
The notation $\matA \succeq \mathbf{0}$ means that the matrix $\matA$ is positive semi-definite.
 The function resulting from the composition of two functions $f$ and $g$ is denoted by $g\circ f$, i.e., $(g\circ f) (x) =g(f(x))$.
 %
For two functions~$f(x)$ and~$g(x)$, the
notation~$f(x) = \bigO(g(x))$, $x\to \infty$, means that
$\lim \sup_{x\to\infty}\bigl|f(x)/g(x)\bigr|<\infty$, and
$f(x) = \littleo(g(x))$, $x\to \infty$, means that $\lim_{x\to\infty}\bigl|f(x)/g(x)\bigr|=0$.
We use $\matI_{a}$ to denote the identity matrix of size $a\times a $, and designate by $\matI_{a,b}$ $(a>b)$ the $a\times b$ matrix containing the first $b$ columns of $\matI_{a}$.
The distribution of a circularly-symmetric complex
Gaussian random vector with covariance matrix~$\matA$ is denoted by $\jpg(\mathbf{0}, \matA)$, the Wishart distribution~\cite[Def.~2.3]{tulino04a} with $\bl$ degrees of freedom and covariance matrix $\mathsf{\matA}$ defined on matrices of size $m\times m$ is denoted by $\mathcal{W}_m(\bl,\mathsf{\matA})$, and the Beta distribution~\cite[Ch.~25]{johnson95-2} is denoted by $\mathrm{Beta}(\cdot,\cdot)$.
The symbol $\posrealset$ stands for the nonnegative real line, $\posrealset^{m} \subset \realset^{\minant}$ is the nonnegative orthant of the $m$-dimensional real Euclidean spaces, and $\realset_{\geq}^{m}\subset\realset_{+}^{m}$ is defined by
\begin{IEEEeqnarray}{rCl}
\realset^{\minant}_{\geq} \define \{\vecx\in \posrealset^{\minant}: x_1\geq\cdots\geq x_\minant\}.
\label{eq:def-R-geq}
\end{IEEEeqnarray}
The indicator function is denoted by $\indfun{\cdot}$, and $[\,\cdot\,]^{+} \define \max\{\,\cdot\,, 0\}$. Finally, $\log(\cdot)$ is the natural logarithm.

Given two distributions $\indist$ and $\outdist$ on a common measurable space $\setW$, we define a randomized test between $\indist$ and $\outdist$ as a random transformation $\testdist_{Z\given W}: \setW\mapsto\{0,1\}$ where $0$ indicates that the test chooses $\outdist$. We shall need the following performance metric for the test between~$\indist$ and~$\outdist$:
\begin{IEEEeqnarray}{rCl}
\label{eq:def-beta}
\beta_\alpha(\indist,\outdist) \define \min\int \testdist_{Z\given W}(1\given w)  \outdist(d w)
\end{IEEEeqnarray}
where the minimum is over all probability distributions $\testdist_{Z\given W}$
satisfying
\begin{IEEEeqnarray}{rCl}
 \int \testdist_{Z\given W} (1\given w) \indist(dw)\geq \alpha.
\end{IEEEeqnarray}

\footnotetext{A channel is reciprocal for a given performance metric (e.g., outage capacity) if substituting $\randmatH$ with $\herm{\randmatH}$ does not change the metric.}

\section{System Model} 
We consider a quasi-static MIMO fading channel with $\txant$ transmit and~$\rxant$ receive antennas.
Throughout this paper, we denote the minimum number of transmit and receive antennas by $\minant$, i.e., $\minant\define \min\{\txant,\rxant\}$.
The channel input-output relation is given by
\begin{IEEEeqnarray}{rCl}
\label{eq:channel_io}
\randmatY &=& \matX \randmatH + \randmatW.
\end{IEEEeqnarray}
Here, $\matX\in\complexset^{\bl\times\txant}$ is the signal transmitted over $\bl$ channel uses;
$\randmatY\in\complexset^{\bl\times\rxant}$ is the corresponding received signal;
the matrix $\randmatH\in\complexset^{\txant\times\rxant}$ contains the complex fading coefficients, which are random but remain constant over the~$\bl$ channel uses; $\randmatW\in\complexset^{\bl\times\rxant}$ denotes the additive noise at the receiver, which is independent of $\randmatH$ and has independent and identically distributed (i.i.d.) $\jpg(0,1)$ entries.

We consider the following four scenarios:
\begin{enumerate}
\item no-CSI: neither the transmitter nor the receiver is aware of the realizations of the fading matrix~$\randmatH$;
\item CSIT: the transmitter knows~$\randmatH$;
\item CSIR: the receiver knows~$\randmatH$;
\item CSIRT: both the transmitter and the receiver know~$\randmatH$.
\end{enumerate}
To keep the notation compact, we shall abbreviate in mathematical formulas the acronyms no-CSI, CSIT, CSIR, and CSIRT as $\nocsi$, $\csit$, $\csir$, and $\csirt$, respectively.
 Next, we introduce the notion of a channel code for each of these four settings.
\begin{dfn}[no-CSI]
\label{dfn:no-csi-code}
 An $(\bl, \NumCode, \error)_{\nocsi}$ code consists of:
\begin{enumerate}[i)]
\item an encoder $\encoder_{\nocsi}$: $\{1,\ldots,\NumCode\} \mapsto \complexset^{\bl\times \txant}$ that maps the {message} $\msg \in \{1,\ldots,\NumCode\}$ to a codeword $\matX \in \{\matC_1,\ldots, \matC_{\NumCode}\}$. The codewords satisfy the power constraint
\begin{IEEEeqnarray}{rCl}
\label{eq:peak-power-constraint}
\fnorm{\matC_i}^2 &\leq& \bl \snr,\quad i=1,\ldots,\NumCode.
\end{IEEEeqnarray}
\item A decoder $\decoder_{\nocsi}$: $\complexset^{\bl\times \rxant } \mapsto\{1,\ldots,\NumCode\}$ satisfying a \emph{maximum probability of error} constraint
\begin{equation}
\label{eq:max-prob-error}
\max\limits_{1\leq j \leq \NumCode}\prob[\decoder_{\nocsi}(\outseqrand) \neq \msg \given \msg =j] \leq \error
\end{equation}
where $\outseqrand$ is the channel output induced by the transmitted codeword $\matX = \encoder_{\nocsi}(j)$ according to~\eqref{eq:channel_io}.
\end{enumerate}
%
%
 \end{dfn}

\begin{dfn}[CSIR]
\label{def:csir-code}
 An $(\bl, \NumCode, \error)_{\csir}$ code consists of:
\begin{enumerate}[i)]
\item an encoder $\encoder_{\nocsi}$: $\{1,\ldots,\NumCode\} \mapsto \complexset^{\bl\times \txant}$ that maps the message $\msg \in \{1,\ldots,\NumCode\}$ to a codeword $\matX \in \{\matC_1,\ldots, \matC_{\NumCode}\}$. The codewords satisfy the power constraint~\eqref{eq:peak-power-constraint}.
\item A decoder $\decoder_{\csir}$: $\complexset^{\bl\times \rxant } \times \complexset^{\txant \times \rxant} \mapsto\{1,\ldots,\NumCode\}$ satisfying
 \begin{equation}
\label{eq:avg-prob-error-csir}
\max\limits_{1\leq j \leq \NumCode}\prob[\decoder_{\csir}(\outseqrand,\randmatH) \neq \msg \given \msg =j] \leq \error.
\end{equation}
\end{enumerate}
%
%
 \end{dfn}

\begin{dfn}[CSIT]\label{def:csit-code} An $(\bl, \NumCode, \error)_{\csit}$ code consists of:
\begin{enumerate}[i)]
\item an encoder $\encoder_{\csit}$: $ \{1,\ldots,\NumCode\}\times \complexset^{\txant\times\rxant} \mapsto \complexset^{\bl\times \txant}$ that maps the message $j \in \{1,\ldots,\NumCode\}$ and the channel $\matH$ to a codeword $\matX = \encoder_{\csit}(j,\matH)$ satisfying
 \begin{IEEEeqnarray}{rCl}
 \label{eq:power-constraint-csit}
   \fnorm{\matX}^2=\fnorm{\encoder_{\csit}(j,\matH)}^2 &\leq& \bl\snr, \notag\\
   &&\forall j=1,\dots, \NumCode,\, \forall \matH \in \complexset^{\txant\times\rxant}. \IEEEeqnarraynumspace
 \end{IEEEeqnarray}
 %
\item A decoder $\decoder_{\nocsi}$: $\complexset^{\bl\times \rxant } \mapsto\{1,\ldots,\NumCode\}$ satisfying~\eqref{eq:max-prob-error}.

\end{enumerate}
%
%
 \end{dfn}

\begin{dfn}[CSIRT]
\label{dfn:csirt-code}
An $(\bl, \NumCode, \error)_{\csirt}$ code consists of:
\begin{enumerate}[i)]
\item an encoder $\encoder_{\csit}$: $ \{1,\ldots,\NumCode\}\times \complexset^{\txant\times\rxant} \mapsto \complexset^{\bl\times \txant}$ that maps the message $j \in \{1,\ldots,\NumCode\}$ and the channel $\matH$ to a codeword $\matX = \encoder_{\csit}(j,\matH)$ satisfying~\eqref{eq:power-constraint-csit}.
\item A decoder $\decoder_{\csir}$: $\complexset^{\bl\times \rxant } \times \complexset^{\txant \times \rxant} \mapsto\{1,\ldots,\NumCode\}$ satisfying~\eqref{eq:avg-prob-error-csir}.
\end{enumerate}
%
%
 \end{dfn}

The maximal achievable rate for the four cases listed above is defined as follows:
%
%
\begin{multline}
\label{eq:def-r-nocsit}
R^{\ast}_l(\bl,\error) \define \sup\mathopen{}\left\{\frac{\log\NumCode}{\bl}: \exists(\bl
,\NumCode, \error)_{l}\text{ code}\right\},\\ l\in\{\nocsi,\csir,\csit,\csirt\}.
\end{multline}
From Definitions~\ref{dfn:no-csi-code}--\ref{dfn:csirt-code}, it follows that
\begin{IEEEeqnarray}{rCl}
\Rnocsi^{\ast}(\bl,\error) &\leq& \Rcsit^{\ast}(\bl,\error) \leq \Rcsirt^{\ast}(\bl,\error)\label{eq:relation-ct-rt}\\
\Rnocsi^{\ast}(\bl,\error) &\leq& \Rcsir^{\ast}(\bl,\error)\leq \Rcsirt^{\ast}(\bl,\error)\label{eq:relation-no-cr}.
\end{IEEEeqnarray}

%

\section{Asymptotic Results and Preview}
\label{sec:error-capacity}
It was noted in~\cite[p.~2632]{biglieri98-10a} that the $\error$-capacity of quasi-static MIMO fading channel does not depend on whether CSI is available at the  receiver.
  Intuitively, this is true because the channel stays constant during the transmission of a codeword, so it can be accurately estimated at the receiver through the transmission of~$\sqrt{\bl}$ pilot symbols with no rate penalty as $\bl\to\infty$. A rigorous proof of this statement follows by our zero-dispersion results (Theorems~\ref{thm:asy-mimo-csirt} and~\ref{thm:zero-dispersion-nocsit}).
 In contrast, if CSIT is available and $\txant >1$, then water-filling over space yields a larger $\error$-capacity~\cite{caire99-05}.
We next define~$C_{\epsilon}$ for both the CSIT and the no-CSIT case.

Let $\insetcov$ be the set of $\txant\times\txant$ positive semidefinite matrices whose trace is upper-bounded by $\snr$, i.e.,
 \begin{IEEEeqnarray}{c}\label{eq:def_set_covariance_matrices}
\insetcov \define \{\matA \in \complexset^{\txant\times\txant}: \matA\succeq \mathbf{0}, \tr(\matA)\leq \snr\}.
\end{IEEEeqnarray}
When CSI is available at the transmitter, the $\error$-capacity $\Ccsit$ is given by~\cite[Prop.~2]{caire99-05}\footnote{More precisely, \eqref{eq:C-epsilon-csit} and~\eqref{eq:C-epsilon-nocsit} hold provided that $\Ccsit$ and $\Cnocsit$ are continuous functions of $\epsilon$~\cite[Th.~6]{verdu94-07a}.\label{footnote:continuity}}
\begin{IEEEeqnarray}{rCl}
\Ccsit &=& \lim\limits_{\bl\to\infty} \Rcsit^{\ast}(\bl,\error)\\
 &=&\lim\limits_{\bl\to\infty}\Rcsirt^\ast(\bl,\error)\\
 &=& \sup\mathopen{}\left\{R: \cdistcsit(R)\leq \error\right\}
\label{eq:C-epsilon-csit}
\end{IEEEeqnarray}
where
\begin{IEEEeqnarray}{rCl}
\cdistcsit(R) \define \prob\mathopen{}\left[\max_{\condcov} \log\det\mathopen{}\left(\matI_\rxant + \herm{\randmatH}\matQ\randmatH\right)< R\right]
\label{eq:cadist_csit}
  \end{IEEEeqnarray}
denotes the outage probability.
Given $\randmatH =\matH$, the function $\log\det\mathopen{}\left(\matI_\rxant + \herm{\matH}\matQ\matH\right)$ in~\eqref{eq:cadist_csit} is maximized by the well-known water-filling power-allocation strategy (see, e.g.,~\cite{telatar99-11a}), which results in
\begin{IEEEeqnarray}{rCl}
\max_{\condcov} \log\det\mathopen{}\left(\matI_\rxant + \herm{\matH}\matQ\matH\right) = \sum\limits_{j=1}^{\minant} \left[\log(\bar{\gamma}\lambda_j)\right]^{+}
\end{IEEEeqnarray}
where the scalars $\lambda_1\geq \cdots\geq \lambda_\minant$ denote the~$\minant$ largest eigenvalues of $\herm{\matH}\matH$, and
$\bar{\gamma}$ is the solution of
\begin{IEEEeqnarray}{rCl}
\sum\limits_{j=1}^{\minant} [\bar{\gamma}- 1 / \lambda_{j}]^{+} =\snr.
\label{eq:def-gamma-bar}
\end{IEEEeqnarray}
%
%
%
In Section~\ref{sec:mimo-csit}, we study quasi-static MIMO channels with CSIT at finite blocklength.
We present an achievability (lower) bound on $\Rcsit^\ast(\bl,\error)$ (Section~\ref{sec:ach-csit-mimo}, Theorem~\ref{thm:actual_ach_bound_csit}) and a converse (upper) bound on $\Rcsirt^\ast(\bl,\error)$ (Section~\ref{sec:converse-csirt}, Theorem~\ref{thm:converse-csirt}).
We show in~Section~\ref{sec:asy-analysis-csit} (Theorem~\ref{thm:asy-mimo-csirt}) that, under mild conditions on the fading distribution, the two bounds match asymptotically up to a $\bigO(\log(\bl)/\bl)$ term.
This allows us to establish the zero-dispersion result~\eqref{eq:zero-dispersion} for the CSIT case.

When CSI is not available at the transmitter, the $\error$-capacity~$\Cnocsit$ is given by~\cite{telatar99-11a,effros10-07}
\begin{IEEEeqnarray}{rCl}
\Cnocsit&=&\lim\limits_{\bl\to\infty} \Rnocsi^{\ast}( \bl,\error)\\
 &=& \lim\limits_{\bl\to\infty}\Rcsir^\ast(\bl,\error)\\
 &=& \sup\{R: \cdistno(R) \leq \error\} \label{eq:C-epsilon-nocsit}
\end{IEEEeqnarray}
where
\begin{IEEEeqnarray}{rCl}
\label{eq:cadist_nocsi}
\cdistno(R) \define \inf_{\condcov}\prob\mathopen{}\left[\log\det\mathopen{}\left(\matI_\rxant + \herm{\randmatH}\matQ\randmatH\right)< R\right]
\end{IEEEeqnarray}
is the outage probability  for the no-CSIT case.
The matrix $\matQ$ that minimizes the right-hand-side (RHS) of~\eqref{eq:cadist_nocsi} is in general not known, making this case more difficult to analyze and our nonasymptotic results less sharp and more difficult to evaluate numerically.
The minimization in~\eqref{eq:cadist_nocsi} can be restricted to all $\matQ$ on the boundary of~$\insetcov$, i.e.,
\begin{IEEEeqnarray}{rCl}
\cdistno(R) = \inf_{\matQ \in \insetcove}\prob\mathopen{}\left[\log\det\mathopen{}\left(\matI_\rxant + \herm{\randmatH}\matQ\randmatH\right)< R\right]
\label{eq:P-out-alt-exp}
\end{IEEEeqnarray}
where
\begin{IEEEeqnarray}{rCl}
\insetcove \define \{\matA \in \complexset^{\txant\times\txant}: \matA \succeq \bm{0}, \tr(\matA) = \snr\}.
\label{eq:def-insetcove}
\end{IEEEeqnarray}
We lower-bound $\Rnocsi^{\ast}(\bl,\error)$ in Section~\ref{sec:ach-nocsi-mimo} (Theorem~\ref{thm:actual_ach_bound_nocsit}), and upper-bound $\Rcsir^{\ast}(\bl,\error)$ in Section~\ref{sec:converse-csir} (Theorem~\ref{thm:converse-csir}).
The asymptotic analysis of the bounds provided in Section~\ref{sec:asy-results-nocsi} (Theorem~\ref{thm:zero-dispersion-nocsit}) allows us to establish~\eqref{eq:zero-dispersion}, although under slightly more stringent assumptions on the fading probability distribution than for the~CSIT~case.

For the i.i.d. Rayleigh-fading model (without CSIT), Telatar~\cite{telatar99-11a} conjectured that the optimal $\matQ$ is of the form\footnote{This conjecture has recently been proved for the multiple-input single-output case~\cite{abbe13-05}.}
\begin{IEEEeqnarray}{rCl}
 \frac{\snr}{\txantop}\, \diag\{\underbrace{1,\ldots,1}_{\txantop},\underbrace{0,\ldots,0}_{\txant-\txantop}\},\quad\,\, 1\leq \txantop\leq \txant
 \label{eq:opt-cov-mat}
\end{IEEEeqnarray}
and that for small $\error$ values or for high SNR values, all available transmit antennas should be used, i.e., $\txantop=\txant$.
We define the $\error$-rate $\Ciso $ resulting from the choice $\matQ=(\snr/\txant)\matI_{\txant}$ as
\begin{IEEEeqnarray}{rCl}
\Ciso \define \sup\{R: \cdistiso(R) \leq \error \}
\end{IEEEeqnarray}
where
\begin{IEEEeqnarray}{rCl}
\cdistiso(R) \define \prob\mathopen{}\left[\log\det\mathopen{}\left(\matI_{\rxant}+ \frac{\snr}{\txant}\herm{\randmatH} \randmatH\right) < R\right].
\label{eq:def-cdist-iso}
\end{IEEEeqnarray}
The $\error$-rate $\Ciso $ is often taken as an accurate lower bound on the actual $\error$-capacity for the case of i.i.d Rayleigh fading and no CSIT.
Motivated by this fact, we consider in Section~\ref{sec:mimo-nocsi} codes with isotropic codewords, i.e., chosen from the set
\begin{IEEEeqnarray}{rCl}
\insetiso \define \left\{\matX \in \complexset^{\bl\times\txant}: \frac{1}{\bl}\herm{\matX} \matX = \frac{\snr}{\txant}\matI_{\txant}\right\}.
\label{eq:def-f-iso}
\end{IEEEeqnarray}
We indicate by $(\bl,\NumCode,\error)_{\iso}$ a code with $\NumCode$ codewords chosen from $\insetiso$ and with a maximal error probability smaller than~$\error$.
For this special class of codes, the maximal achievable rate $\Rnoiso^*(\bl,\error)$ for the no-CSI case and~$\Rcriso^*(\bl,\error)$ for the CSIR case can be characterized more accurately at finite blocklength (Theorem~\ref{thm:converse-csir-iso}) than for the general no-CSI case.
Furthermore, we show in Section~\ref{sec:asy-results-nocsi} (Theorem~\ref{thm:proof-asy-iso}) that under mild conditions on the fading distributions (weaker than the ones required for the general no-CSI case)
\begin{IEEEeqnarray}{rCl}
\{\Rnoiso^{*}(\bl,\error), \Rcriso^{*}(\bl,\error)\}= \Ciso +\bigO\mathopen{}\left(\frac{\log\bl}{\bl}\right). \IEEEeqnarraynumspace
\end{IEEEeqnarray}
%

A final remark on notation. For the single-transmit-antenna case (i.e., $\txant=1$), the $\error$-capacity does not depend on whether CSIT is available or not~\cite[Prop.~3]{caire99-05}. Hence, we shall denote the $\error$-capacity for this case simply as~$\Csimo$.

\section{CSI Available at the Transmitter}
\label{sec:mimo-csit}

\subsection{Achievability}
\label{sec:ach-csit-mimo}
In this section, we consider the case where CSI is available at the transmitter but not at the receiver. Before establishing our achievability bound in Section~\ref{sec:rigious-derivation}, we provide some geometric intuition that will guide us in the choice of the decoder~$\decoder_{\nocsi}$ (see Definition~\ref{def:csit-code}).

\subsubsection{Geometric Intuition}
\label{sec:geometric-intution}
\begin{figure}[t]
	\centering
		\includegraphics[scale=0.71]{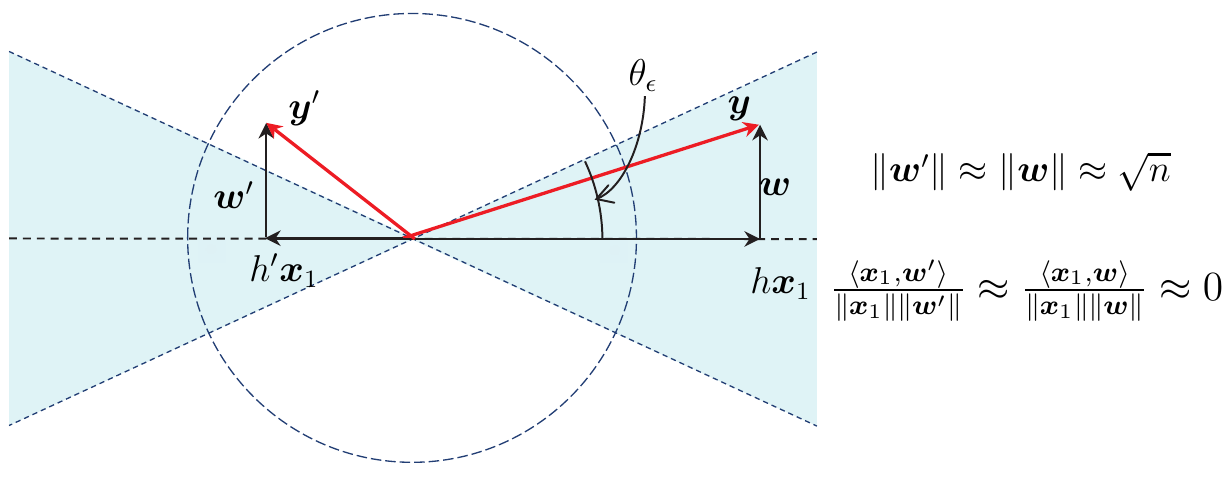}
\caption{\label{fig:sphere} A geometric illustration of the outage event for large blocklength $\bl$.
In the example, the fading realization $h'$ triggers an outage event, $h$ does not.
}
\end{figure}

Consider for simplicity a real-valued quasi-static SISO channel ($\txant=\rxant=1$), i.e., a channel with input-output relation
\begin{equation}\label{eq:siso_real_channel}
\randvecy = H \vecx + \randvecw
\end{equation}
where $\randvecy$, $\vecx$, and $\randvecw$ are $\bl$-dimensional vectors, and $H$ is a (real-valued) scalar.
As reviewed in Section \ref{sec:introduction}, the typical error event for the quasi-static fading channel (in the large blocklength regime) is that the instantaneous channel gain $H^2$ is not large enough to support the desired rate $R$, i.e., $\frac{1}{2}\log(1+\snr H^2) <R$ (outage event).
For the channel in~\eqref{eq:siso_real_channel}, the $\error$-capacity $C_\error$, i.e., the largest rate~$R$ for which the probability that the channel is in outage is less than $\error$, is given by
\begin{equation}
C_{\error} = \sup \mathopen{}\left\{R: \prob\mathopen{}\left[\frac{1}{2}\log(1+\snr H^2)< R\right]\leq \error \right\}.
\label{eq:C-epsilon-siso}
\end{equation}
Roughly speaking, the decoder of a $C_\error$-achieving code may commit an error only when the channel is in outage.
Pick now an arbitrary codeword $\vecx_1$ from the hypersphere
$\{\vecx\in \realset^{\bl}:\, \|\vecx\|^2=\bl\snr\}$,
and let $\randvecy$ be the received signal corresponding to~$\vecx_1$.
Following~\cite{shannon59}, we analyze the angle $\theta(\transmitcwd, \randrevec)$ between~$\transmitcwd$ and~$\randrevec$ as follows.
By the law of large numbers, the noise vector~$\randvecw$ is approximately orthogonal to~$\vecx_1$ if~$\bl$ is large, i.e.,
\begin{IEEEeqnarray}{rCl}
 \frac{\langle\transmitcwd,\randnoisevec\rangle}{\|\transmitcwd\|\|\randnoisevec\|} \to 0, \quad \bl \to \infty.
 \label{eq:approx-angle}
 \end{IEEEeqnarray}
Also by the law of large numbers, $\|\randnoisevec\|^2/\bl \to 1$ as $\bl\to\infty$.
Hence, for a given~$H$ and for large~$\bl$, the angle $\theta(\transmitcwd, \randrevec)$ can be approximated as
 \begin{IEEEeqnarray}{rCl}
\theta(\transmitcwd, \randrevec) &\approx&  \arcsin \frac{\|\randnoisevec\|}{\sqrt{H^2\|\transmitcwd\|^2 + \|\randnoisevec\|^2}}\\
 &\approx& \arcsin \frac{1}{\sqrt{\snr H^2+1}} \label{eq:angle-app-siso}
 \end{IEEEeqnarray}
where the first approximation follows by~\eqref{eq:approx-angle} and the second approximation follows because $\|\randnoisevec\|^2 \approx \bl$.
It follows from~\eqref{eq:C-epsilon-siso} and~\eqref{eq:angle-app-siso}  that $\theta(\transmitcwd, \randrevec)$ is larger than $\theta_{\error} \define \arcsin(e^{-C_\error})$ in the outage case, and smaller than $\theta_\error$ otherwise (see Fig.~\ref{fig:sphere}).

This geometric argument suggests  the use of a threshold decoder that, for a given received signal~$\randvecy$, declares~$\vecx_i$ to be the transmitted codeword if~$\vecx_i$ is the only codeword for which $\theta(\vecx_i, \randrevec) \leq \theta_\error$. If no codewords or more than one codeword meet this condition, the decoder declares an error. %
Thresholding angles instead of log-likelihood ratios (cf.,~\cite[Th.~17 and Th.~25]{polyanskiy10-05}) appears to be a natural approach when CSIR is unavailable.
Note that the proposed threshold decoder does neither require CSIR nor knowledge of the fading distribution.
As we shall see, it achieves~\eqref{eq:zero-dispersion} and yields a tight achievability bound at finite blocklength, provided that the threshold $\theta_{\epsilon}$ is chosen appropriately.

In the following, we generalize the aforementioned threshold decoder to the MIMO case and present our achievability results.

\subsubsection{The Achievability Bound}
\label{sec:rigious-derivation}
To state our achievability (lower) bound on~$\Rcsit^\ast(\bl,\error)$, we will need the following definition, which extends the notion of angle between real vectors to complex subspaces.
\begin{dfn}
Let $\setA$ and $\setB$ be subspaces in $\complexset^{\bl}$ with $a =\dim(\setA) \leq \dim(\setB) =b$. The \emph{principal angles} $0\leq \theta_1\leq \cdots\leq \theta_{a}\leq {\pi}/{2}$ between $\setA$ and $\setB$ are defined recursively by
\begin{multline}
\label{eq:def-princal-angle}
\cos \theta_k \define \!\!\!\max\limits_{\scriptsize\begin{array}{c}
\veca\in \setA, \vecb\in\setB\colon \|\veca\|=\|\vecb\|=1,
 \\ \langle\veca,\veca_i\rangle=\langle\vecb,\vecb_i\rangle=0, i=1,\ldots,k-1 \end{array}} \!\!\!\!|\langle\veca,\vecb\rangle |,\\
 \quad k=1,\ldots,a.
\end{multline}
Here, $\veca_k$ and $\vecb_k$, $k=1,\ldots,a$,  are the vectors that achieve the maximum in~\eqref{eq:def-princal-angle} at the $k$th recursion.
%
The angle between the subspaces $\setA$ and $\setB$ is defined by
\begin{IEEEeqnarray}{rCl}
\label{eq:def-product-prin-sine}
\sin\{\setA,\setB\}\define \prod\limits_{k=1}^{a}\sin\theta_k.
\end{IEEEeqnarray}
\end{dfn}

With a slight abuse of notation, for two matrices $\matA\in \complexset^{\bl\times a}$ and $\matB\in\complexset^{\bl\times b}$, we abbreviate $\sin\mathopen{}\left\{\spanm(\matA),\spanm(\matB)\right\}$ with
$\sin\{\matA,\matB\} $.
When the columns of $\matA $ and $\matB$ are orthonormal bases for $\spanm(\matA)$ and $\spanm(\matB)$, respectively, we have (see, e.g.,~\cite[Sec.~I]{barg02-09})
\begin{IEEEeqnarray}{rCl}
\sin^2\{\matA,\matB\} &=& \det\mathopen{}\left(\matI -  \herm{\matA}\matB\herm{\matB} \matA\right)\\
&=& \det\mathopen{}\left(\matI -  \herm{\matB}\matA\herm{\matA} \matB\right).
\label{eq:rel-angle-sin-eigen}
\end{IEEEeqnarray}
Some additional properties of the operator $\sin\{\cdot,\cdot\}$ are listed in Appendix~\ref{app:proof-angle-btw-subspaces}.

We are now ready to state our achievability bound.
\begin{thm}\label{thm:actual_ach_bound_csit}
Let $\Lambda_1\geq \cdots\geq\Lambda_\minant$ be the $\minant$ largest eigenvalues of $\randmatH\herm{\randmatH}$.
 For every $0\!<\!\error\!<\!1$ and every $0\!<\!\tau\!<\!\error$, there exists an $(\bl, \NumCode, \error)_{\csit}$ code for the channel~\eqref{eq:channel_io}
 that satisfies
 \begin{IEEEeqnarray}{rCl}
 \label{eq:lb-numcode-csit}
\frac{\log\NumCode}{\bl}  &\geq& \frac{1}{\bl} \log\frac{\tau}{\prob\mathopen{}\big[\prod_{j=1}^{\rxant} B_j \leq \gamma_{\bl}\big] }.
 \end{IEEEeqnarray}
Here, $B_j\sim \mathrm{Beta}(\bl-\txant-j+1,\txant)$, $j =1,\ldots,\rxant$, are independent Beta-distributed random variables, and~\mbox{$\gamma_{\bl}\in[0,1]$} is chosen so that
 \begin{IEEEeqnarray}{rCl}
\IEEEeqnarraymulticol{3}{l}{
 \prob\mathopen{}\bigg[\sin^2 \mathopen{}\Big\{\matI_{\bl,\txant},\sqrt{\bl}\matI_{\bl,\txant}\diag\mathopen{}\Big\{\sqrt{v_1^*\Lambda_1},\ldots,  }\notag\\ \quad\quad\quad\sqrt{v_{\minant}^*\Lambda_\minant},\underbrace{0,\ldots,0}_{\txant-\minant}\Big\}
 + \randmatW \Big\}\leq \gamma_\bl\bigg]  \geq 1-\error+\tau \label{eq:def-gamma-n-ach-csit} \IEEEeqnarraynumspace
 \end{IEEEeqnarray}
where
\begin{IEEEeqnarray}{rCl}
v^\ast_j=[\bar{\gamma} - 1/\Lambda_{j}]^{+}, \,\,\, j=1,\ldots,\rxant
\label{eq:water-filling-power}
\end{IEEEeqnarray}
are the water-filling power gains and $\bar{\gamma}$ is defined in~\eqref{eq:def-gamma-bar}.
\end{thm}
\begin{IEEEproof}
The achievability bound is based on a decoder that operates as follows: it first computes the sine of the angle between the subspace spanned by the received matrix~$\randmatY$ and the subspace spanned by each codeword; then, it chooses the first codeword for which the squared sine of the angle is below~$\gamma_\bl$.
To analyze the performance of this decoder, we apply the $\kappa\beta$ bound~\cite[Th.~25]{polyanskiy10-05} to a physically degraded channel whose output is~$\spanm(\randmatY)$.
 See Appendix~\ref{app:proof-ach-bound-csit} for the complete proof.
\end{IEEEproof}

\subsection{Converse}
\label{sec:converse-csirt}
In this section, we shall assume both CSIR and CSIT. Our converse bound is based on the meta-converse theorem~\cite[Th.~30]{polyanskiy10-05}. Since CSI is available at both the transmitter and the receiver, the MIMO channel~\eqref{eq:channel_io} can be transformed into a set of parallel quasi-static channels. The proof of Theorem~\ref{thm:converse-csirt} below builds on~\cite[Sec.~4.5]{polyanskiy10}, which characterizes the nonasymptotic coding rate of parallel AWGN channels.

\begin{thm}
\label{thm:converse-csirt}
Let $\Lambda_1\geq \cdots\geq\Lambda_\minant$ be the $\minant$ largest eigenvalues of $\randmatH\herm{\randmatH}$, and let $\bm{\Lambda} \define \tp{[\Lambda_1,\ldots,\Lambda_\minant]}$.
Consider an arbitrary power-allocation function $\powallocvec:  \posrealset^{\minant} \mapsto \setV_{\minant}$,
where
\begin{IEEEeqnarray}{rCl}\label{eq:def_setV}
  \setV_{\minant} \define \left\{[p_1,\ldots,p_{\minant}]\in \posrealset^{\minant}:\,\sum\nolimits_{j=1}^{\minant}p_j \leq \snr \right\}.
\end{IEEEeqnarray}
Let
\begin{IEEEeqnarray}{rCl}
\label{eq:info_density_mimo_alt_csirt}
\Lcsirt_{\bl}(\powallocvec,\bm{\Lambda}) &\define& \sum\limits_{i=1}^{\bl}\sum \limits_{j=1}^{\minant} \! \Bigg(\!\log\mathopen{}\big(1+\Lambda_j\powalloc_j(\bm{\Lambda})\big) + 1   \notag\\
&& - \,\left|\sqrt{\Lambda_j\powalloc_j(\bm{\Lambda})} Z_{i,j} - \sqrt{ 1 + \Lambda_j\powalloc_j(\bm{\Lambda})} \right|^2\!\Bigg)\IEEEeqnarraynumspace
\end{IEEEeqnarray}
and
\begin{IEEEeqnarray}{rCl}
\Scsirt_\bl(\powallocvec,\bm{\Lambda}) &\define& \sum\limits_{i=1}^{\bl}\sum\limits_{j=1}^{\minant}  \Bigg(\!\log\mathopen{}\big(1+\Lambda_j\powalloc_j(\bm{\Lambda})\big)  + 1 \notag\\
&&\quad\quad\quad\quad\,\, -\, \frac{\big|\sqrt{\Lambda_j\powalloc_j(\bm{\Lambda})}Z_{ij} - 1\big|^2}{1+\Lambda_j\powalloc_j(\bm{\Lambda})}\!\Bigg)\IEEEeqnarraynumspace
\label{eq:info_density_mimo_csirt}
\end{IEEEeqnarray}
where $\powalloc_j(\cdot)$ is the $j$th coordinate of $\powallocvec(\cdot)$, and $Z_{ij}$, $i=1,\ldots, \bl$, $j=1,\ldots,\minant$, are i.i.d. $\jpg(0,1)$ distributed random variables.
For every $\bl$ and every $0<\error<1$, the maximal achievable rate on the channel~\eqref{eq:channel_io} with CSIRT is upper-bounded by
\begin{IEEEeqnarray}{rCl}
\Rcsirt^{\ast}(\bl,\error) \leq \frac{1}{n}\log \frac{ c_\csirt(\bl) }{\inf\limits_{\powallocvec(\cdot)} \prob[\Lcsirt_\bl(\powallocvec,\bm{\Lambda}) \geq \bl \gamma_{\bl}(\powallocvec) ]}
\label{eq:thm-converse-rcsirt}
\end{IEEEeqnarray}
where
\begin{IEEEeqnarray}{rCl}
\label{eq:def-n-func-converse-csirt}
c_\csirt(\bl) &\define& \left(\frac{(\bl-1)^\bl e^{-(\bl-1)}}{\Gamma(\bl)} +\frac{ \Gamma(\bl,\bl-1)}{\Gamma(\bl)} \right)^\minant \notag\\
&& \times \,\Ex{\randmatH}{\det(\matI_{\txant}+\snr \randmatH\herm{\randmatH})}
\end{IEEEeqnarray}
and the scalar $\gamma_{\bl}(\powallocvec)$ is the solution of
\begin{IEEEeqnarray}{rCl}
 \label{eq:thm-converse-def-gamma-n}
 \prob[ \Scsirt_\bl(\powallocvec,\bm{\Lambda}) \leq  \bl \gamma_{\bl}(\powallocvec)] =\error.
 \end{IEEEeqnarray}
The infimum on the RHS of~\eqref{eq:thm-converse-rcsirt} is taken over all power allocation functions $\powallocvec: \posrealset^{\minant}\mapsto \setV_{\minant}$.
\begin{IEEEproof}
See Appendix~\ref{app:proof-converse-cisrt}.
\end{IEEEproof}
\end{thm}

\begin{rem}
\label{rem:converse-csirt-tight}
The infimum on the RHS of~\eqref{eq:thm-converse-rcsirt} makes the converse bound in Theorem~\ref{thm:converse-csirt} difficult to evaluate numerically.
We can further upper-bound the RHS of~\eqref{eq:thm-converse-rcsirt} by lower-bounding $\prob[\Lcsirt_\bl(\powallocvec,\bm{\Lambda}) \geq \bl \gamma_{\bl}(\powallocvec)]$ for each $\powallocvec(\cdot)$ using~\cite[Eq.~(102)]{polyanskiy10-05} and the Chernoff bound. After doing so, the infimum can be computed analytically and the resulting upper bound on $\Rcsirt^\ast(\bl,\error)$ allows for numerical evaluations.  Unfortunately, this bound is in general loose.
\end{rem}

\begin{rem}
As we shall discuss in Section~\ref{sec:converse-csir}, the bound~\eqref{eq:thm-converse-rcsirt} can be tightened and evaluated numerically in the SIMO case or when the codewords are isotropic, i.e., are chosen from the set~$\setF_{\iso}$ in~\eqref{eq:def-f-iso}.
Note that in both scenarios CSIT is not beneficial.
\end{rem}

\subsection{Asymptotic Analysis}
\label{sec:asy-analysis-csit}

Following~\cite[Def. 2]{polyanskiy10-05}, we define the $\error$-dispersion of the
channel~\eqref{eq:channel_io} with CSIT via $\Rcsit^{\ast}(\bl,\error)$ (resp.
$\Rcsirt^{\ast}(\bl,\error)$) as
\begin{multline}
\label{eq:def-dispersion-epsilon}
V_\error^{l} \define \limsup\limits_{\bl\to\infty} \bl\left( \frac{\Ccsit -
R^{\ast}_l(n,\epsilon)}{Q^{-1}(\error)}\right)^2,\\
\,\, \error\in(0,1)\backslash \{1/2\},\, l=\{\csit,\csirt\}.
\end{multline}
Theorem~\ref{thm:asy-mimo-csirt} below characterizes the \error-dispersion of the quasi-static fading channel~\eqref{eq:channel_io} with CSIT.
\begin{thm}
\label{thm:asy-mimo-csirt}
Assume that the fading channel $\randmatH$ satisfies the following conditions:
\begin{enumerate}
\item \label{item:cond-expec-finite} the expectation $\Ex{\randmatH}{\det(\matI_{\txant} + \snr \randmatH\herm{\randmatH})}$ is finite;
\item \label{item:cond-cont-pdf-eig}the joint pdf of the ordered nonzero eigenvalues of $\herm{\randmatH}\randmatH$ exists and is continuously differentiable;
 \item\label{item:cond-pos-cdf} $\Ccsit$ is a point of growth of the outage probability function~\eqref{eq:cadist_csit} , i.e.,\footnote{Note that this condition implies that $\Ccsit$ is a continuous function of $\error$ (see Section~\ref{sec:error-capacity}).}
\begin{IEEEeqnarray}{rCl}
\cdistcsit'\mathopen{}\left(\Ccsit\right) >0.
\label{eq:thm-dispersion-condition2-conv}
\end{IEEEeqnarray}
\end{enumerate}
Then
\begin{IEEEeqnarray}{rCl}
\big\{ \Rcsit^{\ast}(\bl,\error) , \Rcsirt^{\ast}(\bl,\error) \big\} &=& \Ccsit + \bigO\mathopen{}\left(\frac{\log\bl}{n}\right) \label{eq:thm-R-star-expansion-ct}.
\end{IEEEeqnarray}
Hence, the $\error$-dispersion is zero for both the CSIRT and the CSIT case:
\begin{IEEEeqnarray}{rCl}
V_\error^{\csit} = V_\error^{\csirt}=0, \quad\,\,\error\in(0,1)\backslash \{1/2\}.
\end{IEEEeqnarray}
\end{thm}

\begin{IEEEproof}
%
%
To prove~\eqref{eq:thm-R-star-expansion-ct}, we first establish in Appendix \ref{app:proof-asy-csit-conv} the converse result
\begin{IEEEeqnarray}{rCl}
\Rcsirt^\ast(\bl,\error) \leq \Ccsit + \bigO\mathopen{}\left(\frac{\log\bl}{n}\right)
\label{eq:proof-asy-conv-csit}
\end{IEEEeqnarray}
by analyzing the upper bound~\eqref{eq:thm-converse-rcsirt} in the limit $\bl\to\infty$. We next prove  in Appendix~\ref{app:proof-asy-csit-ach} the achievability result
\begin{IEEEeqnarray}{rCl}
\Rcsit^\ast(\bl,\error) \geq \Ccsit + \bigO\mathopen{}\left(\frac{\log\bl}{n}\right)
\label{eq:proof-asy-ach-csit}
\end{IEEEeqnarray}
 by expanding~\eqref{eq:lb-numcode-csit} for $\bl\to\infty$.
 The desired result then follows by~\eqref{eq:relation-ct-rt}.
 \end{IEEEproof}

\begin{rem}
As mentioned in Section~\ref{sec:introduction}, the quasi-static fading channel considered in this paper belongs to the general class of composite or mixed channels, whose $\error$-dispersion is known in some special cases.
Specifically, the dispersion of a mixed channel with two states was derived in~\cite[Th.~7]{polyanskiy11-04}.
This result was extended to channels with finitely many states in~\cite[Th.~4]{tomamichel13-05}.
In both cases, the rate of convergence to the $\error$-capacity is $\bigO(1/\sqrt{\bl})$ (positive dispersion), as opposed to $\bigO(\log(\bl)/\bl)$ in Theorem~\ref{thm:asy-mimo-csirt}.
Our result shows that moving from finitely many to uncountably many states (as in the quasi-static fading case) yields a drastic change in the value of the channel dispersion.
For this reason, our result is not derivable from~\cite{polyanskiy11-04} or \cite{tomamichel13-05}.
\end{rem}

\begin{rem}
It can be shown that the assumptions on the fading matrix in Theorem~\ref{thm:asy-mimo-csirt} are satisfied by most probability distributions used to model MIMO fading channels, such as i.i.d. or correlated Rayleigh, Rician, and Nakagami.
However, the (nonfading) AWGN MIMO channel, which can be seen as a quasi-static fading
channel with fading distribution equal to a step function, does not meet these assumptions and has, in fact, positive dispersion~\cite[Th.~78]{polyanskiy10}.

While zero dispersion indeed may imply fast convergence to $\error$-capacity, this is not true anymore when the probability distribution of the fading matrix approaches a step function, in which case the higher-order terms in the expansion~\eqref{eq:thm-R-star-expansion-ct} become more dominant.
Consider for example a SISO Rician fading channel with Rician factor $K$.
For $\error<1/2$, one can refine~\eqref{eq:thm-R-star-expansion-ct} and show that~\cite{yang13-07}
\begin{IEEEeqnarray}{rCl}
&&C_\error-\frac{\log\bl}{\bl}  - \frac{c_1 \sqrt{K} + c_2}{\bl} + \littleo\mathopen{}\left(\frac{1}{\bl}\right) \leq \Rcsit^\ast(\bl,\error)\notag\\
&&\quad\quad\leq \Rcsirt^\ast(\bl,\error) \leq C_\error + \frac{\log\bl}{\bl} - \frac{\tilde{c}_1 \sqrt{K}+\tilde{c}_2}{\bl} + \littleo\mathopen{}\left(\frac{1}{\bl}\right)\IEEEeqnarraynumspace
\label{eq:eg-rician-bounds}
\end{IEEEeqnarray}
where $c_1$, $c_2$, $\tilde{c}_1$ and $\tilde{c}_2$ are finite constants with $c_1>0$ and $\tilde{c}_1>0$.
As we let the Rician factor~$K$ become large, the fading distribution converges to a step function and the third term in both the left-most lower bound and the right-most upper bound becomes increasingly large in absolute value.

\end{rem}

\subsection{Normal Approximation}
\label{sec:normal-app-csit}

We define the \emph{normal approximation} $\Rnormalcsirt(\bl,\error)$ of  $\Rcsirt^*(\bl,\error)$ as the solution of
\begin{IEEEeqnarray}{rCl}
 \epsilon &=& \Ex{}{Q\mathopen{}\left(\frac{C(\randmatH)- \Rnormalcsirt(\bl,\error)}{\sqrt{V(\randmatH)/\bl}}\right) }.
\label{eq:normal-dist}
\end{IEEEeqnarray}
Here,
  \begin{IEEEeqnarray}{rCl}
    \label{eq:capacity_with_waterfilling}
  C(\matH) = \sum\limits_{j=1}^{\minant} \log(1+ v_j^\ast\lambda_j)
  \end{IEEEeqnarray}
is the capacity of the channel~\eqref{eq:channel_io} when $\randmatH=\matH$  (the water-filling power allocation values $\{v_j^\ast\}$ in~\eqref{eq:capacity_with_waterfilling} are given in~\eqref{eq:water-filling-power} and~$\{\lambda_j\}$ are the eigenvalues of $\herm{\matH}\matH$), and
  \begin{IEEEeqnarray}{rCl}\label{eq:dispersion_gaussian_approx}
  V(\matH) = \minant - \sum\limits_{j=1}^{\minant} \frac{1}{(1+ v_j^\ast\lambda_j)^2}
  \end{IEEEeqnarray}
is the dispersion of the channel~\eqref{eq:channel_io} when $\randmatH=\matH$ \cite[Th.~78]{polyanskiy10}.
 Theorem~\ref{thm:asy-mimo-csirt} and the expansion
\begin{IEEEeqnarray}{rCl}
\Rnormalcsirt(\bl,\error) = \Ccsit +\bigO\mathopen{}\left(\frac{1}{\bl}\right)
\label{eq:expan-RN}
\end{IEEEeqnarray}
(which follows from Lemma~\ref{lem:expectation-phi} in Appendix~\ref{sec:proof_of_averaging_over_channel} and Taylor's theorem) suggest that this approximation is accurate, as confirmed by the numerical results reported in Section~\ref{sec:numerical-results}.
Note that the same approximation has been concurrently proposed in~\cite{molavianJazi13-10}; see also~\cite[Def.~2]{polyanskiy11-04} and~\cite[Sec.~4]{tomamichel13-05} for similar approximations for mixed channels with finitely many states.

\section{CSI Not Available at the Transmitter}
\label{sec:mimo-nocsi}
%

\subsection{Achievability}
\label{sec:ach-nocsi-mimo}

In this section, we shall assume that neither the transmitter nor the receiver have \emph{a priori} CSI. Using the decoder described in~\ref{sec:ach-csit-mimo}, we obtain the following achievability bound.
\begin{thm}\label{thm:actual_ach_bound_nocsit}
Assume that for a given $0<\error <1$ there exists a $\optcov \in \setU_{\txant}$ such that
\begin{IEEEeqnarray}{rCl}
\cdistno(\Cnocsit) &=& \inf_{\condcov}\prob\mathopen{}\left[\log\det\mathopen{}\left(\matI_\rxant + \herm{\randmatH}\matQ\randmatH\right)\leq \Cnocsit\right] \label{eq:C-epsilon-nocsit-comput0}\\
&=&\prob\mathopen{}\left[\log\det\mathopen{}\left(\matI_\rxant + \herm{\randmatH}\optcov\randmatH\right)\leq \Cnocsit\right]\label{eq:C-epsilon-nocsit-comput}
\end{IEEEeqnarray}
i.e., the infimum in~\eqref{eq:C-epsilon-nocsit-comput0} is a minimum.
Then,  for every $0\!<\!\tau\!<\!\error$ there exists an $(\bl, \NumCode, \error)_{\nocsi}$ code
 for the channel~\eqref{eq:channel_io}
 that satisfies
 \begin{IEEEeqnarray}{rCl}
 \label{eq:lb-numcode-nocsi}
 \frac{\log \NumCode}{\bl} &\geq& \frac{1}{\bl} \log\frac{\tau}{\prob\mathopen{}\big[\prod_{j=1}^{\rxant} B_j \leq \gamma_{\bl}\big] }\,.
 \end{IEEEeqnarray}
 Here, $B_j\sim \mathrm{Beta}(\bl-\txantop-j+1,\txantop)$, $j=1,\ldots,\rxant$, are independent Beta-distributed random variables, $\txantop \define \rank(\optcov)$, and $\gamma_{\bl}\in[0,1]$ is chosen so that
 \begin{IEEEeqnarray}{rCl}
  \prob\mathopen{}\left[\sin^2\{\matI_{\bl,\txantop}, \sqrt{\bl}\matI_{\bl,\txantop}\halfcov\randmatH +\randmatW\}\leq \gamma_\bl\right] \geq 1-\error+\tau
 \end{IEEEeqnarray}
with $\halfcov\in \complexset^{\txantop\times \txant}$ satisfying $\herm{\halfcov}\halfcov=\optcov$.
\end{thm}
\begin{IEEEproof}
The proof is identical to the proof of Theorem~\ref{thm:actual_ach_bound_csit}, with the only difference that the precoding matrix $\matP(\randmatH)$ (defined in~\eqref{eq:precoding-matrix}) is replaced by $\sqrt{\bl}\matI_{\bl,\txantop}\halfcov$.
\end{IEEEproof}

The assumption in~\eqref{eq:C-epsilon-nocsit-comput} that the $\error$-capacity-achieving input covariance matrix of the channel~\eqref{eq:channel_io} exists is mild. A sufficient condition for the existence of $\optcov$ is given in the following proposition.
\begin{prop}
\label{prop:continuous}
Assume that $\Ex{}{\fnorm{\randmatH}^2}<\infty$ and that the distribution of $\randmatH $ is absolutely continuous with respect to the Lebesgue measure on $\complexset^{\txant\times\rxant}$.
Then, for every $R\in \posrealset$, the infimum in~\eqref{eq:cadist_nocsi} is a minimum.
\end{prop}
\begin{IEEEproof}
See Appendix~\ref{app:proof-prop-continuous}.
\end{IEEEproof}


For the SIMO case, the RHS of~\eqref{eq:lb-numcode-csit} and the RHS of~\eqref{eq:lb-numcode-nocsi} coincide, i.e.,
\begin{IEEEeqnarray}{rCl}
\big\{\Rcsit(\bl,\error), \Rnocsi(\bl,\error) \big\} &\geq& \frac{1}{\bl} \log\frac{\tau}{\prob\mathopen{}\left[ B \leq \gamma_{\bl}\right] }
\label{eq:simo-ach-bound}
\end{IEEEeqnarray}
 where $B\sim\mathrm{Beta}(\bl-\rxant,\rxant)$, and \mbox{$\gamma_{\bl}\in[0,1]$} is chosen so that
 \begin{IEEEeqnarray}{rCl}
  \prob[\sin^2\{\vece_1, \sqrt{\bl\snr}\vece_1\tp{\randvech} +\randmatW\}\leq \gamma_\bl] \geq 1-\error+\tau.
 \end{IEEEeqnarray}
Here, $\vece_1$ stands for the first column of the identity matrix $\matI_{\bl}$.
The achievability bound~\eqref{eq:simo-ach-bound} follows from~\eqref{eq:lb-numcode-csit} and~\eqref{eq:lb-numcode-nocsi} by noting that the random variable $B$ on the RHS of \eqref{eq:simo-ach-bound} has the same distribution as $\prod\nolimits_{i=1}^{\rxant}B_i$, where $B_i\sim\mathrm{Beta}(\bl-i,1)$, $i=1,\ldots,\rxant$.

\subsection{Converse}
\label{sec:converse-csir}
For the converse, we shall assume CSIR but not CSIT. The counterpart of Theorem~\ref{thm:converse-csirt} is the following result.
\begin{thm}
\label{thm:converse-csir}
Let \insetcove be as in~\eqref{eq:def-insetcove}.
For an arbitrary $\matQ \in \insetcove$, let $\Lambda_1\geq \cdots\geq\Lambda_\minant$ be the ordered eigenvalues of $\herm{\randmatH}\matQ\randmatH$.
Let
\begin{equation}
\label{eq:info_density_mimo_alt_csir}
 \Lcsir_\bl(\matQ)  \define \sum\limits_{i=1}^{\bl}\sum \limits_{j=1}^{\minant} \! \bigg(\!\log(1+\Lambda_j)  + 1 - \bigl|\sqrt{\Lambda_j} Z_{ij} - \sqrt{ 1 + \Lambda_j} \bigr|^2\bigg)
\end{equation}
and
\begin{IEEEeqnarray}{rCl}
\Scsir_\bl(\matQ) \define \sum\limits_{i=1}^{\bl}\sum\limits_{j=1}^{\minant}  \bigg(\!\log(1+\Lambda_j)  + 1- \frac{\big|\sqrt{\Lambda_j}Z_{ij} - 1\big|^2}{1+\Lambda_j}\bigg)\IEEEeqnarraynumspace
\label{eq:info_density_mimo_csir}
\end{IEEEeqnarray}
where $Z_{ij}$, $i=1,\ldots, \bl$, $j=1,\ldots,\minant$, are i.i.d. $\jpg(0,1)$ distributed.
Then, for every $\bl\geq \rxant$ and every $0<\error<1$, the maximal achievable rate on the quasi-static MIMO fading channel~\eqref{eq:channel_io} with CSIR is upper-bounded by
\begin{IEEEeqnarray}{rCl}
\label{eq:thm-converse-rcsir}
\Rcsir^{\ast}(\bl-1,\error) \leq \frac{1}{\bl-1}\log \frac{c_{\csir}(\bl)}{\inf\limits_{\matQ \in \insetcove} \prob[\Lcsir_\bl(\matQ) \geq \bl \gamma_{\bl}(\matQ)]} \, .\IEEEeqnarraynumspace
\end{IEEEeqnarray}
Here,
\begin{IEEEeqnarray}{rCl}
c_{\csir}(\bl) &\define&
\frac{\pi^{\rxant(\rxant-1)}}{\Gamma_{\rxant}(\bl) \Gamma_{\rxant}(\rxant)}\Ex{}{\left(1+\snr\fnorm{\randmatH}^2\right)^{\lfloor(\rxant+1)^2/4 \rfloor} }
\notag\\
&& \times \,\prod\limits_{i=1}^{\rxant}\bigg[ \left(\bl+\rxant-2i\right)^{\bl+\rxant -2i+1}  e^{-(\bl+\rxant -2i)} \notag\\
&&\quad\quad\quad\,\, +\, \Gamma(\bl+\rxant-2i+1, \bl+r-2i) \bigg]\label{eq:def-crn} \IEEEeqnarraynumspace
\end{IEEEeqnarray}
with $\Gamma_{(\cdot)}(\cdot)$ denoting the \emph{complex} multivariate Gamma function~\cite[Eq.~(83)]{james64},
and $\gamma_{\bl}(\matQ)$ is the solution of
\begin{IEEEeqnarray}{rCl}
 \label{eq:thm-converse-def-gamma-n1}
 \prob[ \Scsir_\bl(\matQ) \leq  \bl \gamma_{\bl}(\matQ)] =\error.
 \end{IEEEeqnarray}
\end{thm}
\begin{IEEEproof}
See Appendix~\ref{app:proof-converse-cisr}.
\end{IEEEproof}

The infimum in~\eqref{eq:thm-converse-rcsir} makes the upper bound more difficult to evaluate numerically and to analyze asymptotically up to~$\bigO(\log(\bl)/\bl)$ terms than the upper bound~\eqref{eq:thm-converse-rcsirt} that we established for the CSIT case.
In fact, even the simpler problem of finding the matrix $\matQ$ that minimizes $\lim\limits_{\bl\to\infty}\prob[\Lcsir_\bl(\matQ) \geq \bl \gamma_{\bl}]$ is open.
Next, we consider two special cases for which the bound~\eqref{eq:thm-converse-rcsir} can be tightened and evaluated numerically:  the SIMO case and the case where all codewords are chosen from the set~$\insetiso$.

\subsubsection{SIMO case} For the SIMO case, CSIT is not beneficial~\cite{yang13-07} and the bounds~\eqref{eq:thm-converse-rcsirt} and~\eqref{eq:thm-converse-rcsir} can be tightened as follows.
\begin{thm}
\label{thm:converse}
Let
\begin{IEEEeqnarray}{rCl}
  \Lsimo_\bl &\define& \bl \log(1+\snr G)  + \sum\limits_{i=1}^{\bl}
 \left(1-\bigl| \sqrt{\snr G}Z_i-  \sqrt{1+\snr G} \bigr|^2\!\right)\notag\\
&&\label{eq:info_density_simo_alt}
\end{IEEEeqnarray}
and
\begin{IEEEeqnarray}{rCl}
\Ssimo_\bl &\define& \bl \log(1+\snr G)  +   \sum\limits_{i=1}^{\bl}\left(1-\frac{\big|\sqrt{\snr G}Z_i-1\big|^2}{1+\snr G}\right)\label{eq:info_density_simo}
\end{IEEEeqnarray}
with $G \define \|\randvech\|^2$ and $Z_i$, $i=1,\ldots,\bl$, i.i.d. $\jpg(0,1)$ distributed.
For every $\bl$ and every $0<\error<1$, the maximal achievable rate on the quasi-static fading channel~\eqref{eq:channel_io} with one transmit antenna and with CSIR (with or without CSIT) is upper-bounded by
\begin{IEEEeqnarray}{rCl}
\label{eq:thm-converse-rcsit-simo}
\Rcsir^{\ast}(\bl-1,\error) \leq \Rcsirt^\ast(\bl-1,\error) \leq \frac{1}{n-1} \log \frac{1}{\prob[\Lsimo_\bl \geq \bl \gamma_{\bl}]} \IEEEeqnarraynumspace
\end{IEEEeqnarray}
where $\gamma_{\bl}$ is the solution of
\begin{IEEEeqnarray}{rCl}
 \label{eq:thm-converse-def-gamma-n-simo}
 \prob[ \Ssimo_\bl \leq  \bl \gamma_{\bl}] =\error.
 \end{IEEEeqnarray}
\end{thm}
\begin{IEEEproof}
  See~\cite[Th.~1]{yang13-07}. The main difference between the proof of Theorem~\ref{thm:converse} and the proof of Theorem~\ref{thm:converse-csirt} and~Theorem~\ref{thm:converse-csir} is that the simple bound $\error'\geq 1-1/\NumCode$ on the maximal error probability of the auxiliary channel in the meta-converse theorem~\cite[Th.~30]{polyanskiy10-05} suffices to establish the desired result.
 The more sophisticated bounds reported in Lemma~\ref{lem:converse-q-csirt} (Appendix~\ref{app:proof-converse-cisrt}) and Lemma~\ref{lem:converse-q-channel} (Appendix~\ref{app:proof-converse-cisr}) are not needed.
\end{IEEEproof}

\subsubsection{Converse for~$(\bl,\NumCode,\error)_{\iso}$ codes}
In Theorem~\ref{thm:converse-csir-iso} below, we establish a converse bound on the maximal achievable rate of $(\bl,\NumCode,\error)_{\iso}$ codes introduced in Section~\ref{sec:error-capacity}.
As such codes consist of isotropic codewords chosen from the set $\setF_{\iso}$ in~\eqref{eq:def-f-iso}, CSIT is not beneficial also in this scenario.
\begin{thm}
\label{thm:converse-csir-iso}
%
Let $\Lcsir_\bl(\cdot)$ and $\Scsir_\bl(\cdot)$ be as in~\eqref{eq:info_density_mimo_alt_csir} and~\eqref{eq:info_density_mimo_csir}, respectively.
Then, for every $\bl$ and every $0<\error<1$, the maximal achievable rate $\Rcriso^{\ast}(\bl,\error)$ of $(\bl,\NumCode,\error)_{\iso}$ codes over the quasi-static MIMO fading channel~\eqref{eq:channel_io} with CSIR is upper-bounded~by
\begin{IEEEeqnarray}{rCl}
\Rcriso^{\ast}(\bl,\error) \leq \Rrtiso^{\ast}(\bl,\error)\leq \frac{1}{n}\log \frac{1}{  \prob[\Lcsir_\bl((\snr/\txant) \matI_{\txant}) \geq \bl \gamma_{\bl}]}\notag\\
&& \label{eq:thm-converse-rcsir-iso}
\end{IEEEeqnarray}
where $\gamma_{\bl}$ is the solution of
\begin{IEEEeqnarray}{rCl}
 \label{eq:thm-converse-def-gamma-n-iso}
 \prob[ \Scsir_\bl((\snr/\txant)  \matI_{\txant}) \leq  \bl \gamma_{\bl}] =\error.
 \end{IEEEeqnarray}
\end{thm}
\begin{IEEEproof}
The proof follows closely the proof of~Theorem~\ref{thm:converse-csir}.
As in the SIMO case, the  main difference is that the simple bound $\error'\geq 1-1/\NumCode$ on the maximal error probability of the auxiliary channel in the meta-converse theorem~\cite[Th.~30]{polyanskiy10-05} suffices to establish~\eqref{eq:thm-converse-def-gamma-n-iso}.
\end{IEEEproof}

\subsection{Asymptotic Analysis}
\label{sec:asy-results-nocsi}
To state our dispersion result, we will need the following definition of the gradient $\gradient g$ of a differentiable function $g: \complexset^{\txant\times\rxant} \mapsto \realset$.
Let $\matL \in \complexset^{\txant\times \rxant}$, then we shall write $\gradient g(\matH) =\matL$ if
\begin{IEEEeqnarray}{rCl}
\frac{d}{dt}  g(\matH + t\matA) \Big|_{t=0}= \mathrm{Re}\mathopen{}\left\{\tr\mathopen{}\big(\herm{\matA} \matL\big)\right\},\quad \forall \matA \in \complexset^{\txant\times\rxant}. \IEEEeqnarraynumspace
\label{eq:def-gradient-Crt}
\end{IEEEeqnarray}
Theorem~\ref{thm:zero-dispersion-nocsit} below establishes the zero-dispersion result for the case of no CSIT.
Because of the analytical intractability of the minimization in the converse bound~\eqref{eq:thm-converse-rcsir},  Theorem~\ref{thm:zero-dispersion-nocsit}  requires more stringent conditions on the fading distribution compared to the CSIT case (cf., Theorem~\ref{thm:asy-mimo-csirt}), and its proof is more involved.
\begin{thm}\label{thm:zero-dispersion-nocsit}
    Let $\pdfH$ be the pdf of the fading matrix $\randmatH$.
    Assume that $\randmatH$ satisfies the following conditions:
    \begin{enumerate}
      \item \label{item:th-zd-no-cond1} $\pdfH$ is a smooth function, i.e., it has derivatives of all orders.
      \item \label{item:th-zd-no-cond2} There exists a positive constant $a$ such that
\begin{IEEEeqnarray}{rCl}
\pdfH(\matH) &\leq& a  \fnorm{\matH}^{-2\txant\rxant -\lfloor(\rxant+1)^2/2\rfloor -1} \label{eq:cond-bdd-pdf}\\
\fnorm{\gradient \pdfH(\matH)} &\leq& a \fnorm{\matH}^{-2\txant\rxant -5}.\label{eq:cond-bdd-pdf-deri}
\end{IEEEeqnarray}
\item\label{item:th-zd-no-cond3} The function $\cdistno(\cdot)$ satisfies
\begin{IEEEeqnarray}{rCl}
\liminf\limits_{\delta\to 0} \frac{\cdistno(\Cnocsit + \delta) - \cdistno(\Cnocsit)}{\delta} >0 .
\label{eq:condition-nocsit-subgradient}
\end{IEEEeqnarray}
    \end{enumerate}
    Then,
    \begin{IEEEeqnarray}{rCl}
    \big\{ \Rnocsi^{\ast}(\bl,\error) , \Rcsir^{\ast}(\bl,\error) \big\} &=& \Cnocsit + \bigO\mathopen{}\left(\frac{\log\bl}{n}\right) \label{eq:thm-R-star-expansion-no}.
    \end{IEEEeqnarray}
    Hence, the $\error$-dispersion is zero for both the CSIR and the no-CSI case:
    \begin{IEEEeqnarray}{rCl}
    V_\error^{\nocsi} = V_\error^{\csir}=0, \quad\,\,\error\in(0,1)\backslash \{1/2\}.
    \end{IEEEeqnarray}
\end{thm}
\begin{IEEEproof}
See Appendices~\ref{app:proof-zero-dispersion-nocsit-ub} and \ref{app:proof-asy-ach-nocsit}.
\end{IEEEproof}

\begin{rem}\label{rem:conditions-nocsit}

It can be shown that Conditions~\ref{item:th-zd-no-cond1}--\ref{item:th-zd-no-cond3} in Theorem~\ref{thm:zero-dispersion-nocsit} are satisfied by the probability distributions commonly used to model MIMO fading channels, such as Rayleigh, Rician, and Nakagami.
Condition~\ref{item:th-zd-no-cond2}  requires simply that $f_{\randmatH}$ has a polynomially decaying tail.
Condition~\ref{item:th-zd-no-cond3} plays the same role as~\eqref{eq:thm-dispersion-condition2-conv} in the CSIT case.
The exact counterpart of~\eqref{eq:thm-dispersion-condition2-conv} for the no-CSIT case would~be
\begin{IEEEeqnarray}{rCl}
\cdistno'(\Cnocsit)>0.
\label{eq:condition-nocsit-gradient}
\end{IEEEeqnarray}
However, different from~\eqref{eq:thm-dispersion-condition2-conv}, the inequality~\eqref{eq:condition-nocsit-gradient} does not necessarily hold for the commonly used fading distributions.
Indeed, consider a MISO i.i.d. Rayleigh-fading channel.
As proven in~\cite{abbe13-05}, the $\error$-capacity-achieving covariance matrix for this case is given by~\eqref{eq:opt-cov-mat}.
The resulting outage probability function $\cdistno(\cdot)$ may not be differentiable at the rates $R$ for which the infimum in~\eqref{eq:P-out-alt-exp} is achieved by two input covariance matrices with different number of nonzero entries $t^*$ on the main diagonal.

Next, we briefly sketch how to prove that Condition~\ref{item:th-zd-no-cond3} holds for Rayleigh, Rician, and Nakagami distributions.
Let
\begin{equation}
F_{\matQ}(R) \define \prob[ \log\det\mathopen{}\left(\matI_{\rxant} + \herm{\randmatH} \matQ\randmatH \right) < R].
\label{eq:def-F-matQ}
\end{equation}
Let $\setQ_\error$ be the set of all $\error$-capacity-achieving covariance matrices, i.e.,
\begin{IEEEeqnarray}{rCl}
\setQ_{\error} \define \{\matQ\in \insetcove: F_{\matQ}(\Cnocsit) = \cdistno(\Cnocsit)\}.
\end{IEEEeqnarray}
By Proposition~\ref{prop:continuous}, the set~$\setQ_{\error}$ is non-empty for the considered fading distributions.
It follows from algebraic manipulations that
\begin{IEEEeqnarray}{rCl}
\liminf\limits_{\delta\to 0} \frac{\cdistno(\Cnocsit + \delta) - \cdistno(\Cnocsit)}{\delta} = \inf\limits_{\matQ\in \setQ_\error} F_{\matQ}' (\Cnocsit). \IEEEeqnarraynumspace
\label{eq:lb-subgradient}
\end{IEEEeqnarray}
To show that the RHS of~\eqref{eq:lb-subgradient} is positive, one needs to perform two steps.
First, one shows that the set~$\setQ_\error$ is compact with respect to the metric $d(\matA,\matB) = \fnorm{\matA-\matB}$ and that under Conditions~\ref{item:th-zd-no-cond1} and~\ref{item:th-zd-no-cond2} of Theorem~\ref{thm:zero-dispersion-nocsit}, the function $\matQ\mapsto F_{\matQ}'(\Cnocsit)$ is continuous with respect to the same metric.
By the extreme value theorem~\cite[p.~34]{munkres91-a}, these two properties imply that the infimum on the RHS of~\eqref{eq:lb-subgradient} is a minimum.
Then, one shows that for Rayleigh, Rician, and Nakagami distributions
\begin{IEEEeqnarray}{rCl}
F_{\matQ}'(\Cnocsit) > 0 , \quad \forall \matQ \in \setQ_\error.
\label{eq:condition-f-Q-pos}
\end{IEEEeqnarray}
One way to prove~\eqref{eq:condition-f-Q-pos} is to write $F_{\matQ}'(\Cnocsit)$ in integral form using Lemma~\ref{lem:formula-Stokes} in Appendix~\ref{app:proof-uniform-bdd-pdfu} and to show that the resulting integral is positive.

%
%

\end{rem}

For the SIMO case, the conditions on the fading distribution can be relaxed and the following result holds. 
\begin{thm}
Assume that the pdf of $\|\randvech\|^2$ is continuously differentiable and that the $\error$-capacity~$\Csimo$ is a point of growth for the outage probability function
\begin{IEEEeqnarray}{rCL}
  F(R)=\prob[\log(1+\|\randvech\|^2\snr) <R]
\end{IEEEeqnarray}
i.e., $F'(\Csimo)>0$.
Then,
\begin{IEEEeqnarray}{rCl}
    \big\{ \Rnocsi^{\ast}(\bl,\error) , \Rcsir^{\ast}(\bl,\error) \big\} &=& \Csimo + \bigO\mathopen{}\left(\frac{\log\bl}{n}\right) \label{eq:thm-R-star-expansion-SIMO}.
\end{IEEEeqnarray}
\end{thm}
\begin{IEEEproof}
  In the SIMO case, CSIT is not beneficial~\cite[Th.~5]{yang13-07}.
  Hence, the result follows directly from Theorem~\ref{thm:asy-mimo-csirt} and Proposition~\ref{thm:proof-asy-ach-nocsit} in Appendix~\ref{app:proof-asy-ach-nocsit}.
\end{IEEEproof}

Similarly, for the case of codes consisting of isotropic codewords, milder conditions on the fading distribution are sufficient to establish zero dispersion, as illustrated in the following theorem.
\begin{thm}
\label{thm:proof-asy-iso}
Assume that the joint pdf of the nonzero eigenvalues of $\herm{\randmatH}\randmatH$ is continuously differentiable and that
\begin{IEEEeqnarray}{rCl}
\cdistiso' (\Ciso) >0
\label{eq:thm-disp-cond2-iso}
\end{IEEEeqnarray}
where $\cdistiso$ is the outage probability function given in~\eqref{eq:def-cdist-iso}.
Then, we have
\begin{IEEEeqnarray}{rCl}
\label{eq:asy-formula-iso}
\{\Rnoiso^{\ast}(\bl,\error), \Rcriso^{\ast}(\bl,\error)\} = \Ciso +\bigO\mathopen{}\left(\frac{\log\bl}{n}\right).
\end{IEEEeqnarray}
\end{thm}
\begin{IEEEproof}
See Appendix~\ref{app:proof-asy-iso}.
\end{IEEEproof}

\subsection{Normal Approximation}
For the general no-CSIT MIMO case, the unavailability of a closed-form expression for the $\error$-capacity $\Cnocsit$ in~\eqref{eq:C-epsilon-nocsit} prevents us from obtaining a normal approximation for the maximum coding rate at finite blocklength.
However, such an approximation can be obtained  for the SIMO case and for the case of isotropic codewords.
In both cases, CSIT is not beneficial and the outage capacity can be characterized in closed form.

For the SIMO case, the normal approximation follows directly from \eqref{eq:normal-dist}--\eqref{eq:dispersion_gaussian_approx} by setting $\minant=1$, $v_1^\ast=\rho$ and noting that $\lambda_1=\|\vech\|^2$.

For $(\bl,\NumCode,\error)_{\iso}$ codes, the normal approximation $\Rnormaliso(\bl,\error)$ to the  maximal achievable rate $\Rcriso^*(\bl,\error)$  is obtained as the solution of
\begin{IEEEeqnarray}{rCl}
 \epsilon &=& \Ex{}{Q\mathopen{}\left(\frac{ C_{\iso}(\randmatH)- \Rnormaliso(\bl,\error) }{\sqrt{V_{\iso}(\randmatH)/\bl} }\right) }.
\label{eq:normal-dist-iso}
\end{IEEEeqnarray}
Here,
  \begin{IEEEeqnarray}{rCl}
  C_{\iso}(\matH) = \sum\limits_{j=1}^{\minant} \log(1+ \snr\lambda_j/\txant)
  \end{IEEEeqnarray}
 and
  \begin{IEEEeqnarray}{rCl}
 V_{\iso}(\matH) = \minant - \sum\limits_{j=1}^{\minant} \frac{1}{(1+ \snr\lambda_j/\txant)^2}
  \end{IEEEeqnarray}
where~$\{\lambda_j\}$ are the eigenvalues of $\herm{\matH}\matH$.
A comparison between $\Rnormaliso(\bl,\error)$ and the bounds~\eqref{eq:lb-numcode-nocsi} and~\eqref{eq:thm-converse-rcsir-iso} is provided in the next section.
\section{Numerical Results}

\subsection{Numerical Results}
\label{sec:numerical-results}
\begin{figure}[t]
	\centering
		\includegraphics[scale=0.82]{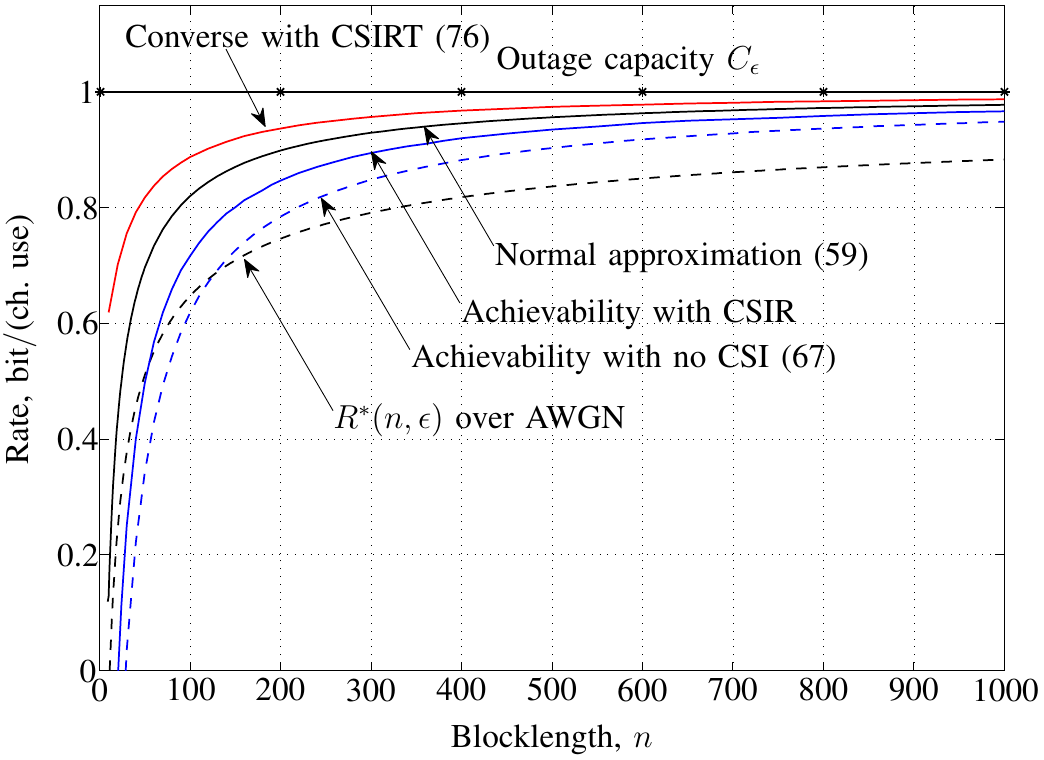}
\vspace{-1.5mm}
\caption{Achievability and converse bounds for a quasi-static SIMO Rician-fading channel with $K$-factor equal to $20$ dB, two receive antennas, $\text{SNR}=-1.55 $ dB, and \mbox{$\epsilon=10^{-3}$}. Note that in the SIMO case $\Ccsit=\Cnocsit=C_{\epsilon}$. \label{fig:bounds-simo}}
\end{figure}

In this section, we compute the bounds reported in~Sections~\ref{sec:mimo-csit} and~\ref{sec:mimo-nocsi}.
Fig.~\ref{fig:bounds-simo} compares $\Rnormalcsirt(\bl,\error)$ with the achievability bound~\eqref{eq:simo-ach-bound} and the converse bound (\ref{eq:thm-converse-rcsit-simo}) for a quasi-static SIMO fading channel with two receive antennas.
The channels between the transmit antenna and each of the two receive antennas are Rician-distributed with $K$-factor equal to $20$ dB.
The two channels are assumed to be independent.
We set $\error=10^{-3}$ and choose $\snr=-1.55$~dB so that $\Csimo =1$ bit$/$(ch. use).
We also plot a lower bound on $\Rcsirt^\ast(\bl,\error)$ obtained by using the~$\kappa\beta$ bound~\cite[Th.~25]{polyanskiy10-05} and assuming CSIR.\footnote{Specifically, we took $\inset = \{\vecx\in \complexset^{\bl} :  \|\vecx\|^2 = \bl\snr\}$, and $\outdist_{\randmatY\randvech} = \indist_{\randvech} \prod\nolimits_{j=1}^{\bl} Q_{\randvecy_j \given \randvech} $ where $Q_{\randvecy_j \given \randvech = \vech} = \jpg(\mathbf{0}, \matI_{\rxant} + \snr \vech \herm{\vech})$.}
%
For reference, Fig.~\ref{fig:bounds-simo} shows also the approximation~\eqref{eq:approx_R_introduction} for $R^\ast(\bl,\error)$ corresponding to an AWGN channel with $C =1$ bit$/$(ch. use),
 replacing the term $\bigO(\log(\bl)/\bl)$ in (\ref{eq:approx_R_introduction}) with $\log(\bl)/(2\bl)$~\cite[Eq.~(296)]{polyanskiy10-05}\cite{tan13-11}.\footnote{\label{footnote:complex-gaussian}The approximation reported in~\cite[Eq.~(296)]{polyanskiy10-05},\cite{tan13-11} holds for a real AWGN channel.
Since a complex AWGN channel with blocklength $\bl$ can be identified as a real AWGN channel with the same SNR and blocklength~$2\bl$,  the approximation~\cite[Eq.~(296)]{polyanskiy10-05},\cite{tan13-11} with $C=\log(1+\snr)$ and $V=\frac{\snr^2+2\snr}{(1+\snr)^2}$ is accurate for the complex case.}
The blocklength required to achieve $90\%$ of the $\error$-capacity of the quasi-static fading channel is in the range $[120,320]$ for the CSIRT case and in the range $[120,480]$ for the no-CSI case.
For the AWGN channel, this number is approximately~$1420$. Hence, for the parameters chosen in Fig.~\ref{fig:bounds-simo}, the prediction (based on zero dispersion) of fast convergence to capacity is validated.
The gap between the normal approximation $\Rnormalcsirt(\bl,\error)$ defined implicitly in~\eqref{eq:normal-dist} and both the achievability (CSIR) and the converse bounds is less than $0.02$ bit$/$(ch.~use) for blocklengths larger than $400$.

Note that although the AWGN curve in Fig.~\ref{fig:bounds-simo} lies below the achievability bound for the quasi-static fading channel, this does not mean that ``fading helps''.
In Fig.~\ref{fig:bounds-simo}, we chose the SNRs so that both channels have the same \error-capacity.
This results in the received power for the quasi-static case being $1.45$~dB larger than that for the AWGN case.

\begin{figure}[t]
	\centering
		\includegraphics[scale=0.82]{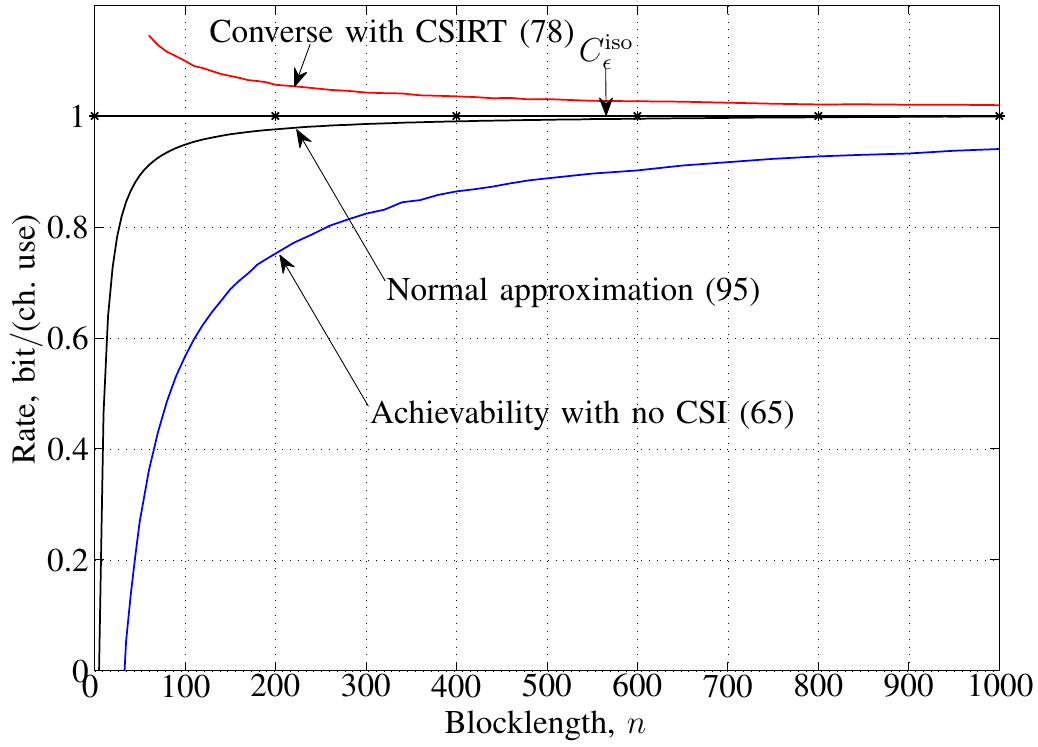}
\caption{Achievability and converse bounds for $(\bl,\NumCode,\error)_{\iso}$ codes over a quasi-static MIMO Rayleigh-fading channel with two transmit and three receive antennas, $\text{SNR}=2.12 $ dB, and \mbox{$\epsilon=10^{-3}$}. \label{fig:bounds-mimo}}
\end{figure}

In Fig.~\ref{fig:bounds-mimo}, we compare the normal approximation $\Rnormaliso(\bl,\error)$ defined (implicitly) in~\eqref{eq:normal-dist-iso} with the achievability bound~\eqref{eq:lb-numcode-nocsi} and the converse bound \eqref{eq:thm-converse-rcsir-iso} on the maximal achievable rate with $(\bl,\NumCode,\error)_{\iso}$ codes over a quasi-static MIMO fading channel with $\txant=2$ transmit and $\rxant=3$ receive antennas. The channel between each transmit-receive antenna pair is Rayleigh-distributed, and the channels between different transmit-receive antenna pairs are assumed to be independent.
We set $\error=10^{-3}$ and choose $\snr=2.12$~dB so that $\Ciso =1$ bit$/$(ch. use).
For this scenario, the blocklength required to achieve $90\%$ of  $\Ciso$ is less than $500$, which again demonstrates fast convergence to~$\Ciso$.

\subsection{Comparison with coding schemes in LTE-Advanced}

\begin{figure}[t]
	\centering
		\includegraphics[scale=0.8]{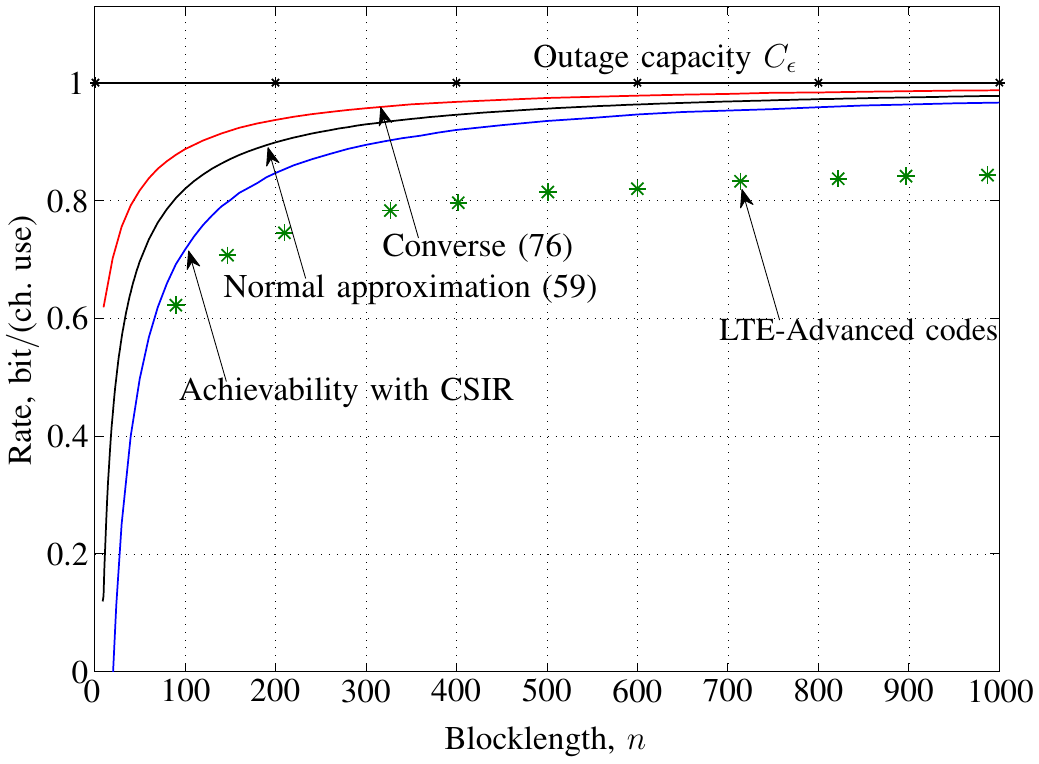}
\caption{Comparison between achievability and converse bounds and the rate achievable with the coding schemes in LTE-Advanced.
 We consider a quasi-static SIMO Rician-fading channel with $K$-factor equal to $20$ dB, two receive antennas, $\text{SNR}=-1.55 $ dB, \mbox{$\epsilon=10^{-3}$}, and CSIR.
The star-shaped markers indicate the rates achievable by the turbo codes in LTE-Advanced (QPSK modulation and 10 iterations of a max-log-MAP decoder~\cite{robertson95}).
\label{fig:codes-simo}}
\end{figure}

The bounds reported in Sections~\ref{sec:mimo-csit} and~\ref{sec:mimo-nocsi} can be used to benchmark the coding schemes adopted in current standards.
In Fig.~\ref{fig:codes-simo}, we compare the performance of the coding schemes used in LTE-Advanced~\cite[Sec.~5.1.3.2]{3gpp-ts36212} against the achievability and converse bounds for the same scenario as in Fig.~\ref{fig:bounds-simo}.
 Specifically, Fig.~\ref{fig:codes-simo} illustrates  the performance of the family of turbo codes chosen in LTE-Advanced for the case of QPSK modulation.
 The decoder employs a max-log-MAP decoding algorithm~\cite{robertson95} with 10 iterations.
We further assume that the decoder has perfect CSI.
For the AWGN case, it was observed in~\cite[Fig.~12]{polyanskiy10-05} that about half of the gap between the rate achieved by
the best available channel codes\footnote{The codes used in~\cite[Fig.~12]{polyanskiy10-05} are a certain family of multiedge low-density parity-check (LDPC) codes.}
and capacity is due to the $1/\sqrt{\bl}$ penalty in~\eqref{eq:approx_R_introduction}; the other half is due to the suboptimality of the codes.
From Fig.~\ref{fig:codes-simo}, we conclude that for quasi-static fading channels the finite-blocklength penalty is significantly reduced because of the zero-dispersion effect. However, the penalty due to the code suboptimality remains. In fact, we see that the gap between the rate achieved by the LTE-Advanced turbo codes and the normal approximation $\Rnormalcsirt(\bl,\error)$ is approximately constant up to a blocklength of $1000$.

\begin{figure}[t]
	\centering
		\includegraphics[scale=0.8]{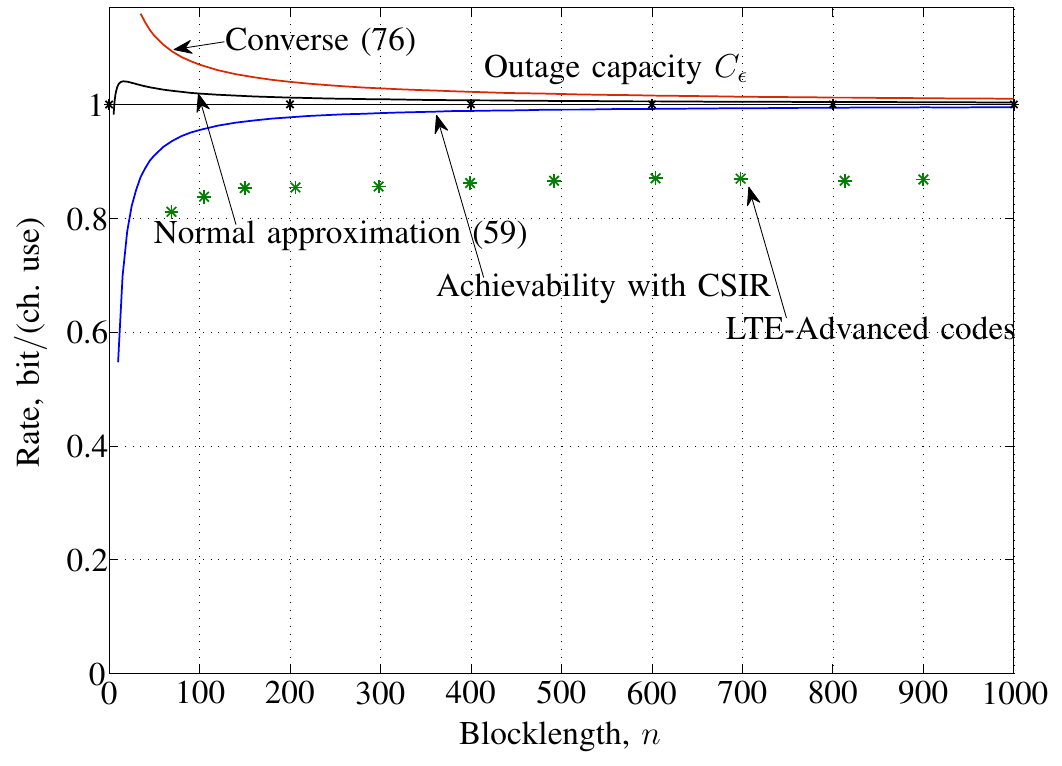}
\caption{Comparison between achievability and converse bounds and rate achievable with the coding schemes in LTE-Advanced.
 We consider a quasi-static SIMO Rayleigh-fading channel with two receive antennas, $\text{SNR} = 2.74$ dB, \mbox{$\epsilon = 0.1$}, and CSIR.
The star-shaped markers indicate the rates achievable by the turbo codes in LTE-Advanced (QPSK modulation and 10 iterations of a max-log-MAP decoder~\cite{robertson95}).
\label{fig:codes-rayleigh}}
\vspace{-2mm}
\end{figure}

LTE-Advanced uses hybrid automatic repeat request (HARQ)  to compensate for packets loss due to outage events.
When HARQ is used, the block error rate that maximizes the average throughput is about $10^{-1}$~\cite[p.~218]{sesia11}.
The performance of LTE-Advanced codes for $\error =10^{-1}$ is analyzed in Fig.~\ref{fig:codes-rayleigh}.
We set $\snr=2.74$~dB and consider Rayleigh fading (the other parameters are as in Fig.~\ref{fig:codes-simo}).
Again, we observe that there is a constant gap between the rate achieved by LTE-Advanced turbo codes and $\Rnormalcsirt(\bl,\error)$.

\section{Conclusion}
In this paper, we established achievability and converse bounds on the maximal achievable rate $R^\ast(\bl,\error)$ for a given blocklength $\bl$ and error probability $\error$ over quasi-static MIMO fading channels.
We proved that (under some mild conditions on the fading distribution) the channel dispersion is zero for all four cases of CSI availability.
The bounds are easy to evaluate
when CSIT is available, when the number of transmit antennas is one, or when the code has isotropic codewords.
In all these cases the outage-capacity-achieving distribution is known.

The numerical results reported in Section~\ref{sec:numerical-results} demonstrate that, in some scenarios, zero dispersion implies fast convergence to $C_\error$ as the blocklength increases.
This suggests that the outage capacity is a valid performance metric for communication systems with stringent latency constraints operating over quasi-static fading channels.
We developed an easy-to-evaluate approximation of $R^\ast(\bl,\error)$ and demonstrated its accuracy by comparison to our achievability and converse bounds.
Finally, we used our bounds to benchmark the performance of the coding schemes adopted in the LTE-Advanced standard.
Specifically, we showed that for a blocklength between $500$ and $1000$ LTE-Advanced codes achieve about $85\%$ of the maximal coding rate.

\begin{appendices}

\section{Auxiliary Lemmas Concerning the Product of Sines of Principal Angles}
\label{app:proof-angle-btw-subspaces}
In this appendix, we state two properties of the product of principal sines defined in~\eqref{eq:def-product-prin-sine}, which will be used in the proof of~Theorem~\ref{thm:asy-mimo-csirt} and of Proposition~\ref{thm:proof-asy-ach-nocsit}.
The first property, which is referred to in \cite{miao92} as ``equalized Hadamard inequality'', is stated in Lemma~\ref{lem:equalize-Hadamard} below.
\begin{lemma}
\label{lem:equalize-Hadamard}
Let $\matA=[\matA_1,\matA_2]\in\complexset^{\bl\times(a_1+a_2)}$, where $\matA_1\in\complexset^{\bl\times a_1}$ and $\matA_2\in\complexset^{\bl\times a_2}$. If $\rank(\matA_1)=a_1$ and $\rank(\matA_2)=a_2$, then
\begin{IEEEeqnarray}{rCl}
\det(\herm{\matA}\matA) =\det(\herm{\matA_1}\matA_1)\det(\herm{\matA_2}\matA_2) \sin^2\{\matA_1,\matA_2\}.
\label{eq:lemma-equal-hadamard}
\end{IEEEeqnarray}
\end{lemma}

\begin{IEEEproof}
The proof follows by extending~\cite[Th.~3.3]{afriat57} to the complex case.
\end{IEEEproof}

The second property provides an upper bound on $\sin\{\setA,\setB\}$ that depends on the angles between the basis vectors of the two subspaces.
\begin{lemma}
\label{lem:angle-btw-subspaces}
Let $\setA$ and $\setB$ be subspaces of $\complexset^{\bl}$ with $\dim(\setA)=a$ and $\dim(\setB) =b$.
Let $\{\veca_1,\ldots,\veca_{a}\}$ be an orthonormal basis for $\setA$, and let $\{\vecb_1,\ldots,\vecb_{b}\}$ be an arbitrary basis (not necessarily orthonormal) for $\setB$. Then
\begin{IEEEeqnarray}{rCl}
\label{eq:lemma-angle}
\sin\{\setA,\setB\} \leq \prod\limits^{\min\{a,b\}}_{j=1} \sin \{\veca_j,\vecb_j\}.
\end{IEEEeqnarray}
\end{lemma}
\begin{IEEEproof}
To keep notation simple, we define the following function, which maps a complex matrix $\matX$ of arbitrary size to its volume:
\begin{IEEEeqnarray}{rCl}
\vol(\matX) \define \sqrt{\det(\herm{\matX}\matX)}.
\end{IEEEeqnarray}
Let $\matA=[\veca_1,\ldots,\veca_{a}]\in\complexset^{\bl\times a}$ and $\matB = [\vecb_1,\ldots,\vecb_{b}]\in\complexset^{\bl\times b}$.
%
If the vectors $\veca_1,\ldots,\veca_{a},\vecb_1,\ldots,\vecb_{b}$ are linearly dependent, then the LHS of~\eqref{eq:lemma-angle} vanishes, in which case~\eqref{eq:lemma-angle} holds trivially.
In the following, we therefore assume that the vectors $\veca_1,\ldots,\veca_{a},\vecb_1,\ldots,\vecb_{b}$ form a linearly independent set.
Below, we prove Lemma~\ref{lem:angle-btw-subspaces} for the case $a\leq b$.
The proof for the case $a>b$ follows from similar steps.

Using Lemma~\ref{lem:equalize-Hadamard}, we get the following chain of (in)equalities:
\begin{IEEEeqnarray}{rCl}
\IEEEeqnarraymulticol{3}{l}{
\sin\{\matA,\matB\}}\notag\\
&=&\frac{\vol([\matA,\matB])}{\vol(\matA)\vol(\matB)} \\
&=& \frac{\vol([\matA,\matB])}{\vol(\matB)} \label{eq:app-comput-volAB-1}\\
&=& \frac{1}{\vol(\matB)} \underbrace{\|\veca_1\|}_{=1} \vol\mathopen{}\big([\veca_2,\ldots,\veca_a,\matB]\big) \notag \\
 &&\cdot\, \sin\mathopen{}\big\{\veca_1, [\veca_{2},\ldots,\veca_a, \matB]\big\}\label{eq:app-comput-volAB-20}\\
&\vdots&\notag\\
&=& \frac{1}{\vol(\matB)} \left(\prod \limits_{i=1}^{a}  \sin\mathopen{}\big\{\veca_i, [\veca_{i+1},\ldots,\veca_a, \matB]\big\}\!\right)  \vol(\matB)\label{eq:app-comput-volAB-2} \quad \\
&\leq & \prod \limits_{i=1}^{a} \sin\{\veca_i,\vecb_i\}\label{eq:app-comput-volAB-3}.
\end{IEEEeqnarray}
Here,~\eqref{eq:app-comput-volAB-1} holds because the columns of $\matA$ are orthonormal and, hence, $\det(\herm{\matA}\matA) =1$;~\eqref{eq:app-comput-volAB-20} and \eqref{eq:app-comput-volAB-2} follow from Lemma~\ref{lem:equalize-Hadamard}; \eqref{eq:app-comput-volAB-3} follows because
\begin{IEEEeqnarray}{rCL}
 \sin\mathopen{}\big\{\veca_i, [\veca_{i+1},\ldots,\veca_a, \matB]\big\} \leq \sin\{\veca_i,\vecb_i\}.
\end{IEEEeqnarray}
\end{IEEEproof}

\section{Proof of Theorem~\ref{thm:actual_ach_bound_csit} (CSIT Achievability Bound)}
\label{app:proof-ach-bound-csit}

Given $\randmatH =\matH$, we perform a singular value decomposition (SVD) of $\matH$ to obtain
\begin{IEEEeqnarray}{rCl}
\matH = \matL \matSigma \herm{\matV}
\label{eq:ach-csit-svd}
\end{IEEEeqnarray}
where $\matL \in \complexset^{\txant\times\txant}$ and $\matV\in\complexset^{\rxant\times\rxant}$ are unitary matrices, and $\matSigma\in\complexset^{\txant\times\rxant}$ is a (truncated) diagonal matrix of dimension $\txant\times\rxant$, whose diagonal elements $\sqrt{\lambda_1},\ldots,\sqrt{\lambda_{\minant}}$, are the ordered singular values of $\matH$.
It will be convenient to define the following $\txant\times\txant$ precoding matrix for each~$\matH$:
\begin{IEEEeqnarray}{rCl}
\label{eq:precoding-matrix}
\matP(\matH) \define \diag\{\sqrt{\bl v^\ast_1},\ldots,\sqrt{\bl v^\ast_{\minant}},\underbrace{0,\ldots,0}_{\txant -\minant}\} \herm{\matL}.
\end{IEEEeqnarray}
%

We consider a code whose codewords $\matX_{j}(\randmatH)$, $j = 1,\ldots,\NumCode$, have the following structure
\begin{IEEEeqnarray}{rCl}
\matX_{j}(\randmatH) = \mathsf{\Phi}_j \matP(\randmatH), \quad \mathsf{\Phi}_j\in \setS_{\bl,\txant}
\end{IEEEeqnarray}
where $\setS_{\bl,\txant}\define \{\matA\in\complexset^{\bl\times\txant}: \herm{\matA}\matA = \matI_{\txant}\}$ denotes the set of all $\bl\times\txant$ unitary matrices, (i.e., the complex \emph{Stiefel manifold}). 
 As $\{\mathsf{\Phi}_j\}$ are unitary, the codewords satisfy the power constraint~\eqref{eq:power-constraint-csit}.
 Motivated by the geometric considerations reported in Section~\ref{sec:geometric-intution}, we consider for a given input $\matX(\randmatH) = \mathsf{\Phi} \matP(\randmatH)$ a physically degraded version of the channel~\eqref{eq:channel_io}, whose output is given by
\begin{IEEEeqnarray}{rCl}
\label{eq:channel-io-degraded}
\Omega_{\randmatY} = \spanm(\mathsf{\Phi}\matP(\randmatH) \randmatH + \randmatW).
\end{IEEEeqnarray}
Note that the subspace $\Omega_{\randmatY}$ belongs with probability one to the \emph{Grassmannian manifold} $\setG_{\bl,\rxant}$, i.e., the set of all $\rxant$ dimensional subspaces in $\complexset^{\bl}$.
Because~\eqref{eq:channel-io-degraded} is a physically degraded version of~\eqref{eq:channel_io}, the rate achievable on~\eqref{eq:channel-io-degraded} is a lower bound on the rate achievable on~\eqref{eq:channel_io}.

To prove the theorem, we apply the $\kappa\beta$ bound~\cite[Th.~25]{polyanskiy10-05} to the channel~\eqref{eq:channel-io-degraded}.
Following~\cite[Eq.~(107)]{polyanskiy10-05}, we define the following measure of performance for the composite hypothesis test between an \emph{auxiliary} output distribution $Q_{\Omega_{\randmatY}}$ defined on the subspace $\Omega_{\randmatY}$ and the collection of channel-output distributions $\{P_{\Omega_{\randmatY} \given \mathbb{\Phi}=\mathsf{\Phi}}\}_{\mathsf{\Phi}\in\setS_{\bl,\txant}}$:
\begin{IEEEeqnarray}{rCl}
\kappa_{\tau}(\setS_{\bl,\txant}, Q_{\Omega_{\randmatY}} ) \define \inf \int \testdist_{Z\given \Omega_{\randmatY}}(1\given  \Omega_{\matY})  \outdist_{\Omega_{\randmatY}}(d \Omega_{\matY}) \IEEEeqnarraynumspace
\label{eq:def-kappa-tau}
\end{IEEEeqnarray}
where the infimum is over all probability distributions $\testdist_{Z\given \Omega_{\randmatY}}: \setG_{\bl,\txant} \mapsto \{0,1\}$ satisfying
\begin{IEEEeqnarray}{rCl}
\, \int \testdist_{Z\given \Omega_{\randmatY}}(1\given  \Omega_{\matY})   P_{\Omega_{\randmatY} \given \mathbb{\Phi}=\mathsf{\Phi}}(d \Omega_{\matY})\geq \tau,\quad \forall \mathsf{\Phi}\in\setS_{\bl,\txant}. \IEEEeqnarraynumspace
\label{eq:cond-def-kappa}
\end{IEEEeqnarray}
By~\cite[Th.~25]{polyanskiy10-05}, we have that for every auxiliary distribution $Q_{\Omega_{\randmatY}} $
\begin{IEEEeqnarray}{rCl}
\NumCode \geq \frac{\kappa_{\tau}(\setS_{\bl,\txant}, Q_{\Omega_{\randmatY}})}{\sup_{\mathsf{\Phi}\in \setS_{\bl,\txant} }\beta_{1-\error+\tau}(\indist_{\Omega_{\randmatY}\given\mathbb{\Phi}=\mathsf{\Phi}}, \outdist_{\Omega_{\randmatY}}) }
\label{eq:kappa-beta-bound-subspace}
\end{IEEEeqnarray}
where $\beta_{(\cdot)}(\cdot,\cdot)$ is defined in~\eqref{eq:def-beta}.
We next lower-bound the RHS of~\eqref{eq:kappa-beta-bound-subspace} to obtain an expression that can be evaluated numerically.
Fix a $\mathsf{\Phi} \in\setS_{\bl,\txant}$ and let
\begin{IEEEeqnarray}{rCl}
\label{eq:threshold_tester_Z}
\angletest_{\mathsf{\Phi}}(\Omega_{\randmatY}  ) &=& \indfun{\sin^2\{ \spanm(\mathsf{\Phi}), \Omega_{\randmatY}\}\leq \gamma_\bl}
\end{IEEEeqnarray}
where $\gamma_\bl\in[0,1]$ is chosen so that
\begin{IEEEeqnarray}{rCl}
\label{eq:definition_gamma_x}
 \indist_{\Omega_{\randmatY} | \mathbb{\Phi} = \mathsf{\Phi}}[\angletest_{\mathsf{\Phi}}(\Omega_{\randmatY}  ) =1] \geq 1-\error+\tau.
\end{IEEEeqnarray}
 Since the noise matrix $\randmatW$ is isotropically distributed, the probability distribution of the random variable $\sin^2\{ \spanm(\mathsf{\Phi}), \Omega_{\randmatY}\}$ (where $\Omega_{\randmatY} \sim \indist_{\Omega_{\randmatY} | \mathbb{\Phi} = \mathsf{\Phi}}$) does not depend on $\mathsf{\Phi}$.
Hence, the chosen $\gamma_{\bl}$ satisfies~\eqref{eq:definition_gamma_x} for all $\mathsf{\Phi}\in\setS_{\bl,\txant}$.
Furthermore, $\angletest_{\mathsf{\Phi}}(\Omega_{\randmatY}  )$ can be viewed as a hypothesis test between $\indist_{\Omega_{\randmatY}\given\mathbb{\Phi}=\mathsf{\Phi}}$ and $ \outdist_{\Omega_{\randmatY}}$. Hence, by definition
\begin{IEEEeqnarray}{rCl}
\beta_{1-\error+\tau}(\indist_{\Omega_{\randmatY}\given\mathbb{\Phi}=\mathsf{\Phi}}, \outdist_{\Omega_{\randmatY}}) \leq \outdist_{\Omega_{\randmatY}}[ \angletest_{\mathsf{\Phi}}( \Omega_{\randmatY} ) =1 ]
\label{eq:csit-ach-beta-ub}
\end{IEEEeqnarray}
for every $\mathsf{\Phi} \in \setS_{\bl, \txant}$.

We next evaluate the RHS of~\eqref{eq:csit-ach-beta-ub}, taking as the auxiliary output distribution the uniform distribution on $\setG_{\bl,\rxant}$, which we denote by~$Q^{\mathrm{u}}_{\Omega_{\randmatY}}$.
%
With this choice, $Q^{\mathrm{u}}_{\Omega_{\randmatY}}[ \sin^2\{\spanm(\mathsf{\Phi}),\Omega_{\randmatY}\} \leq \gamma_\bl]$ does not depend on $\mathsf{\Phi}\in\setS_{\bl,\txant}$.
To simplify calculations, we can therefore set $\mathsf{\Phi} =\matI_{\bl,\txant}$.
Observe that under $Q^{\mathrm{u}}_{\Omega_{\randmatY}}$, the squares of the sines of the principle angles between $\spanm(\matI_{\bl,\txant})$ and~$\Omega_{\randmatY}$ have the same distribution as the eigenvalues of a \emph{complex}  multivariate Beta-distributed matrix $\randmatB \sim \mathrm{Beta}_{\rxant}(\bl-\txant,\txant)$~\cite[Sec.~2]{absil06}. By~\cite[Cor.~1]{roh06}, the distribution of $\det \randmatB$ coincides with the distribution of $\prod_{i=1}^{\rxant} B_i$, where $\{B_i\}$, ${i=1,\ldots,\rxant}$, are independent with $B_i\sim \mathrm{Beta}(\bl-\txant-i+1, \txant)$.
Using this result to compute the RHS of~\eqref{eq:csit-ach-beta-ub} we obtain
\begin{IEEEeqnarray}{rCl}
\label{eq:upper_bound_on_beta_part}
\sup_{\mathsf{\Phi}\in \setS_{\bl,\txant} }\beta_{1-\error+\tau}(\indist_{\Omega_{\randmatY}\given\mathbb{\Phi}=\mathsf{\Phi}}, \outdist_{\Omega_{\randmatY}}) \leq
\prob\mathopen{}\left[\prod_{j=1}^{\rxant} B_j \leq \gamma_{\bl}\right]\IEEEeqnarraynumspace
\end{IEEEeqnarray}
 where $\gamma_{\bl}$ satisfies
  \begin{IEEEeqnarray}{rCl}
\label{eq:def-gamma-n-ach-csit-temp}
  \prob\mathopen{}\Big[\sin^2 \mathopen{}\big\{\matI_{\bl,\txant},\matI_{\bl,\txant}\matP(\randmatH)\randmatH + \randmatW \big\} \leq \gamma_\bl\Big] \geq 1-\error+\tau. \IEEEeqnarraynumspace
 \end{IEEEeqnarray}
Note that~\eqref{eq:def-gamma-n-ach-csit-temp} is equivalent to~\eqref{eq:def-gamma-n-ach-csit}. Indeed
 \begin{IEEEeqnarray}{rCl}
\IEEEeqnarraymulticol{3}{l}{
\prob\mathopen{}\left[\sin^2\mathopen{}\Big\{\matI_{\bl,\txant},\sqrt{\bl}\matI_{\bl,\txant}\matP(\randmatH)\randmatH + \randmatW \Big\}\leq \gamma_\bl\right]} \notag\\
 &=&\prob\mathopen{}\Big[ \sin^2\mathopen{}\Big\{\matI_{\bl,\txant},\sqrt{\bl}\matI_{\bl,\txant}
 \diag\mathopen{}\big\{\sqrt{v_1^{\ast} \Lambda_1},\ldots,\sqrt{v_\minant^{\ast}\Lambda_{\minant}},\quad\quad\quad\notag\\
 &&\hfill \underbrace{0,\ldots,0}_{\txant-\minant} \big\}\herm{\randmatV}  + \randmatW \Big\} \leq \gamma_\bl \Big]\label{eq:ach-csit-angle0}\IEEEeqnarraynumspace\\
 &=& \prob\mathopen{}\Big[ \sin^2\mathopen{}\Big\{\matI_{\bl,\txant}, \sqrt{\bl}\matI_{\bl,\txant}
 \diag\mathopen{}\big\{\sqrt{v_1^{\ast}  \Lambda_1},\ldots,\sqrt{v_\minant^{\ast} \Lambda_{\minant}},\notag\\
 &&\hfill \underbrace{0,\ldots,0}_{\txant-\minant} \big\} + \randmatW\randmatV \Big\}\leq \gamma_\bl \Big]\label{eq:ach-csit-angle1} \IEEEeqnarraynumspace\\
  &=&\prob\mathopen{}\Big[ \sin^2\mathopen{}\Big\{\matI_{\bl,\txant}, \sqrt{\bl}\matI_{\bl,\txant}
 \diag\mathopen{}\big\{\sqrt{v_1^{\ast}  \Lambda_1},\ldots,\sqrt{v_\minant^{\ast}\Lambda_{\minant}},\notag\\
 &&\hfill \underbrace{0,\ldots,0}_{\txant-\minant} \big\} + \randmatW\Big\}\leq \gamma_\bl \Big]\label{eq:ach-csit-angle2}\IEEEeqnarraynumspace
 \end{IEEEeqnarray}
 where~$\randmatV$ contains the right singular vectors of~$\randmatH$ (see~\eqref{eq:ach-csit-svd}).
  Here,~\eqref{eq:ach-csit-angle0} follows from~\eqref{eq:precoding-matrix};
  \eqref{eq:ach-csit-angle1} follows because right-multiplying a matrix $\matA$ by a unitary matrix does not change the subspace spanned by the columns of $\matA$ and hence, it does not change $\sin\{\cdot,\cdot\}$;
    \eqref{eq:ach-csit-angle2} follows because~$\randmatW$ is isotropically distributed and hence $\randmatW\randmatV$ has the same distribution as $\randmatW$.

To conclude the proof, it remains to show that
\begin{IEEEeqnarray}{rCl}
\kappa_{\tau}(\setS_{\bl,\txant}, Q^{\mathrm{u}}_{\Omega_{\randmatY}}) \geq \tau.\label{eq:u-bound-kappa-csit}
\end{IEEEeqnarray}
Once this is done, the desired lower bound~\eqref{eq:lb-numcode-csit} follows by using the inequality~\eqref{eq:upper_bound_on_beta_part} and~\eqref{eq:u-bound-kappa-csit} in~\eqref{eq:kappa-beta-bound-subspace}, by taking the logarithm of both sides of~\eqref{eq:kappa-beta-bound-subspace}, and by dividing by the blocklength $\bl$.

To prove~\eqref{eq:u-bound-kappa-csit}, we replace~\eqref{eq:cond-def-kappa} with the less stringent constraint that
\begin{IEEEeqnarray}{rCl}
\label{eq:refined-kappa-beta-constraint1}
\Ex{\indist_{\mathbb{\Phi}}^{\mathrm{u}}}{  \int \testdist_{Z\given \Omega_{\randmatY}}(1\given  \Omega_{\matY})   P_{\Omega_{\randmatY}\given \mathbb{\Phi}}(d \Omega_{\matY}) }  \geq \tau
\end{IEEEeqnarray}
where $\indist_{\mathbb{\Phi}}^{\mathrm{u}}$ is the uniform input distribution on $\setS_{\bl,\txant}$.
Since replacing~\eqref{eq:cond-def-kappa} by~\eqref{eq:refined-kappa-beta-constraint1} enlarges the feasible region of the minimization problem~\eqref{eq:def-kappa-tau}, we obtain an infimum in~\eqref{eq:def-kappa-tau} (denoted by $\kappa^{\mathrm{u}}_\tau(\setS_{\bl,\txant}, Q^{\mathrm{u}}_{\Omega_{\randmatY}})$) that is no larger than $\kappa_{\tau}(\setS_{\bl,\txant}, Q^{\mathrm{u}}_{\Omega_{\randmatY}})$.
The key observation is that the uniform distribution $\indist_{\mathbb{\Phi}}^{\mathrm{u}}$ induces an isotropic distribution on~$\randmatY$.
This implies that the induced distribution on~$\Omega_{\randmatY}$ is the uniform distribution on~$\setG_{\bl,\rxant}$, i.e., $\outdist^{\mathrm{u}}_{\Omega_{\randmatY}}$.
 Therefore, it follows that
\begin{IEEEeqnarray}{rCl}
\IEEEeqnarraymulticol{3}{l}{\int \testdist_{Z\given \Omega_{\randmatY}}(1\given  \Omega_{\matY})  \outdist^{\mathrm{u}}_{\Omega_{\randmatY}}(d \Omega_{\matY})
}\notag\\
\quad  &=& \Ex{\indist_{\mathbb{\Phi}}^{\mathrm{u}}}{  \int \testdist_{Z\given \Omega_{\randmatY}}(1\given  \Omega_{\matY})   P_{\Omega_{\randmatY}\given \mathbb{\Phi}}(d \Omega_{\matY}) }\\
  &\geq& \tau
\end{IEEEeqnarray}
for all distributions $\indist_{Z\given \Omega_{\randmatY}}$ that satisfy (\ref{eq:refined-kappa-beta-constraint1}).
This proves~\eqref{eq:u-bound-kappa-csit}, since
\begin{IEEEeqnarray}{rCl}
\label{eq:kappa-tau-value}
\kappa_{\tau}(\setS_{\bl,\txant}, Q^{\mathrm{u}}_{\Omega_{\randmatY}}) \geq \kappa^{\mathrm{u}}_\tau(\setS_{\bl,\txant}, Q^{\mathrm{u}}_{\Omega_{\randmatY}}) \geq \tau.
\end{IEEEeqnarray}
%

\section{Proof of Theorem~\ref{thm:converse-csirt} (CSIRT Converse Bound)}
\label{app:proof-converse-cisrt}

When CSI is available at both the transmitter and the receiver, the MIMO channel~\eqref{eq:channel_io} can be transformed into the following set of $\minant$ parallel quasi-static channels
\begin{IEEEeqnarray}{rCl}
\label{eq:channel-io-para}
\randvecy_i = \vecx_i \sqrt{\Lambda_i} +  \randvecw_i,\quad\quad i=1,\ldots,\minant
\end{IEEEeqnarray}
by performing a singular value decomposition~\cite[Sec.~3.1]{telatar99-11a}.
Here, $\Lambda_1\geq \cdots\geq\Lambda_\minant$ denote the $\minant$ largest eigenvalues of~$\randmatH\herm{\randmatH}$, and $\randvecw_i\sim\jpg(\mathbf{0},\matI_\bl)$, $i=1,\ldots,\minant$, are independent noise vectors.

%

Next, we establish a converse bound for the channel~\eqref{eq:channel-io-para}. Let $\matX = [\vecx_1 \cdots \vecx_{\minant}]$ and fix an $(\bl,\NumCode,\error)_{\csirt}$ code.
Note that~\eqref{eq:power-constraint-csit} implies
\begin{IEEEeqnarray}{rCl}
\label{eq:power-constraint-csirt-para}
\sum\limits_{i=1}^{\minant} \|\vecx_i\|^2\leq \bl\snr.
\end{IEEEeqnarray}
To simplify the presentation, we assume that the encoder $\encoder_{\csit}$ is deterministic.
Nevertheless, the theorem holds also if we allow for randomized encoders.
We further assume that the encoder~$\encoder_{\csit}$ acts on the pairs $(j, \bm{\lambda})$ instead of $(j,\matH)$ (cf., Definition~\ref{def:csit-code}).
The channel~\eqref{eq:channel-io-para} and the encoder~$\encoder_{\csit}$ define a random transformation $\indist_{\randmatY,\bm{\Lambda}\given\msg}$ from the message set $\{1,\ldots,\NumCode\}$ to the space $\complexset^{\bl\times\minant} \times \posrealset^{\minant}$:
\begin{IEEEeqnarray}{rCl}
 \indist_{\randmatY,\bm{\Lambda}\given\msg}
&=&\indist_{\bm{\Lambda}} \indist_{\randmatY\given\bm{\Lambda},\msg}\label{eq:converse-deri-indist2}
\end{IEEEeqnarray}
where $\randmatY=[\randvecy_1,\ldots,\randvecy_m]$ and
\begin{IEEEeqnarray}{rCl}
\indist_{\randmatY\given\bm{\Lambda}=\bm{\lambda},\msg =j}\define  \indist_{\randmatY\given\bm{\Lambda}=\bm{\lambda} , \randmatX=\encoder_{\csit}(j,\bm{\lambda}) }.
\end{IEEEeqnarray}
We can think of $\indist_{\randmatY,\bm{\Lambda}\given\msg}$ as the channel law associated with
\begin{equation}
\label{eq:channel-io-new-csirt}
\msg \, {\longrightarrow} \,\randmatY,\bm{\Lambda}.
\end{equation}
To upper-bound $\Rcsirt^\ast (\bl,\error)$, we use the meta-converse theorem~\cite[Th.~30]{polyanskiy10-05} on the channel~\eqref{eq:channel-io-new-csirt}.
We start by associating to each codeword $\matX$ a power-allocation vector $\tilde{\powallocvec}(\matX)$ whose entries $\tilde{\powalloc}_i(\matX)$ are
\begin{IEEEeqnarray}{rCl}
 \tilde{\powalloc}_i(\matX) \define \frac{1}{\bl} \|\vecx_i\|^2,\quad i=1,\ldots,\minant.
\end{IEEEeqnarray}
We take as auxiliary channel $\outdist_{\randmatY , \bm{\Lambda}\given\msg}=\indist_{\bm{\Lambda}}\outdist_{\randmatY\given \bm{\Lambda}, \msg}$,
where
\begin{IEEEeqnarray}{rCl}
\label{eq:def-q-channel-mimo-csirt}
\outdist_{\randmatY\given \bm{\Lambda} = \bm{\lambda}, \msg  =  j} = \prod\limits_{i=1}^{\minant} \outdist_{\randvecy_i\given \bm{\Lambda} = \bm{\lambda}, \msg  =  j}
\end{IEEEeqnarray}
and
\begin{IEEEeqnarray}{rCL}
  \outdist_{\randvecy_i\given \bm{\Lambda} = \bm{\lambda}, \msg  =  j}=\jpg\mathopen{}\Big( \veczero , \bigl[1+ (\tilde{v}_i\circ\encoder_\csit(j,\bm{\lambda})) \lambda_i\bigr] \matI_{\bl}\Big). \IEEEeqnarraynumspace
\end{IEEEeqnarray}
By~\cite[Th.~30]{polyanskiy10-05}, we obtain
\begin{IEEEeqnarray}{rCl}
\label{eq:meta-converse-cisrt}
\min_{j\in\{1,\ldots,\NumCode\}} \beta_{1-\error}(\indist_{\randmatY\bm{\Lambda}\given\msg =j},\outdist_{\randmatY\bm{\Lambda}\given\msg =j}) \leq 1-\error'
\end{IEEEeqnarray}
where $\error'$ is the maximal probability of error over $\outdist_{\randmatY, \bm{\Lambda} \given\msg}$.
 We shall prove Theorem~\ref{thm:converse-csirt} in the following two steps: in Appendix~\ref{sec:lower-bound-beta-csirt},  we evaluate $\beta_{1-\error}(\indist_{\randmatY\bm{\Lambda}\given\msg =j},\outdist_{\randmatY\bm{\Lambda}\given\msg =j})$; in Appendix~\ref{sec:converse-Q-csirt}, we relate $\error'$ to $\Rcsirt^\ast(\bl,\error)$ by establishing a converse bound on the auxiliary channel~$\outdist_{\randmatY ,\bm{\Lambda} \given\msg}$.

\subsubsection{Evaluation of $\beta_{1-\error}$}
\label{sec:lower-bound-beta-csirt}
Let $j^\ast$ be the message that achieves the minimum in~\eqref{eq:meta-converse-cisrt}, let $\encoder^\ast_{\csit}(\bm{\lambda}) \define \encoder_{\csit}(j^\ast,\bm{\lambda})$,
and let
\begin{IEEEeqnarray}{rCl}
\beta_{1-\error}(\encoder^\ast_{\csit}) &\define& \beta_{1-\error}(\indist_{\randmatY ,\bm{\Lambda} \given\msg =j^\ast} , \outdist_{\randmatY ,\bm{\Lambda} \given\msg =j^\ast}) . \label{eq:evaluate-beta} 
\end{IEEEeqnarray}
Using~\eqref{eq:evaluate-beta}, we can rewrite \eqref{eq:meta-converse-cisrt} as
\begin{IEEEeqnarray}{rCl}\label{eq:meta-converse-cisrt-eqv}
\beta_{1-\error}(\encoder^\ast_{\csit}) \leq 1-\error'.
\end{IEEEeqnarray}
Let now
\begin{IEEEeqnarray}{rCl}
r(\encoder^\ast_{\csit};\randmatY,\bm{\Lambda}) \define \log\frac{ d\indist_{\randmatY ,\bm{\Lambda} \given\msg =j^\ast}}{ d\outdist_{\randmatY ,\bm{\Lambda} \given\msg =j^\ast}}.
\label{eq:def-r-csit}
\end{IEEEeqnarray}
Note that, under both $\indist_{\randmatY ,\bm{\Lambda} \given\msg =j^\ast}$ and $\outdist_{\randmatY ,\bm{\Lambda} \given\msg =j^\ast}$, the random variable $r(\encoder^\ast_{\csit};\randmatY,\bm{\Lambda})$ has absolutely continuous cumulative distribution function (cdf) with respect to the Lebesgue measure.
By the Neyman-Pearson lemma~\cite[p.~300]{neyman33a}
\begin{IEEEeqnarray}{rCl}
\beta_{1-\error}(\encoder^\ast_{\csit}) =  \outdist_{\randmatY ,\bm{\Lambda} \given\msg =j^\ast} [r(\encoder^\ast_{\csit} ;\randmatY,\bm{\Lambda}) \geq \bl\gamma_\bl(\encoder^\ast_{\csit})]\label{eq:beta-dist} \IEEEeqnarraynumspace
\end{IEEEeqnarray}
where $\gamma_\bl(\encoder^\ast_{\csit})$ is the solution of
\begin{IEEEeqnarray}{rCl}
 \indist_{\randmatY ,\bm{\Lambda} \given\msg =j^\ast}[r(\encoder^\ast_{\csit};\randmatY,\bm{\Lambda} ) \leq \bl\gamma_\bl(\encoder^\ast_{\csit})] =\error.
 \label{eq:condition-satisfied-gamma-n-csirt}
\end{IEEEeqnarray}
%
Let now $\powallocvec \define \tilde{\powallocvec} \circ \encoder^\ast_{\csit}$. Because of the power constraint~\eqref{eq:power-constraint-csirt-para}, $\powallocvec$ is a mapping from $\{1,\ldots,\NumCode\}$ to the set~$\setV_\minant$ defined in~\eqref{eq:def_setV}.
Furthermore, under $\outdist_{\randmatY ,\bm{\Lambda} \given\msg =j^\ast}$, the random variable $r(\encoder^\ast_{\csit};\randmatY,\bm{\Lambda} )$ has the same distribution as $\Lcsirt_\bl(\powallocvec, \bm{\Lambda})$ in~\eqref{eq:info_density_mimo_alt_csirt}, and under $\indist_{\randmatY ,\bm{\Lambda} \given\msg =j^\ast}$, it has the same distribution as $\Scsirt_\bl(\powallocvec,\bm{\Lambda})$ in~\eqref{eq:info_density_mimo_csirt}.
Thus,~\eqref{eq:meta-converse-cisrt-eqv} is equivalent to
\begin{IEEEeqnarray}{rCL}\label{eq:meta-converse-final-form}
  \prob[\Lcsirt_\bl(\powallocvec,\bm{\Lambda}) \geq \bl \gamma_{\bl}(\powallocvec)]\leq 1-\epsilon'
\end{IEEEeqnarray}
where $\gamma_{\bl}(\powallocvec)$ is the solution of~\eqref{eq:thm-converse-def-gamma-n}.
Note that this upper bound depends on the chosen code only through the induced power allocation function $\powallocvec$.
To conclude, we take the infimum of the LHS of~\eqref{eq:meta-converse-final-form} over all power allocation functions $\powallocvec$ to obtain a bound that holds for all $(\bl,\NumCode,\error)_{\csirt}$ codes.

\subsubsection{Converse on the auxiliary channel} %
\label{sec:converse-Q-csirt}
We next relate $\error'$ to $\Rcsirt^\ast(\bl,\error)$.
The following lemma, whose proof can be found at the end of this appendix, serves this purpose.

\begin{lemma}
\label{lem:converse-q-csirt}
For every code with $\NumCode$ codewords and blocklength $\bl$,
the maximum probability of error $\error'$ over the channel $\outdist_{\randmatY ,\bm{\Lambda} \given\msg}$ satisfies
\begin{IEEEeqnarray}{rCl}
1-\error'\leq \frac{c_{\csirt}(\bl)}{\NumCode}
\label{eq:converse-q-csirt}
\end{IEEEeqnarray}
where $c_{\csirt}(\bl)$ is given in~\eqref{eq:def-n-func-converse-csirt}.
\end{lemma}


Using Lemma~\ref{lem:converse-q-csirt}, we obtain
\begin{IEEEeqnarray}{rCl}
\label{eq:beta-mimo-csirt-relate}
\inf\limits_{\powallocvec(\cdot)} \prob[ \Lcsirt_\bl(\powallocvec,\bm{\Lambda} ) \geq \bl \gamma_\bl(\powallocvec)] \leq \frac{c_{\csirt}(\bl)}{\NumCode}.
\end{IEEEeqnarray}
The desired lower bound~\eqref{eq:thm-converse-rcsirt} follows by taking the logarithm on both sides of~\eqref{eq:beta-mimo-csirt-relate} and dividing by \bl.

\paragraph*{Proof of Lemma~\ref{lem:converse-q-csirt}}
By~\eqref{eq:def-q-channel-mimo-csirt}, given $\bm{\Lambda} =\bm{\lambda}$, the output of the channel $\outdist_{\randmatY, \bm{\Lambda} \given\msg}$ depends on the input $\msg$ only through $\randvecs \define \tilde{\powallocvec} \circ \encoder_{\csit}(\msg,\bm{\lambda})$, i.e., through the norm of each column of the codeword matrix $\encoder_{\csit}(\msg,\bm{\lambda})$.
Let $\randvecu \define \tilde{\powallocvec} (\randmatY)$.
In words, the entries of $\randvecu$ are the square of the norm of the columns of $\randmatY$ normalized by the blocklength $\bl$.
It follows that $(\randvecu,\bm{\Lambda})$ is a sufficient statistic for the detection of $\msg$ from $(\randmatY,\bm{\Lambda})$.
Hence, to lower-bound $\error'$ and establish~\eqref{eq:converse-q-csirt}, it suffices to lower-bound the maximal error probability over the channel ${\outdist}_{\randvecu , \bm{\Lambda} \given \randvecs}$ defined by
\begin{IEEEeqnarray}{rCl}
U_i = \frac{1+ S_i\Lambda_i}{\bl} \sum\limits_{l=1}^{\bl}|W_{i,l}|^2,\quad i=1,\ldots,\minant.
\label{eq:app-q-dist-eqv-csirt}
\end{IEEEeqnarray}
Here, $U_i$ denotes the $i$th entry of $\randvecu$, the random variables $\{W_{i,l}\}$ are i.i.d. $ \jpg(0,1)$-distributed, and the input $\randvecs=[S_1\ldots S_\minant]$ has nonnegative entries whose sum does not exceed $\rho$, i.e., \mbox{$\randvecs\in\setV_{\minant}$}.
Note that, given  $S_i$ and $\Lambda_i$, the random variable $U_i$ in~\eqref{eq:app-q-dist-eqv-csirt} is Gamma-distributed, i.e., its pdf $q_{U_i\given S_i, \Lambda_i}$ is given by
\begin{IEEEeqnarray}{rCl}
\IEEEeqnarraymulticol{3}{l}{  q_{U_i\given S_i, \Lambda_i}(u_i\given s_i, \lambda_i) }\notag\\
\quad &=&\frac{n^n}{(1+s_i\lambda_i)^n\Gamma(n)}u_i^{n-1}\exp\mathopen{}\left(-\frac{n u_i}{1+s_i\lambda_i}\right).
\end{IEEEeqnarray}
Furthermore, the random variables $U_1,\dots,U_m$ are conditionally independent given $\randvecs$ and $\bm{\Lambda}$.

We shall use that $q_{U_i\given S_i, \Lambda_i}$ can be upper-bounded as
\begin{IEEEeqnarray}{rCl}
\IEEEeqnarraymulticol{3}{l}{ q_{U_i\given S_i, \Lambda_i}(u_i\given s_i,\lambda_i)}\notag\\
&\leq&  g_i(u_i,\lambda_i)\\
&\define& \left\{\!\!
           \begin{split}
             &\frac{\bl(\bl-1)^{\bl-1} }{\Gamma(\bl)}e^{-(\bl-1)}, & \hbox{if $u_i\leq\frac{\bl-1}{\bl} (1+\snr \lambda_i)$} \\
             &\frac{\bl^\bl u_i^{\bl-1}e^{ -\bl u_i/(1+\snr\lambda_i)} }{ \Gamma(\bl) (1+\snr \lambda_i)^{\bl-1}}  , & \hbox{if $u_i> \frac{\bl-1}{\bl}(1+\snr \lambda_i)$}
           \end{split}
         \right. \,\,\,\,\,\,\,\label{eq:density-qbar-ub2-rt}
\end{IEEEeqnarray}
which follows because $1+s_i\lambda_i\leq 1+\snr\lambda_i$, and because $q_{U_i\given S_i, \Lambda_i}$ is a unimodal function with maximum at
\begin{equation}
 u_i=\frac{n-1}{n}(1+s_i\lambda_i).
\end{equation}
The bound in~\eqref{eq:density-qbar-ub2-rt} is useful because it is integrable and does not depend on the input $s_i$.

Consider now an arbitrary code $\{\vecc_1(\bm{\Lambda}),\ldots,\vecc_\NumCode(\bm{\Lambda})\} \subset \setV_{\minant}$ for the channel ${\outdist}_{\randvecu, \bm{\Lambda} \given \randvecs}$.
Let $\setD_j(\bm{\Lambda})$, $j=1,\dots,\NumCode$, be the (disjoint) \emph{decoding sets} corresponding to the \NumCode codewords $\{\vecc_j(\bm{\Lambda})\}$.
Let $\error'_{\mathrm{avg}}$ be the \emph{average} probability of error over the channel ${\outdist}_{\randvecu ,\bm{\Lambda} \given \randvecs}$.
We have
\begin{IEEEeqnarray}{rCl}
1-\error'&\leq&
1-\error'_{\mathrm{avg}}\\
&=&\frac{1}{\NumCode}\Ex{\bm{\Lambda}}{  \sum\limits_{j=1}^{\NumCode} \int\nolimits_{\setD_j(\bm{\Lambda})} q_{\randvecu\given\randvecs,\bm{\Lambda}}(\vecu\given\vecc_j(\bm{\Lambda}), \bm{\Lambda})d\vecu} \IEEEeqnarraynumspace\label{eq:converse-q-step1-rt}\\
&\leq& \frac{1}{\NumCode}\Ex{\bm{\Lambda}}{\sum\limits_{j=1}^{\NumCode} \int\nolimits_{\setD_j(\bm{\Lambda})}\left(\prod\limits_{i=1}^{\minant} g_i(u_i,\Lambda_i)\right) d\vecu}  \label{eq:converse-q-step2-rt}\\
&=& \frac{1}{\NumCode}\Ex{\bm{\Lambda}}{\int\nolimits_{  \posrealset^{\minant}  } \left(\prod\limits_{i=1}^{\minant} g_i(u_i,\Lambda_i)\right) d\vecu}  \label{eq:converse-q-step3-rt}\\
&=& \frac{1}{\NumCode}\Ex{\bm{\Lambda}}{ \prod\limits_{i=1}^{\minant}\int\nolimits_{0}^{+\infty} g_i(u_i,\Lambda_i) d u_i}  \label{eq:converse-q-step4-rt}
\end{IEEEeqnarray}
where~\eqref{eq:converse-q-step2-rt} follows from~\eqref{eq:density-qbar-ub2-rt}, and where \eqref{eq:converse-q-step3-rt} follows because $g_i(u_i, \Lambda_i)$ is independent of the message $j$ and because $\bigcup\nolimits_{j=1}^{\NumCode} \setD_j(\bm{\Lambda}) =  \posrealset^{\minant}$.
After algebraic manipulations, we obtain
\begin{IEEEeqnarray}{rCl}
\IEEEeqnarraymulticol{3}{l}{
\int\nolimits_{0}^{\infty} g_i(u_i, \lambda_i) d u_i}\notag\\
\quad  &=& \frac{(1+\snr\lambda_i)}{\Gamma(\bl)}\left[(\bl-1)^\bl e^{-(\bl-1)} + \Gamma(\bl,\bl-1)\right]. \IEEEeqnarraynumspace
\label{eq:app-converse-q-integ-rt}
\end{IEEEeqnarray}
Here, $\Gamma(\cdot,\cdot)$ denotes the \emph{(upper) incomplete Gamma function}~\cite[Sec.~6.5]{abramowitz72}.
Substituting~\eqref{eq:app-converse-q-integ-rt} into \eqref{eq:converse-q-step4-rt}, we finally obtain that for every code $\{c_1(\bm{\Lambda}),\ldots,c_{\NumCode}(\bm{\Lambda})\}\subset \setV_{\minant}$,
\begin{IEEEeqnarray}{rCl}
1-\error' &\leq& \frac{1}{\NumCode} \left(\frac{(\bl-1)^\bl e^{-(\bl-1)}}{\Gamma(\bl)} +\frac{ \Gamma(\bl,\bl-1)}{\Gamma(\bl)} \right)^\minant \notag\\
&&\times \,\Ex{}{\prod\limits_{i=1}^{\minant}(1+\snr\Lambda_i)}\\
&=& \frac{c_{\csirt}(\bl)}{\NumCode}.\label{eq:lb-error-avg-Q-rt}
\end{IEEEeqnarray}
This proves Lemma~\ref{lem:converse-q-csirt}.

\section{Proof of the Converse Part of Theorem~\ref{thm:asy-mimo-csirt}}
\label{app:proof-asy-csit-conv}
As a first step towards establishing~\eqref{eq:proof-asy-conv-csit}, we relax the upper bound~\eqref{eq:thm-converse-rcsirt} by lower-bounding its denominator.
Recall that by definition (see Appendix~\ref{sec:lower-bound-beta-csirt})
\begin{IEEEeqnarray}{rCl}
  \prob[\Lcsirt_\bl(\powallocvec,\bm{\Lambda}) \geq \bl \gamma_{\bl}(\powallocvec)] &=& \beta_{1-\error}(\indist_{\randmatY ,\bm{\Lambda} \given \msg =j^\ast} , \outdist_{\randmatY ,\bm{\Lambda} \given\msg =j^\ast}).\notag\\
  && \label{eq:recall_def_L}
\end{IEEEeqnarray}
We shall use the following inequality: for every $\eta>0$~\cite[Eq.~(102)]{polyanskiy10-05}
\begin{IEEEeqnarray}{c}
\label{eq:inequality_beta-poly}
 \beta_{1-\epsilon}(\indist,\outdist)\geq \frac{1}{\eta}\left(1-\indist \mathopen{}\left[\frac{d\indist}{d\outdist}\geq \eta \right]-\epsilon  \right).
 \end{IEEEeqnarray}
Using~\eqref{eq:inequality_beta-poly} with $\indist=\indist_{\randmatY ,\bm{\Lambda} \given \msg =j^\ast}$, $\outdist=\outdist_{\randmatY ,\bm{\Lambda} \given\msg =j^\ast}$, $\eta=e^{n\gamma}$, and recalling that (see Appendix~\ref{sec:lower-bound-beta-csirt})
\begin{IEEEeqnarray}{rCl}
  1-\indist_{\randmatY ,\bm{\Lambda} \given \msg =j^\ast}\mathopen{}\left[\frac{d\indist_{\randmatY ,\bm{\Lambda} \given \msg =j^\ast}}{d\outdist_{\randmatY ,\bm{\Lambda} \given \msg =j^\ast}}\geq e^{n\gamma}\right] &=& \prob[ \Scsirt_\bl(\powallocvec,\bm{\Lambda})\leq n\gamma]\notag\\
  &&
\end{IEEEeqnarray}
we obtain that for every $\gamma>0$
\begin{IEEEeqnarray}{rCl}
\IEEEeqnarraymulticol{3}{l}{
\beta_{1-\error}\mathopen{}\big(\indist_{\randmatY ,\bm{\Lambda} \given \msg =j^\ast} , \outdist_{\randmatY ,\bm{\Lambda} \given\msg =j^\ast}\big)}\notag\\
 \quad &\geq& e^{-n\gamma}\left(\prob[ \Scsirt_\bl(\powallocvec,\bm{\Lambda})\leq n\gamma] -\error\right).\label{eq:bound_on_L_term}
\end{IEEEeqnarray}
Using~\eqref{eq:bound_on_L_term} and the estimate
\begin{IEEEeqnarray}{rCl}
\log c_{\csirt}(\bl) = \frac{\minant}{2}\log \bl + \bigO(1)
\label{eq:conv-csirt-const-term}
\end{IEEEeqnarray}
(which follows from~\eqref{eq:def-n-func-converse-csirt}, Assumption~\ref{item:cond-expec-finite} in Theorem~\ref{thm:asy-mimo-csirt}, and from algebraic manipulations), we upper-bound the RHS of~\eqref{eq:thm-converse-rcsirt} as
\begin{IEEEeqnarray}{rCl}
\Rcsirt^{\ast}(\bl,\error) &\leq& \gamma - \frac{1}{\bl}\log\mathopen{}\Big( \inf\limits_{\powallocvec(\cdot)} \prob[ \Scsirt_\bl(\powallocvec,\bm{\Lambda})\leq n\gamma] -\error \Big)\notag\\
 &&+ \,\frac{\minant}{2}\frac{\log\bl}{\bl} +  \bigO\mathopen{}\left(\frac{1}{\bl}\right). \label{eq:ub-rcsirt-original}
\end{IEEEeqnarray}

To conclude the proof we show that for every $\gamma$ in a certain neighborhood of $\Ccsit$ (recall that $\gamma$ is a free optimization parameter),
\begin{IEEEeqnarray}{rCl}
\inf\limits_{\powallocvec(\cdot)}\prob[\Scsirt_\bl(\powallocvec,\bm{\Lambda}) \leq  \bl \gamma] &\geq& \cdistcsit(\gamma)  +  \bigO\mathopen{}\left(\frac{1}{\bl}\right)
\label{eq:conv-asy-bd-avg-csit-inf-v}
\end{IEEEeqnarray}
where~$\cdistcsit(\cdot)$ is the outage probability defined in~\eqref{eq:cadist_csit} and the $\bigO\mathopen{}\left({1}/{\bl}\right)$ term is uniform in $\gamma$.
The desired result~\eqref{eq:proof-asy-conv-csit} follows then by substituting~\eqref{eq:conv-asy-bd-avg-csit-inf-v} into~\eqref{eq:ub-rcsirt-original}, setting $\gamma$ as the solution of
\begin{IEEEeqnarray}{rCl}
\cdistcsit(\gamma)  -\error +  \bigO\mathopen{}\left(1/\bl\right) &=& {1}/{\bl}
\label{eq:choose-gamma-csit}
\end{IEEEeqnarray}
and by noting that this $\gamma$ satisfies
\begin{IEEEeqnarray}{rCl}
\gamma = \Ccsit + \bigO\mathopen{}\left(1/\bl\right)
\label{eq:asy-exp-gamma-csit}
\end{IEEEeqnarray}
i.e., it belongs to the desired neighborhood of \Ccsit for sufficiently large $\bl$.
Here,~\eqref{eq:asy-exp-gamma-csit} follows by a Taylor series expansion~\cite[Th.~5.15]{rudin76} of $\cdistcsit(\gamma)$ around $\Ccsit$, and because $\cdistcsit(\Ccsit) =\error$ and $\cdistcsit'(\Ccsit)>0$ by assumption.

In the reminder of this appendix, we will prove~\eqref{eq:conv-asy-bd-avg-csit-inf-v}. Our proof consists of the three steps sketched below.

\paragraph*{Step 1} 
\label{par:step_1}
Given $\powallocvec$ and $\bm{\Lambda}$, the random variable $\Scsirt_\bl(\powallocvec,\bm{\Lambda})$ (see~\eqref{eq:info_density_mimo_csirt} for its definition) is the sum of $\bl$ i.i.d. random variables with mean
\begin{IEEEeqnarray}{rCl}
\murt(\powallocvec,\bm{\Lambda}) \define \sum\limits_{j=1}^{\minant}\log\mathopen{}\big(1+\Lambda_j\powalloc_j(\bm{\Lambda})\big)
\label{eq:def-mean-conv-csirt}
\end{IEEEeqnarray}
and variance
\begin{IEEEeqnarray}{rCl}
\sigmart^2(\powallocvec, \bm{\Lambda}) \define \minant - \sum\limits_{j=1}^{\minant}\frac{1}{\big(1+\Lambda_j\powalloc_j(\bm{\Lambda})\big)^2}.
\label{eq:def-var-conv-csirt}
\end{IEEEeqnarray}
Fix an arbitrary power allocation function $\powallocvec(\cdot)$, and assume that $\bm{\Lambda}=\bm{\lambda}$.
Let
\begin{equation}
u(\powallocvec, \bm{\lambda}) \define \frac{\gamma-\murt(\powallocvec,\bm{\lambda})}{\sigmart(\powallocvec,\bm{\lambda})}.
\label{eq:def-u-v-lambda}
\end{equation}
Using the Cramer-Esseen theorem (see Theorem~\ref{thm:refine-be} below), we show in Appendix~\ref{sec:proof_of_eq:kramer_esseen_step} that
\begin{IEEEeqnarray}{rCL}\label{eq:kramer_esseen_step}
  \prob[ \Scsirt_\bl(\powallocvec,\bm{\Lambda})  \leq  \bl\gamma \given  \bm{\Lambda}=\bm{\lambda}] \geq  q_\bl( u(\powallocvec, \bm{\lambda}) ) +\frac{ \constthree }{\bl}
 \label{eq:conv-csirt-lb-prob-s-cond}
\end{IEEEeqnarray}
where
\begin{IEEEeqnarray}{rCl}
q_{\bl}(x) \define Q(-\sqrt{\bl}x) - \frac{[1-\bl x^2]^{+}e^{-\bl x^2/2}}{6\sqrt{\bl}}
 \label{eq:def-q-n-y}
\end{IEEEeqnarray}
and $\constthree$ is a finite constant independent of $\bm{\lambda}$, $\powallocvec$ and $\gamma$.
\paragraph*{Step 2} 
\label{par:step_2}
We make the RHS of~\eqref{eq:kramer_esseen_step} independent of $\powallocvec$ by minimizing  $q_\bl( u(\powallocvec, \bm{\lambda}) )$ over $\powallocvec$.
Specifically, we establish in Appendix~\ref{sec:proof_of_minimization_over_power_allocation} the following result: for every $\gamma$ in a certain neighborhood of $\Ccsit$, we have that
\begin{IEEEeqnarray}{rCl}
\prob[\Scsirt_\bl(\powallocvec,\bm{\Lambda}) \leq  \bl \gamma \given \bm{\Lambda}=\bm{\lambda}] &\geq& q_\bl(\ulower(\bm{\lambda})) + \frac{\constthree}{\bl}\IEEEeqnarraynumspace
 \label{eq:conv-csirt-lb-prob-s-cond2}
\end{IEEEeqnarray}
where $\ulower(\bm{\lambda})$ is defined in~\eqref{eq:conv-bdd-u-csirt-final}.
%

\paragraph*{Step 3} 
\label{par:step_3}
We average~\eqref{eq:conv-csirt-lb-prob-s-cond2} over $\bm{\Lambda}$ and establish in Appendix~\ref{sec:proof_of_averaging_over_channel}
the bound~\eqref{eq:conv-asy-bd-avg-csit-inf-v}.
This concludes the proof.

\subsection{Proof of~\eqref{eq:kramer_esseen_step}} 
\label{sec:proof_of_eq:kramer_esseen_step}


%
%

We  need the following version of the Cramer-Esseen Theorem.\footnote{The Berry-Esseen Theorem used in \cite{polyanskiy10-05} to prove (\ref{eq:approx_R_introduction}) yields an asymptotic expansion in~\eqref{eq:conv-asy-bd-avg-csit-inf-v} up to a $\bigO(1/\sqrt{\bl})$ term.
This is not sufficient here, since we need an expansion up to a  $\bigO(1/\bl)$ term (see~\eqref{eq:conv-asy-bd-avg-csit-inf-v}).}
%
%
\begin{thm}
\label{thm:refine-be}
Let $X_1,\ldots,X_\bl$ be a sequence of i.i.d. real random variables having zero mean and unit variance.
Furthermore, let
\begin{IEEEeqnarray}{rCl}
\CF(t)&\define&\Ex{}{e^{itX_1}}\,\,\text{and}\,\,
F_\bl(\xi) \define \prob\mathopen{}\left[\frac{1}{\sqrt{\bl}} \sum\limits_{j=1}^{\bl}X_j \leq \xi\right].\IEEEeqnarraynumspace
\end{IEEEeqnarray}
If $\Ex{}{|X_1|^4} < \infty$ and if $\sup_{|t|\geq \zeta} |\CF(t)| \leq \ConstThm$ for some $\ConstThm <1$, where $\zeta\define1/({12\Ex{}{|X_1|^3}})$, then for every $\xi$ and $\bl$
\begin{IEEEeqnarray}{rCl}
\IEEEeqnarraymulticol{3}{l}{
\left|F_\bl(\xi) - Q(-\xi) - k_1(1-\xi^2)e^{-\xi^2/2}\frac{1}{\sqrt{\bl}}\right|
}\notag\\ \quad
&\leq& k_2\left\{\frac{\Ex{}{|X_1|^4} }{\bl}+ \bl^6\left(k_0+\frac{1}{2\bl}\right)^{\bl}\right\}.\IEEEeqnarraynumspace
\label{eq:thm-osipov-refine}
\end{IEEEeqnarray}
Here, $k_1 \define \Ex{}{X_1^3}/(6\sqrt{2\pi})$, and $k_2$ is a positive constant independent of $\{X_i\}$ and $\xi$.

\end{thm}
\begin{IEEEproof}
The inequality~\eqref{eq:thm-osipov-refine} is a consequence of the tighter inequality reported in~\cite[Th.~VI.1]{petrov75}.
\end{IEEEproof}

Let
\begin{IEEEeqnarray}{rCl}
\label{eq:defn_tj}
T_l(\powallocvec,\bm{\Lambda})\define \frac{1}{\sigmart(\powallocvec, \bm{\Lambda}) }\sum\limits_{j=1}^{\minant} \left(1-\frac{\big|\sqrt{\Lambda_j v_j(\bm{\Lambda})}Z_{l,j}-1 \big|^2}{1+ \Lambda_j v_j(\bm{\Lambda})}\right)\IEEEeqnarraynumspace
\end{IEEEeqnarray}
where $Z_{l,j}$, $l=1,\dots,n$ and $j=1,\dots,m$, are i.i.d. $\jpg(0,1)$ distributed.
The random variables $T_1,\ldots, T_\bl$ have zero mean and unit variance, and are conditionally i.i.d. given~$\bm{\Lambda}$.
Furthermore, by construction
\begin{IEEEeqnarray}{rCl}
\prob\mathopen{}\left[\Scsirt_\bl(\powallocvec, \bm{\Lambda}) \leq \bl \gamma \right] &=& \prob\mathopen{}\left[\frac{1}{\sqrt{\bl}}\sum\limits_{l=1}^{\bl}   T_l(\powallocvec,\bm{\Lambda}) \leq \sqrt{\bl} u(\powallocvec, \bm{\Lambda})  \right]\notag\\
&&\label{eq:converse_info_densi_formula}
\end{IEEEeqnarray}
where $u(\powallocvec, \bm{\Lambda})$ was defined in~\eqref{eq:def-u-v-lambda}.
We next show that the conditions under which Theorem~\ref{thm:refine-be} holds are satisfied by the random variables $\{T_l\}$.

We start by noting that if $\lambda_j v_j(\bm{\lambda})$, $j=1,\dots,\minant$, are identically zero, then $\Scsirt_\bl(\powallocvec, \bm{\Lambda})=0$, so~\eqref{eq:conv-csirt-lb-prob-s-cond} holds trivially.
Hence, we will focus on the case where $\{\lambda_j v_j(\bm{\lambda})\}$ are not all identically zero.
Let
\begin{IEEEeqnarray}{rCl}
 \CF_{T_l}(t)\define \Ex{}{e^{it T_l}\big| \bm{\Lambda} =\bm{\lambda}} \IEEEeqnarraynumspace
 \end{IEEEeqnarray}
 and
 \begin{equation}
\zeta \define \frac{1}{12\Ex{}{|T_l|^3 \big|  \bm{\Lambda} =\bm{\lambda} }}.
 \end{equation}
We next show that there exists a $\const_0<1$ such that $ \sup_{|t|>\zeta} |\CF_{T_l}(t)|\leq k_0$
  for every $\bm{\lambda}\in\posrealset^{\minant}$ and every function~$\powallocvec(\cdot)$.
We start by evaluating $\zeta$.
For every $\bm{\lambda}\in\posrealset^{\minant}$ and every~$\powallocvec(\cdot)$ such that $\lambda_j v_j(\bm{\lambda})$, ${1\leq j\leq \minant}$, are not identically zero, it can be shown through algebraic manipulations that
%
\begin{IEEEeqnarray}{rCl}
\label{eq:fourth_norm}
\Ex{}{|T_l|^4 \big|  \bm{\Lambda} =\bm{\lambda}} \leq 9.
\end{IEEEeqnarray}
By Lyapunov's inequality~\cite[p.~18]{petrov75}, this implies that
\begin{IEEEeqnarray}{rCl}
\Ex{}{|T_l|^3 \big|  \bm{\Lambda} =\bm{\lambda} } \leq \Big(\Ex{}{|T_l|^4 \big|   \bm{\Lambda} =\bm{\lambda} }\Big)^{3/4} \leq 9^{3/4}.\IEEEeqnarraynumspace
\end{IEEEeqnarray}
Hence,
\begin{equation}
\label{eq:app-zeta-bound}
\zeta = \frac{1}{12\Ex{}{|T_l|^3 \big|  \bm{\Lambda} =\bm{\lambda} } }\geq\frac{1}{12 \times 9^{3/4} }\define \zeta_0.
\end{equation}
By~\eqref{eq:app-zeta-bound}, we have
\begin{IEEEeqnarray}{rCl}
\sup\limits_{|t|>\zeta} \big|\CF_{T_l}(t)\big| \leq \sup\limits_{|t|>\zeta_0} \big|\CF_{T_l}(t)\big|
\label{eq:conv-asy-bd-v_T}
\end{IEEEeqnarray}
where $\zeta_0$ does not depend on $\bm{\lambda}$ and $\powallocvec$.
Through algebraic manipulations, we can further show that the RHS of~\eqref{eq:conv-asy-bd-v_T} is upper-bounded by
\begin{IEEEeqnarray}{rCl}\label{eq:second_condition}
\sup\limits_{|t|>\zeta_0} \big|\CF_{T_l}(t)\big|  \leq \frac{1}{\sqrt{1+\zeta_0^2/\minant}} \define \ConstThm<1.
 \end{IEEEeqnarray}
The inequalities~\eqref{eq:fourth_norm} and~\eqref{eq:second_condition} imply that the conditions in Theorem~\ref{thm:refine-be} are met.
Hence, we conclude that, by Theorem~\ref{thm:refine-be}, for every $\bl$, $\bm{\lambda}$, and $\powallocvec(\cdot)$,
\begin{IEEEeqnarray}{rCl}
\IEEEeqnarraymulticol{3}{l}{
\prob\mathopen{}\left[\left.\frac{1}{\sqrt{\bl}} \sum\limits_{l=1}^{\bl} T_l \leq \sqrt{\bl}    u(\powallocvec, \bm{\lambda}) \right| \bm{\Lambda}=\bm{\lambda}\right]
- Q\mathopen{}\left(-\sqrt{\bl}u(\powallocvec, \bm{\lambda})\right)}\notag\\
\quad&\geq& \frac{\Ex{}{T_l^3 \given  \bm{\Lambda} =\bm{\lambda}}}{6\sqrt{2\pi} \sqrt{\bl}} (1-\bl u(\powallocvec, \bm{\lambda})^2)e^{-\bl u(\powallocvec,\bm{\lambda})^2/2} - \frac{9 \const_2}{\bl} \notag\\
&&-\,  k_2 \bl^6\left(k_0+\frac{1}{2\bl}\right)^{\bl}\IEEEeqnarraynumspace\label{eq:app-estimate-prob-minus-phi}
\end{IEEEeqnarray}
where~$ u(\powallocvec, \bm{\lambda})$ was defined in~\eqref{eq:def-u-v-lambda}.
The inequality \eqref{eq:conv-csirt-lb-prob-s-cond}  follows then by noting that
\begin{IEEEeqnarray}{rCl}
\label{eq:third-moment-conv}
0 \geq \Ex{}{T_l^3 \Big|  \bm{\Lambda} =\bm{\lambda}} \geq-\sqrt{2\pi}
\end{IEEEeqnarray}
and that
\begin{IEEEeqnarray}{rCl}
\label{eq:lemma-last-term-cramer-esseen}
\sup\limits_{\bl\geq 1} \bl\left( k_2 \bl^6 \left( k_0+\frac{1}{2 \bl}\right)^{\bl}\right) < \infty.
\end{IEEEeqnarray}
%

%
%
%

%
\subsection{Proof of~\eqref{eq:conv-csirt-lb-prob-s-cond2}} 
\label{sec:proof_of_minimization_over_power_allocation}


For every fixed $\bm{\lambda}$, we minimize $q_\bl(u(\powallocvec, \bm{\lambda} ))$ on the RHS of~\eqref{eq:conv-csirt-lb-prob-s-cond} over all power allocation functions $\powallocvec(\cdot)$.
  With a slight abuse of notation, we use $\powallocvec\in\setV_{\minant}$ (where $\setV_{\minant}$ was defined in~\eqref{eq:def_setV}) to denote the vector $\powallocvec(\bm{\lambda})$ whenever no ambiguity arises.
  Since the function $q_\bl(x)$ in~\eqref{eq:def-q-n-y} is monotonically increasing in~$x$, the vector $\powallocvec\in\setV_{\minant}$ that minimizes $q_\bl(u(\powallocvec, \bm{\lambda} ))$ is the solution of
\begin{IEEEeqnarray}{rCl}
\min_{\powallocvec \in \setV_{\minant}} \,\,u(\powallocvec,\bm{\lambda}).
\label{eq:opt-pow-prob}
\end{IEEEeqnarray}
The minimization in~\eqref{eq:opt-pow-prob} is difficult to solve since  $u(\powallocvec,\bm{\lambda})$ is neither convex nor concave in $\powallocvec$.
For our purposes, it suffices to obtain a lower bound on~\eqref{eq:opt-pow-prob}, which is given in~Lemma~\ref{lem:diff-sigma-rt} below.
Together with~\eqref{eq:conv-bdd-u-csirt-final} and the monotonicity of $q_\bl(\cdot)$, this then yields~\eqref{eq:conv-csirt-lb-prob-s-cond2}.

\begin{lemma}
\label{lem:diff-sigma-rt}
Let $\bm{v}^\ast$, $\murt(\powallocvec,\bm{\lambda})$, $\sigmart(\powallocvec,\bm{\lambda})$, and $u(\powallocvec,\bm{\lambda})$ be as in~\eqref{eq:water-filling-power},~\eqref{eq:def-mean-conv-csirt},~\eqref{eq:def-var-conv-csirt}, and~\eqref{eq:def-u-v-lambda}, respectively.
Moreover, let $\murt^\ast(\bm{\lambda}) \define \murt(\bm{v}^{\ast},\bm{\lambda} )$ and
$\sigmart^\ast(\bm{\lambda})\define  \sigmart(\bm{v}^\ast, \bm{\lambda})$.
Then, there exist $ \delta>0 $, $ \tilde{\delta}>0 $ and $\const<\infty $ such that for every $\gamma\in(\Ccsit-\tilde{\delta},\Ccsit + \tilde{\delta})$
 \begin{IEEEeqnarray}{rCl}
 \IEEEeqnarraymulticol{3}{l}{
\min\limits_{\powallocvec \in\setV_{\minant}} u(\powallocvec,\bm{\lambda}) }\notag\\
 \,\,&\geq& \ulower(\bm{\lambda}) \define \left\{
                \begin{array}{ll}
                \!\! \delta/\sqrt{\minant}, &\hbox{if $\murt^\ast(\bm{\lambda}) \leq \gamma-\delta$}\\
                \!\! \dfrac{\gamma - \murt^\ast(\bm{\lambda})}{\sigmart^\ast(\bm{\lambda}) + \const(\gamma - \murt^\ast(\bm{\lambda}))}, & \hbox{if $|\gamma - \murt^\ast(\bm{\lambda})|< \delta$}\\
                 \!\!-\infty, & \hbox{if $\murt^\ast(\bm{\lambda}) \geq \gamma+ \delta$}.
                \end{array}
                                  \right.\notag\\
 && \label{eq:conv-bdd-u-csirt-final}
\end{IEEEeqnarray}
%
%
\end{lemma}
\begin{IEEEproof}
See Appendix~\ref{app:proof-lemma-diff-sigma-rt}.
\end{IEEEproof}


%
%
%
%
%
%

\subsection{Proof of~\eqref{eq:conv-asy-bd-avg-csit-inf-v}} 
\label{sec:proof_of_averaging_over_channel}

We shall need the following lemma, which concerns the speed of convergence of $\prob[B> A/\sqrt{\bl} ]$ to $\prob[B>0]$ as $\bl\to\infty$ for two independent random variables  $A$ and $B$.

\begin{lemma}
\label{lem:expectation-phi}
Let $A$ be a real random variable with zero mean and unit variance.
Let $B$ be a real random variable independent of~$A$ with continuously differentiable pdf $\pdff_B$. 
Then
\begin{IEEEeqnarray}{rCl}\label{eq:lemma_convergence_speed}
  \left|\prob\mathopen{}\left[B\geq \frac{A}{\sqrt{\bl}} \right] -\prob[B\geq 0] \right|\leq \frac{1}{\bl}\Big(\frac{2}{\delta^2} +  \frac{\auxconstone}{\delta} + \frac{\auxconstone}{2}\Big)\IEEEeqnarraynumspace
\end{IEEEeqnarray}
where
\begin{equation}
\auxconstone \define \sup\limits_{t\in(-\delta,\delta)}\max\mathopen{}\big\{|f_{B}(t)|,|f_B'(t)|\big\}
\end{equation}
 and $\delta>0$ is chosen so that~$\auxconstone$ is finite.

%
\end{lemma}

\begin{IEEEproof}
See Appendix~\ref{app:proof-lem-expectation-phi}.
\end{IEEEproof}

To establish~\eqref{eq:conv-asy-bd-avg-csit-inf-v}, we lower-bound $\Ex{}{q_\bl(\ulower(\bm{\Lambda}))}$ on the RHS of~\eqref{eq:conv-csirt-lb-prob-s-cond2} using Lemma~\ref{lem:expectation-phi}.
This entails technical difficulties since the pdf of $\ulower(\bm{\Lambda})$ is not continuously differentiable due to the fact that the water-filling solution~\eqref{eq:water-filling-power} may give rise to different numbers of active eigenmodes for different values of~$\bm{\lambda}$. %
To circumvent this problem, we partition $\realset^{\minant}_{\geq}$ into~$\minant$ non-intersecting subregions $\setW_j$, $j=1,\dots,\minant$~\cite[Eq.~(24)]{caire99-05}
\begin{multline}
\setW_j \define
 \Big\{\vecx\in \realset^{\minant}_{\geq}:
 \frac{1}{x_{j+1}}> \frac{1}{j}\sum\limits_{l=1}^{j}\frac{1}{x_l} + \frac{\snr}{j} \geq \frac{1}{x_j}\Big\},\\ j=1,\ldots,\minant-1
\label{eq:ach-asy-csit-def-setw}
\end{multline}
and
\begin{IEEEeqnarray}{rCl}
\setW_\minant \define \Big\{\vecx\in \realset^{\minant}_{\geq}: \frac{1}{\minant}\sum\limits_{l=1}^{\minant}\frac{1}{x_l} + \frac{\snr}{\minant} \geq \frac{1}{x_\minant}\Big\}.
\label{eq:ach-asy-csit-def-setw-m}
\end{IEEEeqnarray}
In the interior of $\setW_j$, $j=1,\ldots,\minant$, the pdf of $\hat{u}(\bm{\Lambda})$ is continuously differentiable.
Note that $\bigcup_{j=1}^{\minant} \setW_j =\realset^{\minant}_{\geq}$.
For every $\veclambda\in\setW_j$,
 the water-filling solution gives exactly $j$ active eigenmodes, i.e.,
 \begin{IEEEeqnarray}{rCl}
 v_1^\ast(\bm{\lambda})\geq \cdots \geq v_j^\ast(\bm{\lambda})>v_{j+1}^{\ast}(\bm{\lambda})=\cdots=v_{\minant}^\ast(\bm{\lambda})=0. \IEEEeqnarraynumspace
\label{eq:expression-gamma-star}
\end{IEEEeqnarray}
Let
\begin{IEEEeqnarray}{rCl}
\setK_\delta \define \Big\{\veclambda \in \realset^{\minant}_{\geq}:\, |\gamma - \murt^\ast(\veclambda)|< \delta\Big\}.
\label{eq:def-set-K-delta}
\end{IEEEeqnarray}
Using~\eqref{eq:def-set-K-delta} and the sets $\{\setW_j\}$, we express $\Ex{}{q_\bl(\ulower(\bm{\Lambda}))}$ as
\begin{IEEEeqnarray}{rCl}
\IEEEeqnarraymulticol{3}{l}{
 \Ex{}{q_\bl(\ulower(\bm{\Lambda}))} }\notag\\
\quad &=& \Ex{}{q_\bl(\ulower(\bm{\Lambda})) \indfun{ \bm{\Lambda}\notin \setK_{\delta}}} \notag\\
&&+ \,
 \sum\limits_{j=1}^{\minant}  \Ex{}{ q_\bl(\ulower(\bm{\Lambda}))  \indfun{ \bm{\Lambda}\in \setK_{\delta} \cap \Int(\setW_{j})}}\IEEEeqnarraynumspace
 \label{eq:conv-asy-bd-avg-lambda2}
\end{IEEEeqnarray}
where $\Int(\cdot)$ denotes the interior of a given set.
To obtain~\eqref{eq:conv-asy-bd-avg-lambda2}, we used that $\bm{\Lambda}$ lies in $ \bigcup_{j=1}^{\minant}\Int(\setW_j)$ almost surely, which holds because the joint pdf of $\{\Lambda_j\}_{j=1}^{\minant}$ exists by assumption and because the boundary of $\setW_j$ has zero Lebesgue measure.

We next lower-bound the two terms on the RHS of~\eqref{eq:conv-asy-bd-avg-lambda2} separately.
We first consider the first term. When $\murt^\ast(\veclambda) \geq \gamma+\delta$, we have $\ulower(\veclambda)=-\infty$ and $q_\bl\mathopen{}\big(u_1(\veclambda)\big)=0$;
when $\murt^\ast(\veclambda) \leq \gamma-\delta$, we have $\ulower(\veclambda) = \delta/\sqrt{\minant}$ and
\begin{IEEEeqnarray}{rCl}
q_\bl \mathopen{}\big(\ulower(\veclambda)\big) =  Q\mathopen{}\left(-\sqrt{\bl}\frac{\delta}{\sqrt{\minant}}\right)  -\frac{[1-\bl\delta^2/\minant]^{+}e^{-\bl\delta^2/(2\minant)}}{6\sqrt{\bl}}.\notag\\
\label{eq:bound-qbl-ustar-1}
\end{IEEEeqnarray}
Assume without loss of generality that $\bl\geq \minant/\delta^2$ (recall that we are interested in the asymptotic regime $\bl\to\infty$).
In this case, the second term on the RHS of~\eqref{eq:bound-qbl-ustar-1} is zero. Hence,
\begin{IEEEeqnarray}{rCl}
 \IEEEeqnarraymulticol{3}{l}{
 \Ex{}{q_\bl(\ulower(\bm{\Lambda})) \indfun{ \bm{\Lambda}\notin \setK_{\delta}}}   } \notag\\
\quad & = & Q\mathopen{}\left(-\sqrt{\bl}\frac{\delta}{\sqrt{\minant}}\right) \prob\mathopen{}\big[\murt^\ast(\bm{\Lambda}) \leq \gamma -\delta  \big] \\
&\geq & \prob\mathopen{}\big[\murt^\ast(\bm{\Lambda}) \leq \gamma -\delta \big] - e^{-\bl \delta^2/(2\minant)} .\label{eq:p-0-rt}
\end{IEEEeqnarray}
Here,~\eqref{eq:p-0-rt} follows because $Q(-t)\geq 1-e^{-t^2/2}$ for all $t\geq 0 $ and because $\prob[\murt^\ast(\bm{\Lambda}) \leq \gamma-\delta ]\leq 1$.
%
%

We next lower-bound the second term on the RHS of~\eqref{eq:conv-asy-bd-avg-lambda2}.
If $\prob[ \bm{\Lambda}\in \setK_{\delta} \cap \Int(\setW_{j})] =0$, we have
\begin{IEEEeqnarray}{rCl}
\Ex{}{ q_\bl(\ulower(\bm{\Lambda}))  \indfun{ \bm{\Lambda}\in \setK_{\delta} \cap \Int(\setW_{j})}} = 0
\label{eq:exp-qn-ul-zero}
\end{IEEEeqnarray}
since $q_\bl(\cdot)$ is bounded. We thus assume in the following that $\prob[ \bm{\Lambda}\in \setK_{\delta} \cap \Int(\setW_{j})] >0$.
Let $\randulower$ denote the random variable $\ulower(\bm{\Lambda})$.
To emphasize that $\randulower$ depends on $\gamma$ (see~\eqref{eq:conv-bdd-u-csirt-final}), we write $\randulower(\gamma)$ in place of $\randulower$ whenever necessary.
Using this definition and~\eqref{eq:def-q-n-y}, we obtain
\begin{IEEEeqnarray}{rCl}
\IEEEeqnarraymulticol{3}{l}{
 \Ex{}{ q_\bl(\randulower)  \indfun{ \bm{\Lambda}\in \setK_{\delta} \cap \Int(\setW_{j})}}}\notag\\
&=&\bigg( \Ex{}{Q(-\sqrt{\bl}\randulower)\given \bm{\Lambda } \in \setK_{\delta} \cap \Int(\setW_{j})} \notag\\
&&-\,\frac{1}{6\sqrt{\bl}}\Ex{}{\big[1-\bl \randulower^2\big]^{+}e^{-\bl \randulower^2 /2}\Big| \bm{\Lambda } \in \setK_{\delta} \cap \Int(\setW_{j})}\bigg)\notag\\
&&\times \, \prob\mathopen{}\big[\bm{\Lambda}\in \setK_{\delta} \cap \Int(\setW_{j})   \big].\IEEEeqnarraynumspace
\label{eq:p-j-rt}
\end{IEEEeqnarray}
Observe that the transformation
\begin{IEEEeqnarray}{rCl}
(\lambda_1,\ldots,\lambda_j,\gamma) \mapsto (\ulower(\bm{\lambda}),\lambda_2,\ldots,\lambda_j,\gamma)
\end{IEEEeqnarray}
is one-to-one and twice continuously differentiable with nonsingular Jacobian for $\bm{\lambda} \in \setK_{\delta} \,\cap\, \Int(\setW_{j})$, i.e., it is a diffeomorphism of class $C^2$~\cite[p.~147]{munkres91-a}.
Consequently, the conditional pdf $f_{\randulower(\gamma) \given \bm{\Lambda}\in \setK_{\delta} \cap \Int(\setW_{j}) }(t)$ of $\randulower(\gamma)$ given $\bm{\Lambda}\in \setK_{\delta} \cap \Int(\setW_{j})$ as well as its first derivative are jointly continuous functions of $\gamma$ and $t$.
Hence, they are bounded on bounded sets.
It thus follows that for every $j\in\{1,\ldots,\minant\}$, every $\gamma\in(\Ccsit-\altdelta, \Ccsit+\altdelta)$ (where~$\altdelta$ is given by Lemma~\ref{lem:diff-sigma-rt}), and every $\altdelta_1>0$, there exists a $\auxconsttwo <\infty$ such that the conditional pdf $f_{\randulower(\gamma)\given \bm{\Lambda}\in \setK_{\delta} \cap \Int(\setW_{j})}$
and its derivative satisfy
\begin{IEEEeqnarray}{rCl}
&&\sup_{t\in [-\altdelta_1, \altdelta_1]} \sup\limits_{\gamma \in (\Ccsit-\altdelta, \Ccsit+\altdelta )}\! \big|\pdff_{\randulower(\gamma)\given \bm{\Lambda}\in \setK_{\delta} \cap \Int(\setW_{j})}(t)\big| \leq \auxconsttwo\label{eq:lemma-ub-pdfu} \IEEEeqnarraynumspace\\
&&\sup_{t\in [-\altdelta_1, \altdelta_1]}\sup\limits_{\gamma \in (\Ccsit-\altdelta, \Ccsit+\altdelta )}\! \big|\pdff_{\randulower(\gamma)\given \bm{\Lambda}\in \setK_{\delta} \cap \Int(\setW_{j})}'(t)\big| \leq \auxconsttwo \label{eq:lemma-ub-pdfud}.\IEEEeqnarraynumspace
\end{IEEEeqnarray}
We next apply Lemma~\ref{lem:expectation-phi} with $A$ being a standard normal random variable and $B$ being the random variable $\randulower$ conditioned on $ \bm{\Lambda } \in  \setK_{\delta} \cap \Int(\setW_{j})$.
This implies that there exists a finite constant~$\auxconstthree$ independent of~$\gamma$ and~$\bl$ such that the first term on the RHS of~\eqref{eq:p-j-rt} satisfies
\begin{IEEEeqnarray}{rCl}
\IEEEeqnarraymulticol{3}{l}{
\Ex{}{Q\mathopen{}\big(-\sqrt{\bl}\randulower(\gamma)\big)\big|  \bm{\Lambda } \in \setK_{\delta} \cap \Int(\setW_{j})}}\notag\\
\quad &\geq& \prob\mathopen{}\big[ \murt^\ast(\bm{\Lambda}) \leq \gamma  \given \bm{\Lambda}\in \setK_{\delta} \cap \Int(\setW_{j})\big ] +  \frac{\auxconstthree}{\bl}.
\label{eq:exp-q-nu}
\end{IEEEeqnarray}
We next bound the second term on the RHS of~\eqref{eq:p-j-rt} for $\bl \geq \tilde{\delta}_1^{-2} $ as
\begin{IEEEeqnarray}{rCl}
\IEEEeqnarraymulticol{3}{l}{
\frac{1}{6\sqrt{\bl}}\Ex{}{\big[1-\bl \randulower^2\big]^{+}e^{-\bl \randulower^2 /2} \Big| \bm{\Lambda } \in \setK_{\delta} \cap \Int(\setW_{j})}}\notag\\
\quad &\leq & \frac{\auxconsttwo }{6\sqrt{\bl}}\int\nolimits_{-1/\sqrt{\bl}}^{1/\sqrt{\bl}} (1-\bl t^2)e^{-\bl t^2/2} dt\label{eq:ub-ex-1-minus-nu-2}\\
&=& \frac{\auxconsttwo}{3\sqrt{ e} \bl}\label{eq:ub-ex-1-minus-nu-3}
\end{IEEEeqnarray}
where~\eqref{eq:ub-ex-1-minus-nu-2} follows from~\eqref{eq:lemma-ub-pdfu}.
Substituting~\eqref{eq:exp-q-nu} and~\eqref{eq:ub-ex-1-minus-nu-3} into~\eqref{eq:p-j-rt} we obtain
\begin{IEEEeqnarray}{rCl}
\IEEEeqnarraymulticol{3}{l}{
  \Ex{}{ q_\bl(\randulower)  \indfun{ \bm{\Lambda}\in \setK_{\delta} \cap \Int(\setW_{j})}}}\notag\\
\quad   &\geq& \prob\mathopen{}\big[ \mu^\ast(\bm{\Lambda}) \leq \gamma,   \bm{\Lambda}\in \setK_{\delta} \cap \Int(\setW_{j})\big] +  \frac{\auxconstfour}{\bl}\IEEEeqnarraynumspace
\label{eq:p-j-rt-final}
\end{IEEEeqnarray}
for some finite $\auxconstfour$ independent of $\gamma$ and $\bl$.
%
Using~\eqref{eq:p-0-rt},~\eqref{eq:exp-qn-ul-zero} and~\eqref{eq:p-j-rt-final} in~\eqref{eq:conv-asy-bd-avg-lambda2}, and substituting~\eqref{eq:conv-asy-bd-avg-lambda2} into~\eqref{eq:conv-csirt-lb-prob-s-cond2}, we conclude that
\begin{IEEEeqnarray}{rCL}
\prob[\Scsirt_\bl(\powallocvec,\bm{\Lambda}) \leq  \bl \gamma]   &\geq& \prob[ \mu^\ast(\bm{\Lambda}) \leq \gamma ] + \bigO\mathopen{}\left(\frac{1}{n}\right) \IEEEeqnarraynumspace\\
&=& \cdistcsit(\gamma) + \bigO\mathopen{}\left(\frac{1}{n}\right)
\end{IEEEeqnarray}
where the $\bigO\mathopen{}\left({1}/{n}\right)$ term is uniform in $\gamma\in(\Ccsit-\deltaup,\Ccsit+\deltaup)$.
Here, the last step follows from~\eqref{eq:def-mean-conv-csirt} and~\eqref{eq:cadist_csit}.

\subsection{Proof of Lemma~\ref{lem:diff-sigma-rt}}
\label{app:proof-lemma-diff-sigma-rt}

For an arbitrary $\bm{\lambda}\in \realset^{\minant}_{\geq}$, the function $\murt(\powallocvec,\bm{\lambda})$ in the numerator of~\eqref{eq:def-u-v-lambda} is maximized by the (unique) water-filling power allocation $v_j=v_j^\ast$ defined in~\eqref{eq:water-filling-power}:
\begin{IEEEeqnarray}{rCl}
\label{eq:max-mu-csirt}
\murt^\ast(\bm{\lambda}) = \max_{\powallocvec\in\setV_{\minant}} \murt(\powallocvec,\bm{\lambda}) = \murt(\bm{v}^{\ast} , \bm{\lambda}).
\end{IEEEeqnarray}
The function~$\sigmart(\powallocvec,\bm{\lambda})$ on the denominator of~\eqref{eq:def-u-v-lambda} can be bounded as
\begin{IEEEeqnarray}{rCl}
\label{eq:bdd-var-csirt}
0\leq \sigmart(\powallocvec,\bm{\lambda})\leq \sqrt{\minant}.
\end{IEEEeqnarray}
Using~\eqref{eq:max-mu-csirt} and~\eqref{eq:bdd-var-csirt} we obtain that for an arbitrary $\delta>0$
\begin{IEEEeqnarray}{rCl}
\min_{\powallocvec \in \setV_{\minant}} \,\,u(\powallocvec,\bm{\lambda}) \geq \left\{
                                    \begin{array}{ll}
                                      {\delta}/{\sqrt{\minant}}, & \hbox{$\murt^\ast(\bm{\lambda})\leq \gamma-\delta$} \\
                                      -\infty, & \hbox{$\murt^\ast(\bm{\lambda})\geq \gamma + \delta$.}
                                    \end{array}
                                  \right.\label{eq:conv-bdd-u-csirt}\IEEEeqnarraynumspace
\end{IEEEeqnarray}

Let $\powallocvec_{\mathrm{min}}$ be the minimizer of $u(\powallocvec,\bm{\lambda})$ for a given $\bm{\lambda}$.
To prove Lemma~\ref{lem:diff-sigma-rt}, it remains to show that there exist $\delta>0$, $ \tilde{\delta}>0 $ and $\const<\infty $ such that for every $\gamma\in(\Ccsit-\tilde{\delta},\Ccsit + \tilde{\delta})$ and every $\bm{\lambda}\in\realset^{\minant}_{\geq }$ satisfying $|\murt^\ast(\bm{\lambda})-\gamma| <\delta$,
\begin{IEEEeqnarray}{rCl}
\min_{\powallocvec \in \setV_{\minant}} \,\,u(\powallocvec,\bm{\lambda})  &=& u(\powallocvec_{\min},\bm{\lambda} )\\
 &\geq& \frac{\gamma - \murt^\ast(\bm{\lambda})}{\sigmart^\ast(\bm{\lambda}) + \const(\gamma - \murt^\ast(\bm{\lambda}))}.
\label{eq:conv-bdd-u-csirt-2nd}
\end{IEEEeqnarray}
Since
\begin{IEEEeqnarray}{rCl}
u(\powallocvec_{\mathrm{min}},\veclambda) = \frac{\gamma-\murt (\powallocvec_{\mathrm{min}},\veclambda)}{ \sigmart (\powallocvec_{\mathrm{min}},\veclambda) } &\geq& \frac{\gamma-\murt^\ast(\veclambda)}{ \sigmart (\powallocvec_{\mathrm{min}},\veclambda) }\label{eq:lb-u-step1}
\end{IEEEeqnarray}
it suffices to show that for every $\gamma\in(\Ccsit-\tilde{\delta},\Ccsit + \tilde{\delta})$ and every $\bm{\lambda}\in\realset^{\minant}_{\geq }$ satisfying $|\murt^\ast(\bm{\lambda})-\gamma| <\delta$, we have
\begin{IEEEeqnarray}{rCl}
|\sigmart (\powallocvec_{\mathrm{min}},\veclambda)- \sigmart^\ast (\veclambda)|\leq \const |\gamma - \mu^\ast(\veclambda)|
\label{eq:bound-diff-sigma-min-star}
\end{IEEEeqnarray}
and that
\begin{IEEEeqnarray}{rCl}
\sigmart^* (\veclambda) -\const |\gamma - \mu^\ast(\veclambda)| >0.
\label{eq:positivity-sigma-star-k}
\end{IEEEeqnarray}
The desired bound~\eqref{eq:conv-bdd-u-csirt-2nd} follows then by lower-bounding $\sigmart(\powallocvec_{\mathrm{min}},\bm{\lambda})$ in~\eqref{eq:lb-u-step1} by $\sigmart^* (\veclambda) -\const |\gamma - \mu^\ast(\veclambda)|$ when $\murt^\ast(\veclambda) \geq \gamma$ and by upper-bounding $\sigmart(\powallocvec_{\mathrm{min}},\bm{\lambda})$ by $\sigmart^* (\veclambda) + \const |\gamma - \mu^\ast(\veclambda)|$  when $\murt^\ast(\veclambda) < \gamma$.

We first establish~\eqref{eq:bound-diff-sigma-min-star}.
By the mean value theorem, there exist $\powalloc'_j$  between $\powalloc^*_j$ and $\powalloc_{\mathrm{min},j}$, $j=1,\ldots,\minant$, such that
\begin{IEEEeqnarray}{rCl}
\IEEEeqnarraymulticol{3}{l}{ \big|\sigmart(\powallocvec_{\mathrm{min}},\bm{\lambda}) -\sigmart^\ast(\bm{\lambda})  \big|}\notag\\
\quad &=& \left|\sum\limits_{j=1}^{\minant} \frac{2\lambda_j}{(1+\lambda_j \powalloc'_j)^3}(\powalloc_{\mathrm{min},j} - \powalloc^*_j)\right| \label{eq:bound-eig-E-00}\\
&\leq &\sum\limits_{j=1}^{\minant} \frac{2\lambda_j}{(1+\lambda_j \powalloc'_j)^3}\left|\powalloc_{\mathrm{min},j} - \powalloc^*_j\right| \\
&\leq & 2\lambda_1 \sum\limits_{j=1}^{\minant}  \left|\powalloc_{\mathrm{min},j} -\powalloc^*_j\right| \label{eq:bound-eig-E-1} \\
&\leq& 2\lambda_1 \sqrt{\minant}\|\powallocvec_{\mathrm{min}} - \powallocvec^*\|.
\label{eq:bound-eig-E-2}
\end{IEEEeqnarray}
%
Here, the last step follows because for every $\veca=[a_1,\ldots,a_\minant]\in \realset^{\minant}$, we have $\sum\nolimits_{j=1}^{\minant} |a_j| \leq \sqrt{\minant}\|\veca\|$.

Next, we upper-bound $\lambda_1$ and $\|\powallocvec_{\mathrm{min}} - \powallocvec^*\|$ separately.
The variable $\lambda_1$ can be bounded as follows.
Because the water-filling power levels $\{v_l^\ast\}$ in~\eqref{eq:water-filling-power} are nonincreasing, we have that
\begin{IEEEeqnarray}{rCl}
\frac{\snr}{\minant} &\leq& v_1^\ast \leq \snr.
\label{eq:bound-gamma-star}
\end{IEEEeqnarray}
Choose $\delta_1>0$ and $\altdelta>0$ such that $\delta_1 + \altdelta < \Ccsit$.
%
Using~\eqref{eq:bound-gamma-star} together with
 \begin{equation}
 \log(1+\lambda_1 v^\ast_1)\leq\murt^\ast(\bm{\lambda}) \leq  \minant\log(1+\lambda_1 v^\ast_1)
 \end{equation}
 and the assumption that $\gamma \in (\Ccsit- \altdelta, \Ccsit+\altdelta)$,   we obtain that whenever $|\murt^\ast(\bm{\lambda}) - \gamma|<\delta_1$
\begin{multline}
\const_{0} \define \frac{1}{\snr} \left(e^{(\Ccsit -\delta_1-\altdelta)/\minant}-1\right) \\
\leq \lambda_1 \leq  \frac{\minant}{\snr}\left(e^{\Ccsit +\delta_1+\altdelta}-1\right) \define \constrt_1 .
\label{eq:bd-lambda1}
\end{multline}

The term $\|\powallocvec_{\mathrm{min}} - \powallocvec^*\|$ can be upper-bounded as follows.
Since~$\powallocvec_{\mathrm{min}}$ is the minimizer of $u(\powallocvec,\bm{\lambda})$, it must satisfy the Karush--Kuhn--Tucker (KKT) conditions~\cite[Sec.~5.5.3]{boyd04}:
\begin{IEEEeqnarray}{rCl}
-\frac{\partial u(\powallocvec,\bm{\lambda} )}{\partial \powalloc_l}\Big|_{\powalloc_l=\powalloc_{\mathrm{min},l}} &=& \eta,  \,\,\,\hbox{$\forall \,l$ for which $\powalloc_{\mathrm{min},l}>0$} \IEEEeqnarraynumspace \label{eq:KKT-cond1}\\
-\frac{\partial u(\powallocvec,\bm{\lambda} )}{\partial \powalloc_l}\Big|_{\powalloc_l=\powalloc_{\mathrm{min},l}} &\leq& \eta, \,\,\,\hbox{$\forall\, l$ for which $\powalloc_{\mathrm{min},l}=0$}\label{eq:KKT-cond2}
\end{IEEEeqnarray}
for some $\eta$.
The derivatives in~\eqref{eq:KKT-cond1} and~\eqref{eq:KKT-cond2} are given by
\begin{IEEEeqnarray}{rCl}
-\frac{\partial u(\powallocvec,\bm{\lambda} )}{\partial \powalloc_l}\Big|_{\powalloc_l=\powalloc_{\mathrm{min},l}} &= &
\left(1+ \frac{\gamma - \murt(\powallocvec_{\mathrm{min}},\bm{\lambda})}{(1+\lambda_l\powalloc_{\mathrm{min},l})^2\sigmart^2(\powallocvec_{\mathrm{min}},\bm{\lambda})}\right)\notag\\
&& \times \frac{1}{(\powalloc_{\mathrm{min},l}+1/\lambda_l) \sigmart(\powallocvec_{\mathrm{min}},\bm{\lambda})}
.\IEEEeqnarraynumspace\,\,\,\,
\end{IEEEeqnarray}
Let $\tilde{\eta} \define 1/ (\sigmart(\powallocvec_{\mathrm{min}},\bm{\lambda}) \eta)$.
Then,~\eqref{eq:KKT-cond1} and~\eqref{eq:KKT-cond2} can be rewritten as
\begin{IEEEeqnarray}{rCl}
\powalloc_{\mathrm{min},l} &=& \left[\tilde{\eta} \left(1+ \frac{\gamma - \murt(\powallocvec_{\mathrm{min}},\bm{\lambda})}{(1+\lambda_l\powalloc_{\mathrm{min},l})^2\sigmart^2(\powallocvec_{\mathrm{min}},\bm{\lambda})}\right)-\frac{1}{\lambda_l}\right]^{+}
\label{eq:KKT-cond21}\IEEEeqnarraynumspace\!
\end{IEEEeqnarray}
where $\tilde{\eta}$ satisfies
\begin{IEEEeqnarray}{rCl}
\sum\limits_{l=1}^{\minant} \left[\tilde{\eta} \left(1+ \frac{\gamma - \murt(\powallocvec_{\mathrm{min}},\bm{\lambda})}{(1+\lambda_l\powalloc_{\mathrm{min},l})^2\sigmart^2(\powallocvec_{\mathrm{min}},\bm{\lambda})}\right)-\frac{1}{\lambda_l}\right]^{+} =\snr.\IEEEeqnarraynumspace
\label{eq:KKT-cond21-cont}
\end{IEEEeqnarray}
Here, the equality in~\eqref{eq:KKT-cond21-cont} follows because $u(\powallocvec, \bm{\lambda})$ is monotonically decreasing in~$\powalloc_j$, which implies that the minimizer~$\powallocvec_{\min}$ of $u(\powallocvec, \bm{\lambda})$ must satisfy $\sum\nolimits_{l=1}^{\minant} \powalloc_{\min,l} = \snr$.
%
%
Comparing~\eqref{eq:KKT-cond21} and~\eqref{eq:KKT-cond21-cont} with~\eqref{eq:water-filling-power} and~\eqref{eq:def-gamma-bar},
we obtain, after algebraic manipulations
\begin{IEEEeqnarray}{rCl}
\|\powallocvec_{\mathrm{min}} - \bm{v}^{\ast} \| &\leq& \constrt_{2} |\gamma - \murt(\powallocvec_{\mathrm{min}},\bm{\lambda})|
\label{eq:second-ub-on-diff-powallocvec}
\end{IEEEeqnarray}
for some $\constrt_{2}<\infty$ that does not depend on $\veclambda$, $\powallocvec_{\mathrm{min}}$, $\powallocvec^*$ and $\gamma$.

To further upper-bound the RHS of~\eqref{eq:second-ub-on-diff-powallocvec},
recall that $\powallocvec_{\mathrm{min}}$ minimizes $u(\powallocvec,\bm{\lambda})=(\gamma-\murt(\powallocvec,\bm{\lambda}))/\sigmart(\powallocvec,\bm{\lambda})$ for a given $\bm{\lambda}$ and that $\murt^\ast(\bm{\lambda}) = \max_{\powallocvec\in\setV_{\minant}} \murt(\powallocvec,\bm{\lambda})$.
Thus, if $\murt^\ast(\bm{\lambda}) \geq \gamma$ then we must have $u(\powallocvec_{\mathrm{min}},\bm{\lambda}) \leq u(\bm{v}^\ast, \veclambda) \leq 0$, which implies that
\begin{IEEEeqnarray}{rCl}
0 \leq  \murt(\powallocvec_{\mathrm{min}},\bm{\lambda}) -\gamma \leq \murt^\ast(\bm{\lambda})  -\gamma.
\label{eq:bound-diff-mu-star-1}
\end{IEEEeqnarray}
If $\murt^\ast(\bm{\lambda}) < \gamma$ then
\begin{IEEEeqnarray}{rCl}
0\leq \frac{\gamma - \murt(\powallocvec_{\mathrm{min}},\bm{\lambda})}{\sqrt{\minant}}  \leq u(\powallocvec_{\mathrm{min}},\bm{\lambda})   \leq \frac{\gamma - \murt^\ast(\bm{\lambda})}{\sigmart^\ast(\bm{\lambda})} \IEEEeqnarraynumspace
\label{eq:bound-u-lambda-vbar}
\end{IEEEeqnarray}
where in the second inequality we used that~$\sigmart(\powallocvec_{\mathrm{min}},\bm{\lambda}) \leq \sqrt{\minant}$ (see~\eqref{eq:bdd-var-csirt}).
Using~\eqref{eq:bound-gamma-star} and~\eqref{eq:bd-lambda1}, we can lower-bound $\sigmart^\ast(\bm{\lambda}) $ as
\begin{IEEEeqnarray}{rCl}
\label{eq:bound-sigma-rt}
\sigmart^\ast(\bm{\lambda}) &\geq& \sqrt{1 - \frac{1}{(1+\lambda_1 \powalloc^*_1)^2}} \\
&\geq& \sqrt{1-\frac{1}{(1+ \snr\constrt_0 /\minant)^2} } \define \constrt_3.
\label{eq:bound-sig-star}
\end{IEEEeqnarray}
Substituting~\eqref{eq:bound-sig-star} into~\eqref{eq:bound-u-lambda-vbar}, we obtain
\begin{IEEEeqnarray}{rCl}
0\leq \gamma - \murt(\powallocvec_{\mathrm{min}},\bm{\lambda}) &\leq& \frac{\sqrt{\minant}}{\const_3}\big[\gamma-\murt^\ast(\bm{\lambda})\big].
\label{eq:bound-diff-mu-star-2}
\end{IEEEeqnarray}
Combining~\eqref{eq:bound-diff-mu-star-2} with~\eqref{eq:bound-diff-mu-star-1} and using that $\sqrt{\minant} / k_3 >1$, we get
\begin{IEEEeqnarray}{rCl}
 \big |\gamma - \murt(\powallocvec_{\mathrm{min}},\bm{\lambda}) \big| \leq \frac{\sqrt{m}}{k_3}\big|\gamma - \murt^\ast(\bm{\lambda})\big|.
\label{eq:conv-asy-bd-mu1-diff}
\end{IEEEeqnarray}
Finally, substituting~\eqref{eq:conv-asy-bd-mu1-diff} into~\eqref{eq:second-ub-on-diff-powallocvec}, then~\eqref{eq:second-ub-on-diff-powallocvec} and~\eqref{eq:bd-lambda1} into~\eqref{eq:bound-eig-E-2},
and writing $ \const \define \constrt_{1}\constrt_{2} \sqrt{\minant}/\const_3 $, we conclude that~\eqref{eq:bound-diff-sigma-min-star} holds for every $\gamma\in(\Ccsit-\tilde{\delta},\Ccsit + \tilde{\delta})$ and every $\bm{\lambda}$ satisfying $|\murt^\ast(\bm{\lambda})-\gamma| <\delta_1$.
To prove~\eqref{eq:positivity-sigma-star-k}, we choose $0<\delta< \min\{\delta_1, \const_3/\const \}$.
It then follows that for every $\veclambda$ satisfying~$|\murt^\ast(\veclambda)-\gamma|<\delta$ we have
\begin{IEEEeqnarray}{rCl}
 \sigmart^\ast(\bm{\lambda}) - \const|\gamma - \murt^\ast(\bm{\lambda})| \geq \const_3 -\const \delta >0.
\label{eq:verify-sigma-positive}
\end{IEEEeqnarray}
Here, in~\eqref{eq:verify-sigma-positive} we used the bound~\eqref{eq:bound-sig-star}. This concludes the proof.

 \subsection{Proof of Lemma~\ref{lem:expectation-phi}}
\label{app:proof-lem-expectation-phi}
By assumption, there exist $\delta>0$ and $\auxconstone<\infty$ such that
 \begin{IEEEeqnarray}{rCl}
 \label{eq:app-bdd-pdfb}
 \sup_{t \in (-\delta,\delta)} \max\mathopen{}\big\{|\pdff_{B}(t)|,|\pdff_{B}'(t)|\big\}\leq \auxconstone.
 \end{IEEEeqnarray}
Let $F_A$ and $\cdfF_B$ be the cdfs of $A$ and $B$, respectively.
We rewrite $\prob[B\geq A/\sqrt{\bl}]$ as follows:
\begin{IEEEeqnarray}{rCl}
\prob[B\geq A/\sqrt{\bl}] &=& \underbrace{\int\nolimits_{|a|\geq\delta\sqrt{\bl}}\prob[B \geq a/\sqrt{\bl}] dF_A(a)}_{\define c_0(n)} \notag\\
&&+\, \int\nolimits_{|a|< \delta\sqrt{\bl}} \underbrace{\prob[B \geq a/\sqrt{\bl}]}_{= 1-\cdfF_B(a/\sqrt{\bl})}dF_A(a).\IEEEeqnarraynumspace
\label{eq:app-prob-x-leq-u}
\end{IEEEeqnarray}
We next expand the argument of the second integral in~\eqref{eq:app-prob-x-leq-u} by applying Taylor's theorem~\cite[Th.~5.15]{rudin76} on $\cdfF_B(a/\sqrt{\bl})$ as follows:
for all $a\in(-\delta \sqrt{\bl},\delta\sqrt{\bl})$
\begin{IEEEeqnarray}{rCl}
1-\cdfF_B(a/\sqrt{\bl})&=& 1-\cdfF_B(0) - \pdff_B(0)\frac{a}{\sqrt{\bl}} - \frac{\pdff'_B(a_0)}{2}\frac{a^2}{\bl}\IEEEeqnarraynumspace
\end{IEEEeqnarray}
for some $a_0 \in (0, a/\sqrt{\bl})$.
Averaging over $A$, we get
\begin{IEEEeqnarray}{rCl}
\IEEEeqnarraymulticol{3}{l}{
\int\nolimits_{|a|<\delta\sqrt{\bl}}1-\cdfF_B(a/\sqrt{\bl})  d F_A(a) } \notag\\
\;\;
&=& \underbrace{(1-\cdfF_B(0))}_{=\prob[B\geq 0]}\prob[|A|< \delta\sqrt{\bl}]\notag\\
&& -\; \frac{\pdff_B(0)}{\sqrt{\bl}}\underbrace{\Ex{}{A \cdot
\indfun{|A| < \delta \sqrt{\bl}}}}_{\define c_1(\bl)} \notag\\
&& -\, \underbrace{\Ex{}{\frac{A^2 \pdff_{B}'(A_0)}{2\bl} \cdot \indfun{|A|<\delta\sqrt{\bl}}}}_{\define c_2(n)}.
\label{eq:taylor_expectation}\IEEEeqnarraynumspace
\end{IEEEeqnarray}
Hence,
\begin{IEEEeqnarray}{rCL}
  %
  \IEEEeqnarraymulticol{3}{l}{\left| \prob[B\geq A/\sqrt{n}] -\prob[B\geq 0]\right|}\\
   &=& \bigg|c_0(n) -\prob[B\geq 0]\cdot\prob[|A|\geq \delta\sqrt{\bl}]\notag\\
   && \quad -\,\frac{\pdff_B(0)}{\sqrt{\bl}}c_1(n)-c_2(n)\bigg| \\
  &\leq & c_0(n) + \prob[|A|\geq \delta\sqrt{\bl}] +\frac{\auxconstone}{\sqrt{\bl}}\abs{c_1(n)} +\abs{c_2(n)}\label{eq:triangle_ineq}\,\,\,\,\\
  &\leq & 2\prob[|A|\geq \delta\sqrt{\bl}] +\frac{\auxconstone}{\sqrt{\bl}}\abs{c_1(n)} +\abs{c_2(n)} \label{eq:triangle_ineq-2}\\
  &\leq & \frac{2}{\delta^2\bl} +\frac{\auxconstone}{\sqrt{\bl}}\abs{c_1(n)} +\abs{c_2(n)}\label{eq:markov_ineq}.
\end{IEEEeqnarray}
Here, in~\eqref{eq:triangle_ineq} we used the triangle inequality together with~\eqref{eq:app-bdd-pdfb} and the trivial bound $\prob[B\geq 0]\leq 1$;
 \eqref{eq:triangle_ineq-2} follows because $ c_0(n) \leq \prob[|A|\geq \delta\sqrt{\bl}]  $;
 \eqref{eq:markov_ineq} follows from Chebyshev's inequality and because $\Ex{}{A^2}=1$ by assumption.

To conclude the proof, we next upper-bound $\abs{c_1(n)}$, and $\abs{c_2(n)}$.
The term $\abs{c_1(\bl)}$ can be bounded as
\begin{IEEEeqnarray}{rCl}
\left|c_1(\bl)\right| 
&=& \big|\Ex{}{A \cdot \indfun{|A|\geq \delta \sqrt{\bl}}}\big| \label{eq:app-bound-x-u-c1-1} \\
 &\leq &\frac{1}{\delta\sqrt{\bl}} \Ex{}{\delta\sqrt{\bl} |A| \cdot \indfun{ |A|\geq \delta \sqrt{\bl} } }\\
 &\leq & \frac{1}{\delta\sqrt{\bl}} \Ex{}{ A^2 \cdot \indfun{|A|\geq \delta \sqrt{\bl}}}\\
&\leq& \frac{1}{\delta\sqrt{\bl}}\label{eq:app-bound-x-u-c1}
\end{IEEEeqnarray}
where \eqref{eq:app-bound-x-u-c1-1} follows because $\Ex{}{A}=0$ by assumption.

Finally, $\abs{c_2(\bl)}$ can be bounded as
\begin{IEEEeqnarray}{rCl}
|c_2(\bl)|&\leq& \Ex{}{\frac{A^2 |\pdff_{B}'(A_0)| }{2\bl} \cdot \indfun{|A|<\delta\sqrt{\bl}}}\\
 &\leq & \Ex{}{A^2\cdot \indfun{|A|<\delta\sqrt{\bl}} } \frac{\auxconstone}{2\bl} \label{eq:app-bound-x-u-c3_b}\\
 &\leq & \frac{\auxconstone}{2\bl}.
 \label{eq:app-bound-x-u-c3}
\end{IEEEeqnarray}
Here, (\ref{eq:app-bound-x-u-c3_b}) follows because the support of $A_0$ is contained in $(0, \delta)$ and from~\eqref{eq:app-bdd-pdfb}.
Substituting \eqref{eq:app-bound-x-u-c1} and~\eqref{eq:app-bound-x-u-c3} into~\eqref{eq:markov_ineq}, we obtain the desired inequality~\eqref{eq:lemma_convergence_speed}.

\section{Proof of the Achievability Part of Theorem~\ref{thm:asy-mimo-csirt}}
\label{app:proof-asy-csit-ach}
%

In order to prove~\eqref{eq:proof-asy-ach-csit}, we study the achievability bound~\eqref{eq:lb-numcode-csit} in the large-\bl limit.
We start by analyzing the denominator on the RHS of~\eqref{eq:lb-numcode-csit}.
Let $\alpha= \bl -\txant -\rxant >0$. Then,
\begin{IEEEeqnarray}{rCl}
\prob\mathopen{}\left[\prod_{i=1}^{\rxant} B_i \leq \gamma_\bl\right] &=& \prob\mathopen{}\left[\prod_{i=1}^{\rxant}B_i^{-\alpha} \geq \gamma_\bl^{-\alpha}\right]\label{eq:app-prob-error2-inner-step0-t}\\
&\leq&  \frac{\Ex{}{\prod_{i=1}^{\rxant} B_i^{-\alpha}}}{\gamma_\bl^{-\alpha}}\label{eq:app-prob-error2-inner-step1-t}\\
&=&\gamma_\bl^{\bl -\txant-\rxant} \prod_{i=1}^{\rxant}\Ex{}{B_i^{-(\bl-\txant -\rxant)}}\label{eq:app-prob-error2-inner-step2-t}
\end{IEEEeqnarray}
where~\eqref{eq:app-prob-error2-inner-step1-t} follows from Markov's inequality, and \eqref{eq:app-prob-error2-inner-step2-t} follows because the $B_1,\ldots,B_{\rxant}$ are independent.
Recalling that $B_i\sim \mathrm{Beta}(\bl-\txant-i+1, \txant)$, we obtain that for every $i\in\{1,\ldots,\rxant\}$
\begin{IEEEeqnarray}{rCl}
\IEEEeqnarraymulticol{3}{l}{
\Ex{}{B_i^{-(\bl-\txant-\rxant)}}
}\notag\\
\quad
&=& \frac{\Gamma(\bl-i+1)}{\Gamma(\bl-\txant-i+1)\Gamma(\txant)}\int\nolimits_{0}^{1} s^{\rxant - i} (1-s)^{\txant-1} ds\IEEEeqnarraynumspace\\
&\leq & \frac{\Gamma(\bl-i+1)}{\Gamma(\bl-\txant-i+1)\Gamma(\txant)}\\
&\leq& \bl^{\txant}\label{eq:app-bound-exp-bi-t}.
\end{IEEEeqnarray}
Substituting (\ref{eq:app-bound-exp-bi-t}) into (\ref{eq:app-prob-error2-inner-step2-t}), we get
\begin{IEEEeqnarray}{rCl}
\prob\mathopen{}\left[\prod_{i=1}^{\rxant} B_i \leq \gamma_\bl\right]&\leq& \bl^{\rxant\txant} \gamma_\bl^{\bl-\txant-\rxant}.\label{eq:asy-analysis-error2-t}
\end{IEEEeqnarray}
Setting $\tau=1/\bl$ and $\gamma_\bl = \exp(-\Ccsit + \bigO(1/\bl))$ in~\eqref{eq:lb-numcode-csit}, and using \eqref{eq:asy-analysis-error2-t}, we obtain
\begin{IEEEeqnarray}{rCl}
\label{eq:ach_logM_MIMO_csit}
\frac{\log \NumCode}{\bl} &\geq&  \Ccsit - (1+\rxant\txant) \frac{\log\bl}{\bl} +\bigO\mathopen{}\left(\frac{1}{n}\right).
\end{IEEEeqnarray}

To conclude the proof, it remains to show that there exists a $\gamma_\bl = \exp(-\Ccsit + \bigO(1/\bl))$ satisfying~\eqref{eq:def-gamma-n-ach-csit}.
 To this end, we note that
 \begin{IEEEeqnarray}{rCl}
\IEEEeqnarraymulticol{3}{l}{
  \prob\mathopen{}\bigg[ \sin^2\mathopen{}\bigg\{\matI_{\bl,\txant}, \sqrt{\bl}\matI_{\bl,\txant}
 \diag\Big\{\sqrt{v_1^{\ast}\Lambda_1},\ldots,\sqrt{v_\minant^{\ast}\Lambda_{\minant}},}\notag\\
 && \hfill  \underbrace{0,\ldots,0}_{\txant-\minant} \Big\}+ \randmatW\bigg\}\leq \gamma_\bl \bigg]    \quad\notag\\
\quad &\geq& \prob\mathopen{}\bigg[ \prod_{j=1}^{\minant}  \sin^2\mathopen{}\left\{ \vece_j, \sqrt{\bl v_j^{\ast}\Lambda_j}\vece_j + \randvecw_j \right\}\leq \gamma_\bl \bigg] \label{eq:ach-csit-angle3}\\
 &=& \prob\mathopen{}\bigg[\prod_{j=1}^{\minant}  \sin^2\mathopen{}\left\{\vece_1, \sqrt{\bl v_j^{\ast}\Lambda_j}\vece_1 + \randvecw_j \right\}\leq \gamma_\bl \bigg].\IEEEeqnarraynumspace\label{eq:ach-csit-angle4}
 \end{IEEEeqnarray}
Here,~\eqref{eq:ach-csit-angle3} follows from Lemma~\ref{lem:angle-btw-subspaces} (Appendix~\ref{app:proof-angle-btw-subspaces}) by letting~$\vece_j$ and~$\randvecw_j$ stand for the $j$th column of~$\matI_{\bl,\txant}$ and $\randmatW$, respectively;~\eqref{eq:ach-csit-angle4} follows by symmetry.
We next note that by~\eqref{eq:lemma-equal-hadamard}, the random variable $\sin^2\{\vece_1, \sqrt{\bl v_j^{\ast}\Lambda_j}\vece_1 + \randvecw_j\}$ has the same distribution as
\begin{IEEEeqnarray}{c}
T_j \define \frac{\sum\nolimits_{i=2}^{\bl} |W_{i,j}|^2}{|\sqrt{\bl v_j^\ast\Lambda_j} + W_{1,j}|^2 + \sum\nolimits_{i=2}^{\bl} |W_{i,j}|^2}.\label{eq:ach-dist-sin-theta-csit}
\end{IEEEeqnarray}
Thus,
 \begin{IEEEeqnarray}{c}
  \prob\mathopen{}\Bigg[\!\prod_{j=1}^{\minant} \! \sin^2\mathopen{}\left\{\vece_1, \!\sqrt{\bl v_j^{\ast}\Lambda_j}\vece_1 + \randvecw_j \right\}\! \leq \gamma_\bl \!\Bigg]
 = \prob\mathopen{}\Bigg[\!\prod\limits_{j=1}^{\minant} \!  T_j\leq \gamma_\bl\! \Bigg].\notag\\
 \label{eq:asy-ach-csit-prob-prod-ts}
 \end{IEEEeqnarray}

To evaluate the RHS of~\eqref{eq:asy-ach-csit-prob-prod-ts}, we observe that by the law of large numbers, the noise term $\frac{1}{\bl}\sum\nolimits_{i=2}^{\bl} |W_{i,j}|^2$ in~\eqref{eq:ach-dist-sin-theta-csit} concentrates around $1$ as $\bl\to\infty$.
Hence, we expect that for all $\gamma > 0$
\begin{IEEEeqnarray}{rCl}
 \prob\mathopen{}\left[\prod\limits_{j=1}^{\minant} T_j\leq \gamma\right] \to \prob\mathopen{}\left[ \prod\limits_{j=1}^{\minant}\frac{1}{v_j^\ast\Lambda_j +1 }\leq \gamma \right]\text{ as } \bl \to\infty.\IEEEeqnarraynumspace
\label{eq:expect-prod-t}
\end{IEEEeqnarray}
We shall next make this statement rigorous by showing that, for all $\gamma$ in a certain neighborhood of~$e^{-\Ccsit}$,
\begin{IEEEeqnarray}{rCL}\label{eq:expansion to be proven}
  \prob\mathopen{}\left[\prod\limits_{j=1}^{\minant} T_j\leq \gamma \right] \geq
  \prob\mathopen{}\left[ \prod\limits_{j=1}^{\minant}\frac{1}{v_j^\ast\Lambda_j +1 }\leq \gamma \right] + \bigO\mathopen{}\left(\frac{1}{n}\right)\IEEEeqnarraynumspace
\end{IEEEeqnarray}
where the term $\bigO(1/n)$ is uniform in $\gamma$.
To this end, we build on Lemma~\ref{lem:expectation-phi} in Appendix~\ref{sec:proof_of_averaging_over_channel}.
The technical difficulty is that the joint pdf of $\Lambda_1 v_1^{\ast},\ldots,\Lambda_\minant v_\minant^{\ast}$ is not continuously differentiable because the functions $\{v_j^{\ast}(\cdot)\}$ are not differentiable on the boundary of the nonintersecting regions $\setW_1,\ldots, \setW_\minant$ defined in~\eqref{eq:ach-asy-csit-def-setw} and~\eqref{eq:ach-asy-csit-def-setw-m}.
To circumvent this problem, we study the asymptotic behavior of~$\{T_j\}$ conditioned on $\bm{\Lambda}\in \Int(\setW_u)$, in which case the joint pdf of $\Lambda_j v_j^\ast(\bm{\Lambda})$, $j=1,\ldots,\minant$, is continuously differentiable.
This comes without loss of generality since~$\bm{\Lambda}$ lies in $\bigcup_{u=1}^{\minant} \mathrm{Int}\mathopen{}\big(\setW_u)$ almost surely (see also Appendix~\ref{sec:proof_of_averaging_over_channel}).

To simplify notation, we use~$T_{j}^{(u)}$ to denote the random variable  $T_j$ conditioned on the event $\bm{\Lambda}\in \Int(\setW_u)$, $u=1,\ldots,\minant$.
We further denote by $\bm{\Lambda}^{(u)}$ and $\altbmLambda^{(u)}$ the random vectors~$\tp{[\Lambda_1,\ldots,\Lambda_u]}$ and $[\Lambda_1 v_1^\ast(\bm{\Lambda}),  \ldots,\Lambda_u v_u^\ast(\bm{\Lambda})\tp{]}$ conditioned on the event $\bm{\Lambda}\in\Int(\setW_u)$, respectively.
Using these definitions, the LHS of~\eqref{eq:expansion to be proven} can be rewritten  as
\begin{IEEEeqnarray}{rCl}
\IEEEeqnarraymulticol{3}{l}{
\prob\mathopen{}\Bigg[\prod\limits_{j=1}^{\minant}   T_j\leq \gamma \Bigg]
}\notag\\
  &=& \sum\limits_{u=1}^{\minant} \Bigg\{ \prob\mathopen{}\Bigg[\prod\limits_{j=1}^{\minant}   T_j\leq \gamma \Big| \bm{\Lambda}\in \Int(\setW_u)\Bigg] \prob[\bm{\Lambda}\in \Int(\setW_u)]\Bigg\}\notag\\
\label{eq:ach-csit-prob-whole-lb-step1} \\
&=&\sum\limits_{u=1}^{\minant} \Bigg\{  \prob\mathopen{}\Bigg[\bigg(\prod\limits_{j=1}^{u}T_j^{(u)}\bigg)  \cdot \underbrace{\bigg(\prod\limits_{j=u+1}^{\minant} \frac{\sum\nolimits_{i=2}^{\bl} |W_{i,j}|^2}{ \sum\nolimits_{i=1}^{\bl} |W_{i,j}|^2}\bigg)}_{\leq 1}
\leq \gamma \Bigg]\notag\\
&&\quad\quad\,\,\times\,  \prob[\bm{\Lambda}\in \Int(\setW_u)]\Bigg\}\IEEEeqnarraynumspace \label{eq:ach-csit-prob-whole-lb-1}\\
&\geq & \sum\limits_{u=1}^{\minant}\Bigg\{ \prob\mathopen{}\Bigg[\prod\limits_{j=1}^{u} T_j^{(u)}  \leq \gamma \Bigg]\prob[\bm{\Lambda}\in \Int(\setW_u)]\Bigg\}.
\label{eq:ach-csit-prob-whole-lb}
\end{IEEEeqnarray}
Here,~\eqref{eq:ach-csit-prob-whole-lb-1} follows because, by~\eqref{eq:expression-gamma-star}, $T_j = (\sum_{i=2}^{\bl} |W_{i,j}|^2) \\ / (\sum_{i=1}^{\bl} |W_{i,j}|^2)$ for $j=u+1,\ldots,\minant$.

The following lemma, built upon Lemma~\ref{lem:expectation-phi}, allows us to establish~\eqref{eq:expansion to be proven}.

\begin{lemma}
\label{lem:prob-prod-cdf}
Let $\randvecg =\tp{[G_1,\ldots,G_u]} \in \realset^{u}_{\geq}$ be a random vector with continuously differentiable joint pdf. Let
\begin{IEEEeqnarray}{rCl}
D_j \define \frac{\sum\nolimits_{i=2}^{\bl} |W_{i,j}|^2}{|\sqrt{\bl G_j} + W_{1,j}|^2 + \sum\nolimits_{i=2}^{\bl} |W_{i,j}|^2},\, j=1,\ldots, u\IEEEeqnarraynumspace
\label{eq:def-D-j}
\end{IEEEeqnarray}
where $W_{i,j}$, $i=1,\ldots, \bl$, $j =1,\ldots, u$,  are i.i.d.$~\jpg(0,1)$-distributed.
Fix an arbitrary $\argpn_0\in(0,1)$.
Then, there exist a $\delta>0$ and a finite constant $\const$ such that
\begin{IEEEeqnarray}{rCl}
\inf_{\argpn\in (\argpn_0-\delta, \argpn_0+\delta)}\!\! \Bigg(\!\prob\mathopen{}\Bigg[\prod\limits_{j=1}^{u} D_j \leq \argpn \Bigg] - \prob\mathopen{}\Bigg[\prod\limits_{j=1}^{u} \frac{1}{1+G_j} \leq \argpn\Bigg]\! \Bigg) > \frac{\const}{\bl}.\notag\\
\IEEEeqnarraynumspace\label{eq:lemma-prod-cdf}
\end{IEEEeqnarray}
\end{lemma}
\begin{IEEEproof}
See Appendix \ref{app:proof-ach-prod-cdf}.
\end{IEEEproof}

Using Lemma~\ref{lem:prob-prod-cdf} with $G_j = \altLambda_j^{(u)} $, it follows that there exist $\delta_u >0$ and $0\leq \const_u <\infty$, such that for every $\gamma \in \big(e^{-\Ccsit-\delta_u},e^{ -\Ccsit +\delta_u }\big)$
\begin{IEEEeqnarray}{rCl}
\prob\mathopen{}\left[\prod\limits_{j=1}^{u} T_{j}^{(u)} \leq \gamma \right] \geq  \prob\mathopen{}\left[\prod\limits_{j=1}^{u}\frac{1}{1+\widetilde{\Lambda}_j^{(u)}} \leq \gamma \right] + \bigO\mathopen{}\left(\frac{1}{\bl}\right).\IEEEeqnarraynumspace\label{eq:ach-csit-prob-region-lb}
\end{IEEEeqnarray}
To show that $\altLambda_j^{(u)}$, $j=1,\ldots,u$, indeed satisfy the conditions in Lemma~\eqref{lem:prob-prod-cdf}, we use~\eqref{eq:expression-gamma-star},~\eqref{eq:water-filling-power}, and~\eqref{eq:def-gamma-bar}, to obtain
\begin{IEEEeqnarray}{rCl}
\label{eq:ach-csit-change-vari-cond}
\altLambda_j^{(u)} =    \frac{\Lambda_j^{(u)} }{u}\left(\snr+\sum\limits_{l=1}^{u}\frac{1}{\Lambda_l^{(u)}} \right)-1,\quad j=1,\ldots,u.\IEEEeqnarraynumspace
\end{IEEEeqnarray}
Since the joint pdf of $\bm{\Lambda}$ is continuously differentiable by assumption, the joint pdf of $\bm{\Lambda}^{(u)}$ is also continuously differentiable. Moreover, it can be shown that the transformation $\bm{\Lambda}^{(u)}   \mapsto \altbmLambda^{(u)}$ defined by~\eqref{eq:ach-csit-change-vari-cond} is a diffeomorphism of class $C^2$~\cite[p.~147]{munkres91-a}.
Therefore, the joint pdf of $\altbmLambda^{(u)}$ is continuously differentiable.

We next use~\eqref{eq:ach-csit-prob-region-lb} in~\eqref{eq:ach-csit-prob-whole-lb} to conclude that for every $\gamma\! \in\! \big(e^{-\Ccsit-\deltalow}, e^{-\Ccsit + \deltalow} \big)$ (where  $\deltalow \define \min\{\delta_1,\ldots,\delta_{\minant}\}$)
\begin{IEEEeqnarray}{rCl}
\IEEEeqnarraymulticol{3}{l}{
\prob\mathopen{}\Bigg[\prod\limits_{u=1}^{\minant}   T_j\leq \gamma \Bigg]}\notag\\
&\geq&  \sum\limits_{u=1}^{\minant}\Bigg\{   \prob\mathopen{}\Bigg[\prod\limits_{j=1}^{u} \frac{1}{1+\widetilde{\Lambda}_j^{(u)}} \leq \gamma \Bigg] \prob[\bm{\Lambda}\in \Int(\setW_u)]\Bigg\} + \bigO\mathopen{}\left(\frac{1}{\bl}\right)  \notag\\
&&\\
&=& \prob\mathopen{}\Bigg[\prod\limits_{j=1}^{\minant} \frac{1}{1 +  \Lambda_j v_j^\ast(\bm{\Lambda})} \leq \gamma\Bigg] + \bigO\mathopen{}\left(\frac{1}{\bl}\right)\\
&=& 1- \prob\mathopen{}\Bigg[ \sum\limits_{j=1}^{\minant}\log(1 +  \Lambda_j v_j^\ast(\bm{\Lambda})) \leq -\log \gamma\Bigg] + \bigO\mathopen{}\left(\frac{1}{\bl}\right) \\
&=& 1- \cdistcsit( -\log \gamma) + \bigO\mathopen{}\left(\frac{1}{\bl}\right)
\label{eq:ach-csit-prob-whole-lb2}
\end{IEEEeqnarray}
where $\cdistcsit(\cdot)$ is given in~\eqref{eq:cadist_csit}.

We next choose $\gamma_\bl$ as the solution of
\begin{IEEEeqnarray}{rCl}
1- \cdistcsit( -\log \gamma_\bl) + \bigO\mathopen{}\left(\frac{1}{\bl}\right)  = 1- \error + \frac{1}{\bl}.
\label{eq:ach-csit-gamma-bl-0}
\end{IEEEeqnarray}
Since $\cdistcsit(\Ccsit)=\error $ and $\cdistcsit'(\Ccsit)>0$, it follow from  Taylor's theorem that
\begin{IEEEeqnarray}{rCl}
-\log \gamma_\bl &=& \Ccsit +\bigO\mathopen{}\left(\frac{1}{\bl}\right) \label{eq:ach-csit-gamma-bl}.
\end{IEEEeqnarray}
So, for sufficiently large \bl, $\gamma_\bl$ in~\eqref{eq:ach-csit-gamma-bl} belongs to the interval $\big(e^{-\Ccsit-\deltalow}, e^{-\Ccsit + \deltalow} \big)$.
Hence, by~\eqref{eq:asy-ach-csit-prob-prod-ts},~\eqref{eq:ach-csit-prob-whole-lb2}, and~\eqref{eq:ach-csit-gamma-bl-0}, this~$\gamma_\bl$ satisfies~\eqref{eq:def-gamma-n-ach-csit}.
This concludes the proof.

%

\subsection{Proof of Lemma \ref{lem:prob-prod-cdf}}
\label{app:proof-ach-prod-cdf}
Choose $\delta>0$ such that $\delta \leq \argpn_0/2$.
Throughout this appendix, we shall use $\constrm$ to indicate a finite constant that does neither depend on $\argpn\in(\argpn_0-\delta,\argpn_0+\delta)$ nor on $\bl$; its magnitude and sign may change at each occurrence.

Let $\gth \define 2/\argpn_0-1$ and let
\begin{IEEEeqnarray}{rCl}
p_1 &\define& \prob\mathopen{}\Bigg[\prod\limits_{j=1}^{u}D_j\leq \argpn \Bigg| G_1\geq \gth \Bigg]  \\
 p_2 &\define& \prob\mathopen{}\Bigg[\prod\limits_{j=1}^{u}D_j\leq \argpn \Bigg| G_1< \gth \Bigg].
\end{IEEEeqnarray}
To prove Lemma~\ref{lem:prob-prod-cdf}, we decompose $\prob\mathopen{}\big[\prod\nolimits_{j=1}^{u}D_j\leq \argpn \big]$ as
\begin{IEEEeqnarray}{rCl}
\prob\mathopen{}\Bigg[\prod\limits_{j=1}^{u}D_j\leq \argpn \Bigg]&=&  p_1 \prob\mathopen{}\left[G_1\geq \gth\right] +  p_2\prob\mathopen{}\left[G_1< \gth \right].\IEEEeqnarraynumspace
\label{eq:ach-prob-prod-leq-gamma-1}
\end{IEEEeqnarray}
The proof consists of the following steps:
\begin{enumerate}
\item\label{item:bounding-p1}  We show in Section~\ref{app:item-bound-p1} that for every $\argpn\in(\argpn_0-\delta, \argpn_0+\delta)$, the term $p_1$ in~\eqref{eq:ach-prob-prod-leq-gamma-1} can be lower-bounded as
\begin{IEEEeqnarray}{rCl}
p_1&\geq&  1-  \frac{\constrm}{\bl}.
\label{eq:ach-lb-p1-4-ol}
\end{IEEEeqnarray}

\item \label{item:altD-1}

 Using Lemma~\ref{lem:expectation-phi} in Appendix~\ref{sec:proof_of_averaging_over_channel}, we show in Section~\ref{app:item-altD-1} that~$p_2$ can be lower-bounded as
\begin{IEEEeqnarray}{rCl}
p_2 &\geq& \prob\mathopen{}\Bigg[ \frac{1}{1 + G_1}\prod\limits_{j=2}^{u} D_j \leq \argpn \Bigg| G_1< \gth \Bigg] - \frac{\constrm}{\bl}. \IEEEeqnarraynumspace
\label{eq:ach-nocsit-lb-prod-first-term-ol}
\end{IEEEeqnarray}
\item\label{item:bounding-p2} Reiterating Step~\ref{item:altD-1} for $D_2,\ldots, D_u$, we conclude that~\eqref{eq:ach-nocsit-lb-prod-first-term-ol} can be further lower-bounded as
\begin{IEEEeqnarray}{rCl}
p_2 
&\geq & \prob\mathopen{}\Bigg[\prod\limits_{j=1}^{u} \frac{1}{1 + G_j}\leq \argpn \Bigg| G_1< \gth\Bigg]- \frac{\constrm}{\bl}.
\label{eq:ach-nocsit-lb-prod-all-term} \IEEEeqnarraynumspace
\end{IEEEeqnarray}
\item\label{item:end-step} Finally, using~\eqref{eq:ach-lb-p1-4-ol} and~\eqref{eq:ach-nocsit-lb-prod-all-term} in~\eqref{eq:ach-prob-prod-leq-gamma-1}, we show in Section~\ref{app:item-end-step} that
\begin{IEEEeqnarray}{rCl}
\prob\mathopen{}\Bigg[\prod\limits_{j=1}^{u} D_j \leq \argpn\Bigg]& \geq & \prob\Bigg[\prod\limits_{j=1}^{u} \frac{1}{1+G_j} \leq \argpn \Bigg] - \frac{\constrm}{\bl}. \IEEEeqnarraynumspace
\label{eq:ach-lb-prod-tj-1-ol}
\end{IEEEeqnarray}
This proves Lemma~\ref{lem:prob-prod-cdf}.

\end{enumerate}

\subsubsection{Proof of~\eqref{eq:ach-lb-p1-4-ol}}
\label{app:item-bound-p1}
%
%
Let $\delta_1$ be an arbitrary real number in $(1/( \argpn_0-\delta), 2/\argpn_0)$ and let $\delta_2\define \sqrt{\gth} - \sqrt{\delta_1-1}>0$.
Let $W_{\bl+1,1}\sim \jpg(0,1)$ be independent of all other random variables appearing in the definition of the $\{D_j\}$ in~\eqref{eq:def-D-j}.
Finally, let $W_{\mathrm{re}} $ denote the real part of $W_{1,1}$.
For every $\argpn\in(\argpn_0-\delta,\argpn_0+\delta)$
\begin{IEEEeqnarray}{rCl}
p_1&\geq& \prob\mathopen{}\left[D_1 \leq \argpn \big| G_1\geq \gth \right]\label{eq:ach-lb-p1-1}\\
&\geq& \prob\mathopen{}\Bigg[\left|\sqrt{\bl G_1} + W_{1,1}\right|^2 \geq \frac{ 1 - \argpn }{\argpn} \sum\limits_{i=2}^{\bl} |W_{i,1}|^2,\notag\\
  &&  \quad\quad W_{\mathrm{re}}\geq -\sqrt{\bl}\delta_2 \Bigg| G_1\geq \gth \Bigg]\\
&\geq& \prob\mathopen{}\left[\bl (\sqrt{G_1}-\delta_2)^2 \geq  \frac{ 1 - \argpn }{\argpn}\sum\limits_{i=2}^{\bl} |W_{i,1}|^2 \Bigg|G_1\geq \gth \right]
\notag\\
&&\times\, \prob\mathopen{}\big[ W_{\mathrm{re}}\geq -\sqrt{\bl}\delta_2\big] \\
&\geq& \prob\mathopen{}\left[\bl (\delta_1-1)  \geq \frac{ 1 - \argpn }{\argpn} \sum\limits_{i=2}^{\bl} |W_{i,1}|^2\right]
\prob\mathopen{}\big[W_{\mathrm{re}}\geq -\sqrt{\bl}\delta_2\big]\notag\\
&& \label{eq:ach-lb-p1-12}\\
&\geq& \prob\mathopen{}\left[\bl (\delta_1-1) \geq  \big(1/(\argpn_0-\delta) -1 \big) \sum\limits_{i=2}^{\bl+1} |W_{i,1}|^2\right]
\notag\\
&&\times\,
\prob\mathopen{}\big[|W_{\mathrm{re}}| \leq \sqrt{\bl}\delta_2\big]
\label{eq:ach-lb-p1-2}\\
&\geq& \left(1 -\frac{1}{\bl} \left(\frac{\delta_1(\argpn_0-\delta)-1}{1-(\argpn_0-\delta)}\right)^2 \right)\left( 1- \frac{1}{2\bl\delta_2^2}\right)\label{eq:ach-lb-p1-3}\\
&\geq& 1- \frac{\constrm}{\bl} . \label{eq:ach-lb-p1-4}
\end{IEEEeqnarray}
Here,~\eqref{eq:ach-lb-p1-1} follows because $D_i\leq 1$, $i=2,\ldots,u$, with probability one (see~\eqref{eq:def-D-j});
\eqref{eq:ach-lb-p1-12} follows because $\delta_1-1 = (\sqrt{\gth} -\delta_2)^2$;
\eqref{eq:ach-lb-p1-2} follows because $\argpn > \argpn_0-\delta$ and because $\sum\nolimits_{i=2}^{\bl+1}|W_{i,1}|^2$ is stochastically larger than $\sum\nolimits_{i=2}^{\bl}|W_{i,1}|^2$;
\eqref{eq:ach-lb-p1-3} follows from Chebyshev's inequality applied to both probabilities in~\eqref{eq:ach-lb-p1-2}. This proves~\eqref{eq:ach-lb-p1-4-ol}.

Before proceeding to the next step, we first argue that, if $\prob[G_1\geq \gth]=1$, then~\eqref{eq:lemma-prod-cdf} follows directly from~\eqref{eq:ach-lb-p1-4}. Indeed, in this case we obtain from~\eqref{eq:ach-lb-p1-4} and~\eqref{eq:ach-prob-prod-leq-gamma-1} that
\begin{IEEEeqnarray}{rCl}
\prob\mathopen{}\Bigg[\prod\limits_{j=1}^{u}D_j  \leq \argpn \Bigg] =p_1 \geq 1-\frac{\constrm}{\bl} .
\label{eq:ach-prob-prod-leq-gamma-spec}
\end{IEEEeqnarray}
We further have, with probability one,
\begin{IEEEeqnarray}{rCl}
\prod\limits_{j=1}^{u}\frac{1}{1+G_j} \leq \frac{1}{1+G_1} \leq \frac{1}{1+\gth} = \frac{\argpn_0}{2} \leq \argpn_0-\delta < \argpn \IEEEeqnarraynumspace
\label{eq:exception-case}
\end{IEEEeqnarray}
which gives
\begin{IEEEeqnarray}{rCl}
\prob\mathopen{}\Bigg[\prod\limits_{j=1}^{u}\frac{1}{1+G_j}  \leq \argpn \Bigg]=1.
\label{eq:ach-prod-prob-G-spec}
\end{IEEEeqnarray}
Subtracting~\eqref{eq:ach-prob-prod-leq-gamma-spec} from~\eqref{eq:ach-prod-prob-G-spec} yields~\eqref{eq:lemma-prod-cdf}.
In the following, we shall focus exclusively on the case  $\prob[G_1\geq \gth]<1$.

\subsubsection{Proof of~\eqref{eq:ach-nocsit-lb-prod-first-term-ol}}
\label{app:item-altD-1}
To evaluate $p_2$ in~\eqref{eq:ach-prob-prod-leq-gamma-1}, we proceed as follows.
 Defining~$\randratio \define \argpn /\prod\nolimits_{j=2}^{u} D_j$, we obtain
\begin{IEEEeqnarray}{rCl}
p_2
&=& \prob\mathopen{}\Bigg[\prod\limits_{j=1}^{u}D_j\leq \argpn \Bigg| G_1< \gth \Bigg] \\
&=& \prob\mathopen{}\left[ D_1 \leq \randratio \big|  G_1< \gth \right]\\
&=& \prob\mathopen{}\left[ D_1  \leq \randratio ,\, \randratio\geq 1\big|  G_1< \gth  \right] \notag\\
&&+\, \prob\mathopen{}\left[ D_1 \leq \randratio, \, \randratio<1\big|  G_1< \gth  \right]\IEEEeqnarraynumspace\\
&=&  \prob\mathopen{}\left[ \randratio \geq 1\big|  G_1< \gth  \right] + \prob\mathopen{}\left[ D_1 \leq \randratio, \, \randratio<1 \big|  G_1< \gth  \right] \notag\\
&& \label{eq:ach-nocsit-lb-prod}
\end{IEEEeqnarray}
where~\eqref{eq:ach-nocsit-lb-prod} follows because
\begin{IEEEeqnarray}{c}
\prob\mathopen{}\left[ D_1  \leq \randratio  \big|\randratio\geq 1 ,\,  G_1< \gth  \right]=1.
\end{IEEEeqnarray}
 The second term on the RHS of~\eqref{eq:ach-nocsit-lb-prod} can be rewritten as
\begin{IEEEeqnarray}{rCl}
\IEEEeqnarraymulticol{3}{l}{
\prob\mathopen{}\left[ D_1 \leq \randratio, \, \randratio<1 \big|  G_1< \gth  \right]
}\notag
\\   \,\,\,
&=&\opE_{\randratio, G_2,\ldots, G_{u}\given  G_1<\gth}\mathopen{}\Big[ \indfun{\randratio < 1} \notag\\
 &&\quad\quad\quad\,\, \times\, \prob\mathopen{}\big[D_1\leq \randratio \big| \randratio, G_2,\ldots, G_u, G_1<\gth \big]\Big].\label{eq:ach-prob-avg-term1}\IEEEeqnarraynumspace
\end{IEEEeqnarray}
%
Since events of measure zero do not affect~\eqref{eq:ach-prob-avg-term1}, we can assume without loss of generality  that the conditional joint pdf of $Z,G_2,\ldots,G_u $ given $G_1 < \gth$ is strictly positive.
To lower-bound~\eqref{eq:ach-prob-avg-term1}, we first bound the conditional probability $\prob\mathopen{}\left[D_1\leq \randratio \big| \randratio, G_2,\ldots, G_u, G_1<\gth \right]$.
Again, let $W_{\mathrm{re}}$ denote the real part of $W_{1,1}$, and let $W_{n+1,1}\sim \jpg(0,1)$ be independent of all other random variables appearing in the definition of the $\{D_j\}$ in~\eqref{eq:def-D-j}.
Following similar steps as in~\eqref{eq:ach-lb-p1-1}--\eqref{eq:ach-lb-p1-4}, we obtain for $Z<1$
\begin{IEEEeqnarray}{rCl}
\IEEEeqnarraymulticol{3}{l}{
\prob\mathopen{}\left[D_1 \leq \randratio \given \randratio, G_2 ,\ldots, G_u, G_1<\gth\right]}\notag\\
\,\,\,\,&=& \prob\mathopen{}\Bigg[ \frac{\sum\nolimits_{i=2}^{\bl} |W_{i,1}|^2 }{\big|\sqrt{\bl G_1} + W_{1,1} \big|^2 + \sum\nolimits_{i=2}^{\bl} |W_{i,1}|^2}
\leq \randratio \Bigg|  \notag\\
&&
\hfill \randratio, G_2, \ldots,G_u, G_1<\gth \Bigg]\IEEEeqnarraynumspace\\
&=& \prob\mathopen{}\Bigg[\left|\sqrt{\bl G_1} + W_{1,1}\right|^2 \geq \big(\randratio^{-1}-1\big)\sum\limits_{i=2}^{\bl} |W_{i,1}|^2 \Bigg| \notag\\
&& \hfill \randratio,  G_2, \ldots, G_u, G_1<\gth  \Bigg]\IEEEeqnarraynumspace\\
& \geq & \prob\mathopen{}\Bigg[ \Big|\sqrt{\bl G_1} + W_{\mathrm{re}} \Big|^2 \geq \big(\randratio^{-1}-1\big)\sum\limits_{i=2}^{\bl+1} |W_{i,1}|^2 \Bigg|\notag\\
&& \hfill \randratio,  G_2,  \ldots, G_u, G_1<\gth \Bigg] \label{eq:ach-cond-prob1-2}\IEEEeqnarraynumspace\\
& \geq &\prob\mathopen{}\Bigg[\sqrt{\bl G_1} \geq - W_{\mathrm{re}} +\sqrt{\randratio^{-1}-1}\sqrt{\sum\nolimits_{i=2}^{\bl+1} |W_{i,1}|^2} \Bigg| \quad \notag\\
&&\hfill \randratio, G_2,\ldots, G_u, G_1<\gth \Bigg].\label{eq:ach-cond-prob1} \IEEEeqnarraynumspace
\end{IEEEeqnarray}
%
%

Next, we lower-bound the RHS of~\eqref{eq:ach-cond-prob1} using Lemma~\ref{lem:expectation-phi} in Appendix~\ref{sec:proof_of_averaging_over_channel}.
Let $\mu_{W}$ and $\sigma_{W}^2$ be the mean and the variance of the random variable $\sqrt{\sum\nolimits_{i=2}^{\bl+1} |W_{i,1}|^2}$.
Let $\randratiotwo\define \sqrt{ \randratio^{-1}-1}$.
Furthermore, let
\begin{IEEEeqnarray}{rCl}
K_1 \define \frac{1}{\sqrt{1/2+ \randratiotwo^2\sigma_W^2}} \Bigg(\!\!-W_{\mathrm{re}} + \randratiotwo\sqrt{\sum\limits_{i=2}^{\bl+1} |W_{i,1}|^2} -\mu_W \randratiotwo \!\Bigg)\notag\\
&& \label{eq:def-K1}
\end{IEEEeqnarray}
and
\begin{IEEEeqnarray}{rCl}
\overline{G}_1\define  \frac{1}{\sqrt{1/2+\randratiotwo^2\sigma_W^2}}  \left(\sqrt{G_1} - \frac{ \mu_W}{\sqrt{\bl} }\randratiotwo\right).
\label{eq:def-overlineG-1}
\end{IEEEeqnarray}
Note that $K_1$ is a zero-mean, unit-variance random variable that is conditionally independent of~$\overline{G}_1$ given $Z_2$.
Using these definitions, we can rewrite the RHS of~\eqref{eq:ach-cond-prob1} as
\begin{IEEEeqnarray}{rCl}
\prob\mathopen{}\left[  \overline{G}_1  \geq K_1 /\sqrt{\bl} \Big| \randratiotwo, G_2,\ldots, G_u, G_1<\gth \right].
\label{eq:ach-prob-equivent-form-1}
\end{IEEEeqnarray}
In order to use Lemma~\ref{lem:expectation-phi}, we need to establish an upper bound on the conditional pdf of ${\overline{G}_1}$ given $\randratiotwo, G_2,\ldots, G_u$ and $G_1<\gth $, which we denote by $\condpdfbarG$, and on its derivative.
As $f_{G_1,\ldots,G_u}$ is continuously differentiable by assumption, $f_{G_1,\ldots,G_u}$ and its partial derivatives are bounded on bounded sets.
Together with the assumption that $\prob[G_1 \geq \gth] <1$,  this implies that the conditional pdf $f_{G_1,\ldots,G_u \given G_1<\gth }$ of $G_1,\ldots,G_u$ given $G_1<\gth$ and its partial derivatives are all bounded on $[0,\gth)^{u}$. Namely, for every $\{x_1,\dots,x_u\} \in [0,\gth)^u$ and every $i\in\{1,\ldots,u\}$
\begin{IEEEeqnarray}{rCl}
f_{G_1,\ldots,G_u\given G_1<\gth }(x_1,\ldots,x_u)  &\leq& \constrm \label{eq:app-bound-pdf-G1}\\
\left|\frac{\partial f_{G_1,\ldots,G_u\given G_1<\gth }(x_1,\ldots,x_u)}{\partial x_i}\right| &\leq& \constrm \label{eq:app-bound-pdf-G1-d}. \IEEEeqnarraynumspace
\end{IEEEeqnarray}
%
%
Let $\condpdfG$ be the conditional pdf of $G_1$ given $G_2,\ldots,G_u$ and $G_1<\gth $, and let $f_{G_2,\ldots, G_u \given G_1<\gth}$ be the conditional pdf of $G_2,\ldots,G_u$ given $G_1<\gth$.
 Then, $\condpdfbarG$ can be bounded as
\begin{IEEEeqnarray}{rCl}
\IEEEeqnarraymulticol{3}{l}{
\condpdfbarG(x\given \ratiotwo, g_2\ldots,g_u)}\notag\\
&=& 2 \condpdfG\mathopen{}\bigg(\Big(\sqrt{1/2+ z_2^2\sigma^2_{W}} x + \ratiotwo\mu_{W}/\sqrt{\bl}\Big)^2\bigg| g_2,\ldots,g_u\! \bigg)
\notag\\
&&\times \, \sqrt{1/2+ \ratiotwo^2\sigma^2_{W}} \left(\sqrt{1/2 + \ratiotwo^2\sigma^2_{W}} x + \ratiotwo\mu_{W}/\sqrt{\bl}\right)\IEEEeqnarraynumspace \label{eq:app-bound-pdf-barG1}\\
&\leq & \frac{\constrm \cdot  \sqrt{\gth} \sqrt{1/2+ \sigma^2_{W}\ratiotwo^2}  }{f_{G_2,\ldots, G_u \given G_1<\gth}(g_2,\ldots,g_u)}\label{eq:app-bound-pdf-barG2}.
\end{IEEEeqnarray}
Here, \eqref{eq:app-bound-pdf-barG1} follows from \eqref{eq:def-overlineG-1}, and \eqref{eq:app-bound-pdf-barG2} follows from~\eqref{eq:app-bound-pdf-G1} and because
we condition on the event that $G_1 < \gth$, so
\begin{IEEEeqnarray}{rCl}
\sqrt{1/2 + \ratiotwo^2\sigma^2_{W}} x + \ratiotwo\mu_{W}/\sqrt{\bl} \leq \sqrt{\gth}.
\end{IEEEeqnarray}
%
 To further upper-bound~\eqref{eq:app-bound-pdf-barG2}, we shall use that $\sigma_W$ and~$\randratiotwo$ are bounded:
\begin{IEEEeqnarray}{rCl}
\sigma_W^2 &=& \bl - \left(\frac{\Gamma(\bl+1/2)}{\Gamma(\bl)}\right)^2\label{eq:app-bound-sigma_W-1}\\
& \leq& 1/4\label{eq:app-bound-sigma_W-3}
\end{IEEEeqnarray}
and
\begin{IEEEeqnarray}{rCl}
\randratiotwo^2 &=&  \randratio^{-1} -1 \\
&\leq& 1/\argpn-1 \label{eq:app-bound-arpgn-2-1}\\
&\leq& (\argpn_0-\delta)^{-1}-1. \label{eq:app-bound-arpgn-2}
\end{IEEEeqnarray}
Here,~\eqref{eq:app-bound-sigma_W-1} follows by using that $\sqrt{2\sum\nolimits_{i=2}^{\bl+1} |W_{i,1}|^2}$ is $\chi$-distributed with $2\bl$ degrees of freedom and by using~\cite[Eq.~(18.14)]{johnson95-1};~\eqref{eq:app-bound-sigma_W-3} follows from~\cite[Sec.~2.2]{qi12-02};~\eqref{eq:app-bound-arpgn-2-1} follows from the definition of $Z$ and because $\prod_{j=2}^{u}D_j \leq 1$. Substituting~\eqref{eq:app-bound-sigma_W-3} and~\eqref{eq:app-bound-arpgn-2} into~\eqref{eq:app-bound-pdf-barG2}, we obtain
\begin{IEEEeqnarray}{rCl}
\condpdfbarG (x\given \ratiotwo, g_2\ldots,g_u) &\leq& \frac{\constrm}{f_{G_2,\ldots, G_u\given G_1<\gth}(g_2,\ldots,g_u)}. \IEEEeqnarraynumspace\label{eq:app-bound-pdf-barG3}
\end{IEEEeqnarray}
Following similar steps, we can also establish that
\begin{IEEEeqnarray}{c}
 \left| \condpdfbarG'(x \given \ratiotwo, g_2\ldots,g_u)\right|\leq\frac{\constrm}{f_{G_2,\ldots, G_u\given G_1<\gth}(g_2,\ldots,g_u)}. \notag\\
  \label{eq:app-bound-pdf-barG-2}
\end{IEEEeqnarray}
Using~\eqref{eq:app-bound-pdf-barG3}--\eqref{eq:app-bound-pdf-barG-2} and Lemma~\ref{lem:expectation-phi}, we obtain that
\begin{IEEEeqnarray}{rCl}
\IEEEeqnarraymulticol{3}{l}{
\prob\mathopen{}\left[  \overline{G}_1  \geq K_1 /\sqrt{\bl} \Big| \randratiotwo ,G_2=g_2, \ldots, G_u=g_u, G_1<\gth\right] } \notag\\
\quad &\geq& \prob\mathopen{}\left[  \overline{G}_1 \geq 0  \Big| \randratiotwo ,G_2=g_2, \ldots, G_u=g_u,G_1<\gth\right] \notag\\
 &&- \, \frac{\constrm}{\bl}\left(1 + \frac{1}{f_{G_2,\ldots, G_u\given G_1<\gth}(g_2,\ldots,g_u)}\right) \label{eq:ach-nocsit-bdd-G1-2}. \IEEEeqnarraynumspace
\end{IEEEeqnarray}
%
%
 Returning to the analysis of~\eqref{eq:ach-prob-avg-term1}, we combine \eqref{eq:ach-cond-prob1},~\eqref{eq:ach-prob-equivent-form-1} and~\eqref{eq:ach-nocsit-bdd-G1-2} to obtain
\begin{IEEEeqnarray}{rCl}
\IEEEeqnarraymulticol{3}{l}{
\prob\mathopen{}\left[ D_1 \leq \randratio, \, \randratio<1 \big|  G_1< \gth  \right]  }\notag\\
&\geq & \opE_{\randratio, G_2,\ldots,G_u \given G_1 < \gth}\Bigg[ \indfun{\randratio< 1} \notag\\
&&\times \Bigg(\prob\mathopen{}\Big[ \overline{G}_1 \geq 0 \Big| \randratio, G_2,\ldots,G_u,G_1<\gth\Big] \notag\\
&&- \, \frac{\constrm}{\bl}\bigg(1 + \frac{1}{f_{G_2,\ldots, G_u\given G_1<\gth}(G_2,\ldots,G_u)}\bigg)\!\Bigg)\Bigg] \label{eq:ach-prob-avg-term2-step1}\\
&\geq & \prob\mathopen{}\left[\frac{1}{1+ \bl G_1/\mu_W^2 }  \leq   \randratio, \randratio<1  \bigg| G_1<\gth \right]- \frac{\constrm}{\bl} \notag\\
&&\times \Biggl(1 + \!\! \int_{0}^{\gth} \!\!\cdots\!\!\int_{0}^{\gth}\!\frac{ f_{G_2,\ldots,G_{u}\given G_1<\gth }(g_2,\ldots,g_{u})}{ f_{G_2,\ldots, G_{u} \given G_1<\gth }(g_2,\ldots,g_{u})}  dg_2 \cdots dg_{u}\!\Biggr) \notag\\
&&\label{eq:ach-prob-avg-term2-step2}\\
&\geq & \prob\mathopen{}\left[\frac{1}{1+  G_1 }  \leq\randratio, \randratio<1 \bigg| G_1<\gth\right] -\frac{\constrm}{\bl}\label{eq:ach-prob-avg-term2}.
\end{IEEEeqnarray}
Here, in~\eqref{eq:ach-prob-avg-term2-step2} we used~\eqref{eq:def-overlineG-1}, that $\indfun{\randratio< 1} \leq 1$, that $G_1,\ldots,G_{u}$ are nonincreasing, and that $\constrm$ in~\eqref{eq:ach-prob-avg-term2-step1} is positive;~\eqref{eq:ach-prob-avg-term2} follows because~\cite[Eq.~(18.14)]{johnson95-1}
\begin{IEEEeqnarray}{rCl}
\mu_W =\frac{\Gamma(\bl+1/2)}{\Gamma(\bl)} \leq \sqrt{\bl}
\end{IEEEeqnarray}
and because the integral on the RHS of~\eqref{eq:ach-prob-avg-term2-step2} is bounded.
Substituting (\ref{eq:ach-prob-avg-term2}) into \eqref{eq:ach-nocsit-lb-prod}, we obtain
\begin{IEEEeqnarray}{rCl}
p_2 &\geq& \prob\mathopen{}\left[\randratio \geq 1 \given G_1<\gth\right] \notag\\
 && +\, \prob\mathopen{}\left[ \frac{1}{1+G_1} \leq \randratio, \randratio<1  \bigg| G_1<\gth\right]-\frac{\constrm}{\bl}\\
&=&  \prob\mathopen{}\left[\frac{1}{1+G_1} \leq \randratio, \randratio \geq  1  \bigg| G_1<\gth\right]
\notag\\
   &&  +\,   \prob\mathopen{}\left[\frac{1}{1+G_1} \leq \randratio, \randratio<1  \bigg| G_1<\gth\right]  -\frac{\constrm}{\bl} \IEEEeqnarraynumspace \label{eq:ach-p2-final-prob-2}\\
&=& \prob\mathopen{}\left[\frac{1}{1+G_1} \leq \randratio \bigg| G_1<\gth \right] -\frac{\constrm}{\bl}\\
&=& \prob\mathopen{}\Bigg[\frac{1}{1 + G_1}\prod\limits_{j=2}^{u} D_j \leq \argpn \Bigg| G_1< \gth \Bigg]  -\frac{\constrm}{\bl}
\end{IEEEeqnarray}
where~\eqref{eq:ach-p2-final-prob-2} follows because $1/(1+G_1) \leq 1$ with probability one. This proves~\eqref{eq:ach-nocsit-lb-prod-first-term-ol}.

\subsubsection{Proof of~\eqref{eq:ach-lb-prod-tj-1-ol}}
\label{app:item-end-step}
Set $p_0\define \prob[G_1\geq \gth]$. Substituting~\eqref{eq:ach-lb-p1-4-ol} and~\eqref{eq:ach-nocsit-lb-prod-all-term} into~\eqref{eq:ach-prob-prod-leq-gamma-1}, we obtain
\begin{IEEEeqnarray}{rCl}
\IEEEeqnarraymulticol{3}{l}{\prob \mathopen{}\Bigg[\prod\limits_{j=1}^{u} D_j \leq \argpn\Bigg]}\notag\\
\,\,& \geq & p_0 +  (1-p_0) \prob\mathopen{}\Bigg[\prod\limits_{j=1}^{u} \frac{1}{1 + G_j}\leq \argpn \Bigg| G_1< \gth\Bigg]  - \frac{\constrm}{\bl}\notag \\
&&\\
&=&  \underbrace{\prob\mathopen{}\Bigg[\prod\limits_{j=1}^{u}\frac{1}{1+G_j} \leq \argpn \Bigg| G_1\geq \gth \Bigg]}_{=1} p_0
\notag\\
&& +\,(1-p_0) \prob\mathopen{}\Bigg[\prod\limits_{j=1}^{u} \frac{1}{1 + G_j}\leq \argpn \Bigg| G_1< \gth\Bigg]
- \frac{\constrm}{\bl} \label{eq:ach-lb-prod-tj-1-0} \IEEEeqnarraynumspace\\
&=& \prob\mathopen{}\Bigg[\prod\limits_{j=1}^{u} \frac{1}{1+G_j} \leq \argpn \bigg] - \frac{\constrm}{\bl} .\label{eq:ach-lb-prod-tj-1}
\end{IEEEeqnarray}
The first factor in~\eqref{eq:ach-lb-prod-tj-1-0} is equal to one because of~\eqref{eq:exception-case}. This proves~\eqref{eq:ach-lb-prod-tj-1-ol} and concludes the proof of Lemma~\ref{lem:prob-prod-cdf}.

\section{Proof of Proposition~\ref{prop:continuous} (Existence of Optimal Covariance Matrix)}
\label{app:proof-prop-continuous}
Since the set~$\setU_{\txant}$ is compact, by the extreme value theorem~\cite[p.~34]{munkres91-a}, it is sufficient to show
that, under the assumptions in the proposition, the function
$\matQ \mapsto \prob\mathopen{}\left[\log\det\mathopen{}\left(\matI_\rxant + \herm{\randmatH}\matQ\randmatH\right)\leq \argpn\right]$
is continuous in $\matQ\in\setU_{\txant}$ with respect to the metric $d(\matA,\matB) =\fnorm{\matA-\matB}$.
%
%
%

Consider an arbitrary sequence $\{\matQ_{l}\}$ in $\setU_{\txant}$ that converges to~$\matQ$.
Then
\begin{IEEEeqnarray}{rCl}
\IEEEeqnarraymulticol{3}{l}{
\det(\matI_\rxant + \herm{\matH}\matQ_l\matH)}\notag\\
 &=& \det(\matI_\rxant + \herm{\matH}\matQ\matH + \herm{\matH}(\matQ_l - \matQ)\matH)\\
&=& \det(\matI_\rxant + \herm{\matH}\matQ\matH) \notag\\
&& \times \det\mathopen{}\left(\matI_\rxant + \herm{\matH}(\matQ_l-\matQ)\matH  (\matI_\rxant + \herm{\matH}\matQ\matH)^{-1} \right)\\
&\leq&  \det(\matI_\rxant + \herm{\matH}\matQ\matH)\notag\\
  &&\times \Big(1 + \fnorm{ \herm{\matH}(\matQ_l-\matQ)\matH  (\matI_\rxant + \herm{\matH}\matQ\matH)^{-1}}\Big)^\rxant \label{eq:bound-ratio-det00}\\
&\leq & \det(\matI_\rxant + \herm{\matH}\matQ\matH) \notag\\
 &&\times \Big(1 + \fnorm{\matQ_l-\matQ} \fnorm{\matH}^2 \fnorm{(\matI_\rxant + \herm{\matH}\matQ\matH)^{-1}}\Big)^\rxant \IEEEeqnarraynumspace\label{eq:bound-ratio-det01}\\
&\leq &\det(\matI_\rxant + \herm{\matH}\matQ\matH) \Big(1+ \fnorm{\matQ_l-\matQ} \fnorm{\matH}^2\sqrt{\rxant}\Big)^\rxant.\label{eq:bound-ratio-det1}
\end{IEEEeqnarray}
Here,~\eqref{eq:bound-ratio-det00} follows from Hadamard's inequality;~\eqref{eq:bound-ratio-det01} follows from the sub-multiplicative property of the Frobenius norm, namely, $\fnorm{\matA\matB}\leq \fnorm{\matA} \fnorm{\matB}$;~\eqref{eq:bound-ratio-det1} follows because $\fnorm{(\matI_\rxant + \herm{\matH}\matQ\matH)^{-1}} \leq \fnorm{\matI_{\rxant}} =\sqrt{\rxant}$.
Similarly, by replacing $\matQ_l$ with $\matQ$ in the above steps, we obtain
\begin{IEEEeqnarray}{rCl}
\IEEEeqnarraymulticol{3}{l}{
\det(\matI_\rxant + \herm{\matH}\matQ\matH) }\notag\\
\quad &\leq& \det(\matI_\rxant + \herm{\matH}\matQ_l\matH) (1+ \fnorm{\matQ_l-\matQ} \fnorm{\matH}^2\sqrt{\rxant})^\rxant.\label{eq:bound-ratio-det2} \IEEEeqnarraynumspace
\end{IEEEeqnarray}
The inequalities~\eqref{eq:bound-ratio-det1} and~\eqref{eq:bound-ratio-det2} imply that
\begin{IEEEeqnarray}{rCl}
\IEEEeqnarraymulticol{3}{l}{\big| \log\det(\matI_\rxant + \herm{\matH}\matQ_l\matH) - \log\det(\matI_\rxant + \herm{\matH}\matQ\matH) \big| }\notag\\
\quad &\leq & \rxant \log (1+ \fnorm{\matQ_l-\matQ} \fnorm{\matH}^2\sqrt{\rxant})\\
&\leq& \rxant^{3/2} \fnorm{\matQ_l-\matQ} \fnorm{\matH}^2.
\end{IEEEeqnarray}
Hence, for every $c>0$
\begin{IEEEeqnarray}{rCl}
\IEEEeqnarraymulticol{3}{l}{
\prob\mathopen{}\Big[\big|\log\det(\matI_\rxant + \herm{\randmatH}\matQ_l\randmatH) -\log\det(\matI_\rxant + \herm{\randmatH}\matQ\randmatH)\big|\geq c  \Big] }\notag\\
\quad &\leq & \prob\mathopen{}\left[ \fnorm{\randmatH}^2 \geq \frac{c}{\rxant^{3/2}} \frac{1}{\fnorm{\matQ_l-\matQ}}\right]\\
& \leq&  \Ex{}{\fnorm{\randmatH}^2} \cdot \fnorm{\matQ_l-\matQ} \frac{\rxant^{3/2}}{c} \label{eq:convergence-prob-logdet1}\\
&\to& 0, \quad\quad\text{as } \matQ_l\to \matQ\label{eq:convergence-prob-logdet}
\end{IEEEeqnarray}
where~\eqref{eq:convergence-prob-logdet1} follows from Markov's inequality and~\eqref{eq:convergence-prob-logdet} follows because, by assumption, $\Ex{}{\fnorm{\randmatH}^2} <\infty$.
Thus, the sequence of random variables $\{\log\det(\matI_\rxant + \herm{\randmatH}\matQ_l\randmatH)  \}$ converges in probability to $\log\det(\matI_\rxant + \herm{\randmatH}\matQ\randmatH)$.
Since convergence in probability implies convergence in distribution, we conclude that
\begin{IEEEeqnarray}{rCl}
\IEEEeqnarraymulticol{3}{l}{
\prob\mathopen{}\left[\log\det\mathopen{}\left(\matI_\rxant + \herm{\randmatH}\matQ_l\randmatH\right)\leq \argpn\right] }\notag\\
\quad &\to & \prob\mathopen{}\left[\log\det\mathopen{}\left(\matI_\rxant + \herm{\randmatH}\matQ\randmatH\right)\leq \argpn\right] \text{ as } \matQ_l\to \matQ
\label{eq:convergence-logdet-l}
\end{IEEEeqnarray}
 for every $\argpn\in \realset$ for which the cdf of  $\log\det(\matI_\rxant + \herm{\randmatH}\matQ\randmatH)$ is continuous~\cite[p.~308]{grimmett01}.
 However, the cdf of  $\log\det(\matI_\rxant + \herm{\randmatH}\matQ\randmatH)$ is continuous for every $\argpn \in \realset$ since the distribution of $\randmatH$ is, by assumption, absolutely continuous with respect to the Lebesgue measure and the function $\matH \mapsto \log\det(\matI_\rxant + \herm{\matH}\matQ\matH)$ is continuous.
Consequently,~\eqref{eq:convergence-logdet-l} holds  for every $\argpn\in \realset$, thus proving Proposition~\ref{prop:continuous}.

\setcounter{subsubsection}{0}

\section{Proof of Theorem~\ref{thm:converse-csir} (CSIR Converse Bound)}
\label{app:proof-converse-cisr}
For the CSIR case, the input of the channel~\eqref{eq:channel_io} is $\matX$ and the output is the pair $(\randmatY,\randmatH)$.
An $(\bl,\NumCode,\error)_{\mathrm{e}}$ code is defined in a similar way as the $(\bl,\NumCode,\error)_{\csir}$ code in Definition~\ref{def:csir-code}, except that each codeword satisfies the power constraint~\eqref{eq:peak-power-constraint} with equality, i.e., each codeword belongs to the set
\begin{equation}
\setF_{\bl, \txant} \define \{\matX \in \complexset^{\bl\times \txant} : \fnorm{\matX}^2  =\bl\snr \}.
\end{equation}
Denote by $R_{\mathrm{e}}^\ast(\bl,\error)$ the maximal achievable rate with an $(\bl,\NumCode,\error)_{\mathrm{e}}$ code.
Then by~\cite[Sec.~XIII]{shannon59} (see also~\cite[Lem.~39]{polyanskiy10-05},
\begin{IEEEeqnarray}{rCl}
\label{eq:relate-R-Re}
\Rcsir^{\ast}(\bl-1,\error) \leq \frac{\bl}{\bl-1} R_{\mathrm{e}}^\ast(\bl,\error).
\end{IEEEeqnarray}
We next establish an upper bound on $R_{\mathrm{e}}^\ast(\bl,\error)$.
Consider an arbitrary $(\NumCode,\bl,\epsilon)_{\mathrm{e}}$ code.
To each codeword $\matX\in \setF_{\bl, \txant}$, we associate a matrix $\covmat(\matX) \in \complexset^{\txant\times\txant}$:
\begin{IEEEeqnarray}{rCl}
\label{eq:def-u-mimo-csir}
\covmat(\matX) \define \frac{1}{\bl}\herm{\matX}\matX.
\end{IEEEeqnarray}
%
%
To upper-bound $R_{\mathrm{e}}^\ast(\bl,\error)$, we use the meta-converse theorem~\cite[Th.~30]{polyanskiy10-05}.
As \emph{auxiliary} channel $\outdist_{\randmatY\randmatH\given \randmatX}$, we take
\begin{IEEEeqnarray}{rCl}
\label{eq:def-q-channel-mimo}
\outdist_{\randmatY\randmatH\given \randmatX}=\indist_{\randmatH}\times \outdist_{\randmatY\given \randmatX\randmatH}
\end{IEEEeqnarray}
where 
\begin{IEEEeqnarray}{rCl}
\label{eq:def-q-channel-mimo-cond}
\outdist_{\randmatY\given \randmatX =  \matX,\randmatH=\matH} = \prod\limits_{i=1}^{\bl} \outdist_{\randvecy_i \given \randmatX =  \matX,\randmatH=\matH}
\end{IEEEeqnarray}
with $\randvecy_i$, $i=1,\dots,n$ denoting the rows of $\randmatY$, and
\begin{IEEEeqnarray}{rCL}\label{eq:def_pdf_rows_Y}
   \outdist_{\randvecy_i \given \randmatX =  \matX,\randmatH=\matH} =\jpg\mathopen{}\left(\mathbf{0} , \matI_{\rxant} + \herm{\matH} \covmat(\matX) \matH\right).
\end{IEEEeqnarray}
By~\cite[Th.~30]{polyanskiy10-05}, we have
\begin{IEEEeqnarray}{rCl}
\inf\limits_{\matX\in \setF_{\bl, \txant}} \beta_{1-\error}\mathopen{}\left(\indist_{\randmatY\randmatH\given \randmatX = \matX}, \outdist_{\randmatY\randmatH\given \randmatX = \matX}\right)\leq 1-\error'
\label{eq:def-beta-X0}
\end{IEEEeqnarray}
where $\error'$ is the \textit{maximal probability of error} of the optimal code with $\NumCode$ codewords over the auxiliary channel~(\ref{eq:def-q-channel-mimo}).
To shorten notation, we define
\begin{IEEEeqnarray}{rCl}
\beta_{1-\error}^\bl(\matX)  \define \beta_{1-\error}\mathopen{}\left(\indist_{\randmatY\randmatH\given \randmatX = \matX}, \outdist_{\randmatY\randmatH\given \randmatX = \matX}\right).
\end{IEEEeqnarray}
To prove the theorem, we proceed as in Appendix~\ref{app:proof-converse-cisrt}: we first evaluate $\beta^\bl_{1-\error}(\matX)$, then we relate $\error'$ to $R_{\mathrm{e}}^\ast(\bl,\error)$ by establishing a converse bound on the channel $\outdist_{\randmatY\randmatH\given \randmatX}$.
\subsubsection{Evaluation of $\beta_{1-\error}(\matX)$}
 \label{sec:lower-bound-beta-csir}
Let $\matG$ be an arbitrary $\bl\times\bl$ unitary matrix. Let $g_{\mathrm{i}}: \setF_{\bl, \txant} \mapsto \setF_{\bl, \txant}$ and $g_{\mathrm{o}}: \complexset^{\bl\times\rxant}\times\complexset^{\txant\times\rxant} \mapsto  \complexset^{\bl\times\rxant}\times\complexset^{\txant\times\rxant}$ be two mappings defined as
\begin{IEEEeqnarray}{rCl}
g_{\mathrm{i}}(\matX) \define \matG\matX \quad\text{and}\quad g_{\mathrm{o}}(\matY,\matH) \define (\matG\matY, \matH).
\end{IEEEeqnarray}
Note that
\begin{IEEEeqnarray}{rCl}
\indist_{\randmatY\randmatH\given\randmatX}(g_{\mathrm{o}}^{-1}(\setE) \given g_{\mathrm{i}}(\matX)) = \indist_{\randmatY\randmatH\given\randmatX}(\setE\given \matX)
\end{IEEEeqnarray}
for all measurable sets $\setE \subset  \complexset^{\bl\times\rxant}\times\complexset^{\txant\times\rxant}$ and $\matX\in\setF_{\bl, \txant}$, i.e., the pair $(g_{\mathrm{i}},g_{\mathrm{o}})$ is a symmetry~\cite[Def.~3]{polyanskiy13} of $\indist_{\randmatY\randmatH\given\randmatX}$.
 Furthermore, \eqref{eq:def-q-channel-mimo-cond} and~\eqref{eq:def_pdf_rows_Y} imply that
\begin{IEEEeqnarray}{rCl}
\outdist_{\randmatY\randmatH\given\randmatX=\matX} = \outdist_{\randmatY\randmatH\given\randmatX=g_{\mathrm{i}}(\matX)}
\end{IEEEeqnarray}
and that $\outdist_{\randmatY\randmatH\given\randmatX=\matX}$ is invariant under $g_{\mathrm{o}}$ for all $\matX\in\insetMIMO$.
Hence, by~\cite[Prop.~19]{polyanskiy13}, we have that
\begin{IEEEeqnarray}{rCl}
\beta_{1-\error}^{\bl} (\matX) = \beta_{1-\error}^{\bl} (g_{\mathrm{i}}(\matX)) = \beta_{1-\error}^{\bl} (\matG\matX).
\label{eq:beta-rotation-invariant}
\end{IEEEeqnarray}
Since $\matG$ is arbitrary, this implies that $\beta_{1-\error}^{\bl} (\matX)$ depends on $\matX$ only through $\matU(\matX)$.
%
%
%
Consider the QR decomposition~\cite[p.~113]{horn85a} of~$\matX$
\begin{equation}
\matX=\matV\matX_0
\label{eq:qr-x}
\end{equation}
where $\matV\in \complexset^{\bl\times\bl}$ is unitary and $\matX_0 \in \complexset^{\bl\times\txant}$ is upper triangular.
By~\eqref{eq:beta-rotation-invariant} and~\eqref{eq:qr-x},
%
%
%
%
%
\begin{IEEEeqnarray}{rCl}
\beta_{1-\error}^\bl(\matX_0) =\beta_{1-\error}^\bl(\matX).
\label{eq:same-beta}
\end{IEEEeqnarray}
Let
\begin{IEEEeqnarray}{rCl}
r(\matX_0;\randmatY\randmatH ) \define \log \frac{d\indist_{\randmatY\randmatH\given \randmatX = \matX_0}}{d\outdist_{\randmatY\randmatH\given \randmatX = \matX_0}}.
\end{IEEEeqnarray}
Under both $\indist_{\randmatY\randmatH\given \randmatX = \matX_0}$ and $\outdist_{\randmatY\randmatH\given \randmatX = \matX_0}$, the random variable $r(\matX_0;\randmatY\randmatH )$ has absolutely continuous cdf with respect to the Lebesgue measure. By the Neyman-Pearson lemma~\cite[p.~300]{neyman33a}
\begin{IEEEeqnarray}{rCl}
\label{eq:neyman_pearson_one-csir}
  \beta_{1-\error}^{\bl}( \matX_0) =\outdist_{\randmatY\randmatH\given \randmatX = \matX_0}\bigl[r(\matX_0;\randmatY\randmatH ) \geq \bl \gamma_{\bl}(\matX_0) \bigr]
\end{IEEEeqnarray}
where $\gamma_{\bl}(\matX_0)$ is the solution of
\begin{IEEEeqnarray}{rCl}
  \indist_{\randmatY\randmatH\given \randmatX = \matX_0} \bigl[r(\matX_0;\randmatY\randmatH ) \leq \bl \gamma_{\bl}(\matX_0)\bigr]=\error.
\end{IEEEeqnarray}
It can be shown that under $\indist_{\randmatY\randmatH\given \randmatX =\matX_0}$, the random variable $r(\matX_0; \randmatY \randmatH)$ has the same distribution as $\Scsir_\bl(\matU(\matX_0))$ in~\eqref{eq:info_density_mimo_csir}, and under $\outdist_{\randmatY\randmatH\given \randmatX =\matX_0}$, it has the same distribution as $\Lcsir_\bl(\covmat(\matX_0))$ in~\eqref{eq:info_density_mimo_alt_csir}.

\subsubsection{Converse on the auxiliary channel}
\label{sec:converse-Q-channel}
To prove the theorem, it remains to lower-bound $\error'$, which is the maximal probability of error over the  auxiliary channel~\eqref{eq:def-q-channel-mimo}.
The following lemma serves this purpose.
\begin{lemma}
\label{lem:converse-q-channel}
For every code with $\NumCode$ codewords and blocklength $\bl\geq \rxant$, the maximum probability of error $\error'$ over the auxiliary channel \eqref{eq:def-q-channel-mimo} satisfies
\begin{IEEEeqnarray}{rCl}
1-\error'\leq \frac{c_{\csir}(\bl)}{\NumCode}
\label{eq:converse-q}
\end{IEEEeqnarray}
where $c_{\csir}(\bl)$ is given in~\eqref{eq:def-crn}. 
\end{lemma}
%

Substituting~\eqref{eq:neyman_pearson_one-csir} into~\eqref{eq:def-beta-X0} and using~\eqref{eq:converse-q}, we then obtain upon minimizing~\eqref{eq:neyman_pearson_one-csir} over all matrices in $ \insetcove$
\begin{IEEEeqnarray}{rCl}
R_{\mathrm{e}}^*(\bl,\error )  \leq \frac{1}{\bl} \frac{c_{\csir}(\bl)}{\inf\limits_{\matQ \in \insetcove} \prob[\Lcsir_\bl(\matQ) \geq \bl \gamma_{\bl}]}.
\label{eq:ineq-Re-n}
\end{IEEEeqnarray}
The final bound~\eqref{eq:thm-converse-rcsir} follows by combining~\eqref{eq:ineq-Re-n} with~\eqref{eq:relate-R-Re} and by noting that the upper bound does not depend on the chosen code.

\paragraph*{Proof of Lemma~\ref{lem:converse-q-channel}}
%
%
%
According to \eqref{eq:def_pdf_rows_Y}, given $\randmatH =\matH$, the output of the auxiliary channel depends on~$\matX$ only through $\covmat(\matX)$.
In the following, we shall omit the argument of $\covmat(\matX)$ where it is immaterial.
Let $\randcmY \define \covmat(\randmatY)$.
Then, $(\randcmY,\randmatH)$ is a sufficient statistic for the detection of $\matX$ from $(\randmatY,\randmatH)$.
 Therefore, to establish~\eqref{eq:converse-q}, it is sufficient to lower-bound the
maximal probability of error~$\error'$ over the equivalent auxiliary channel
\begin{IEEEeqnarray}{rCl}
{\outdist}_{\randcmY\randmatH\given\randmatU} = \indist_{\randmatH}\times {\outdist}_{\randcmY\given\randmatU,\randmatH}
\label{eq:app-q-dist-eqv-tot}
\end{IEEEeqnarray}
where ${\outdist}_{\randcmY\given\randmatU =\matU,\randmatH=\matH}$ is the Wishart distribution~\cite[Def.~2.3]{tulino04a}:
\begin{IEEEeqnarray}{rCl}
{\outdist}_{\randcmY\given\randmatU =\matU,\randmatH=\matH} = \mathcal{W}_\rxant\mathopen{}\left(\bl,\frac{1}{\bl}(\matI_{\rxant} + \herm{\matH}\covmat\matH)\right).
\label{eq:app-q-dist-eqv}
\end{IEEEeqnarray}
Let $\matB\define \matI_{\rxant} + \herm{\matH}\covmat\matH$, and let $q_{\randcmY\given\randmatB}(\covmatY\given \matB)$ be
the pdf associated with~\eqref{eq:app-q-dist-eqv}, i.e.,~\cite[Def.~2.3]{tulino04a}
\begin{IEEEeqnarray}{rCl}
{q}_{\randcmY\given  \randmatB}(\covmatY\given \matB)
 &=& \frac{\det\covmatY^{\bl-\rxant}}{\Gamma_{\rxant}(\bl)\det\bigl(\frac{1}{\bl}\matB\bigr)^\bl} \exp\mathopen{}\left(\!-\tr\mathopen{}\Big(\!\big(\bl^{-1}\matB\big)^{-1}\covmatY\Big)\!\right).\notag\\ && \label{eq:density-wishart}
\end{IEEEeqnarray}
It will be convenient to express ${q}_{\randcmY\given  \randmatB}(\covmatY\given \matB)$ in the coordinate system of the eigenvalue decomposition
\begin{IEEEeqnarray}{rCl}
\label{eq:change-variable-eigend}
\randcmY = \randmatQ\randmatD \herm{\randmatQ}
\end{IEEEeqnarray}
where $\randmatQ\in\complexset^{\rxant\times\rxant}$ is unitary, and $\randmatD$ is a diagonal matrix whose diagonal elements $D_1,\ldots,D_\rxant$ are the eigenvalues of $\randcmY$ in descending order. In order to make the eigenvalue decomposition~\eqref{eq:change-variable-eigend} unique, we assume that the first row of $\randmatQ$ is real and non-negative.  Thus, $\randmatQ$ only lies in a \emph{submanifold} $\widetilde{\setS}_{\rxant,\rxant}$ of the Stiefel manifold~$\setS_{\rxant,\rxant}$.
Using \eqref{eq:change-variable-eigend}, we rewrite~\eqref{eq:density-wishart} as
\begin{IEEEeqnarray}{rCl}
q_{\randmatQ,\randmatD \given \randmatB}(\matQ, \matD \given \matB)
 &=& \frac{\bl^{\rxant \bl} \exp\mathopen{}\left(-\bl \cdot\tr(\matB^{-1}\matQ\matD \herm{\matQ})\right)}{  \Gamma_{\rxant}(\bl)\det\matB^\bl}\notag\\
 && \times \det\matD^{\bl-\rxant} \prod\limits_{i<j}^{\rxant}(d_i-d_j)^2\label{eq:density-eigend}\IEEEeqnarraynumspace
\end{IEEEeqnarray}
where in~\eqref{eq:density-eigend} we used  the fact\ that the Jacobian of the eigenvalue decomposition~\eqref{eq:change-variable-eigend} is $\prod\nolimits_{i<j}^{\rxant}(d_i-d_j)^2$ (see~\cite[Th.~3.1]{edelman89-05a}).

We next establish an upper bound on~\eqref{eq:density-eigend} that is integrable and does not depend on~$\matB$.
To this end, we will bound each of the factors on the RHS of~\eqref{eq:density-eigend}.
To bound the argument of the exponential function, we apply the trace inequality~\cite[Th.~20.A.4]{marshall79} 
\begin{IEEEeqnarray}{rCl}
\tr(\matB^{-1}\matQ\matD \herm{\matQ}) \geq \sum\limits_{i=1}^{\rxant}\frac{d_i}{b_i}
\label{eq:bound-trace-exp}
\end{IEEEeqnarray}
for every unitary matrix $\matQ$, where $b_1\geq \ldots\geq b_\rxant$ are the ordered eigenvalues of $\matB$.
Using~\eqref{eq:bound-trace-exp} in~\eqref{eq:density-eigend} and further upper-bounding the terms $(d_i-d_j)^2$ in~\eqref{eq:density-eigend} with $d_i^2$, we obtain
\begin{IEEEeqnarray}{rCl}
q_{\randmatQ \randmatD\given \randmatB}(\matQ, \matD\given \matB)
&\leq&\frac{\bl^{\rxant\bl}}{\Gamma_{\rxant}(\bl)}\prod\limits_{i=1}^{\rxant} \Bigg\{  \frac{d_i^{\bl+\rxant-2i}}{b_i^{\bl}}\exp\mathopen{}\left(-\bl\frac{d_i}{b_i}\right) \Bigg\}.\notag\\
&& \label{eq:density-eigend-ub1} \IEEEeqnarraynumspace
\end{IEEEeqnarray}
 Since $\matB =\matI_\rxant +\herm{\matH} \covmat\matH$, we have that
\begin{IEEEeqnarray}{rCl}
1\leq b_i &\leq& 1+\tr\mathopen{}\left(\herm{\matH} \covmat\matH\right)\\
&\leq &1 + \fnorm{\matH}^{2}\tr\left( \covmat\right)\label{eq:upper-bd-eigenvalue-snr2}\\
&= & 1 + \fnorm{\matH}^{2}\snr \define \ubb \label{eq:upper-bd-eigenvalue-snr3}
\end{IEEEeqnarray}
where~\eqref{eq:upper-bd-eigenvalue-snr2} follows from the Cauchy-Schwarz inequality and \eqref{eq:upper-bd-eigenvalue-snr3} follows because $\matU \in \insetcove$.
Using~\eqref{eq:upper-bd-eigenvalue-snr3},  we can upper-bound each factor on the RHS of~\eqref{eq:density-eigend-ub1} as follows:
\begin{IEEEeqnarray}{rCl}
\IEEEeqnarraymulticol{3}{l}{
\frac{d_i^{\bl+\rxant-2i}}{b_i^{\bl}}\exp\mathopen{}\left(-\bl\frac{d_i}{b_i}\right)
}\notag\\
&\leq& g_i(d_i)\define \left\{\!\!
           \begin{array}{l}
             \!\!\left(\dfrac{\bl+\rxant-2i}{\bl}\right)^{\bl+\rxant -2i}b_0^{[\rxant-2i]^{+}}  e^{-(\bl+\rxant -2i)},\\
              \hfill  \hbox{if $d_i \leq \frac{\ubb(\bl+\rxant-2i)}{\bl}$} \quad \\
             \!\!\left(\dfrac{ d_i}{\ubb}\right)^{\bl + \rxant-2i}  b_0^{[\rxant -2i]^{+}} e^{-\bl d_i/ \ubb},\\
              \hfill \hbox{if  $d_i > \frac{\ubb(\bl+\rxant-2i)}{\bl}$.}\quad
           \end{array}
         \right. \,\,\,\,\,\,\,\label{eq:density-eigend-ub2}
\end{IEEEeqnarray}

We are now ready to establish the desired converse result for the auxiliary channel~$Q$.
 Consider an arbitrary code for the auxiliary channel~$Q$ with encoding function $\encoder_0:\{1,\ldots,\NumCode\}  \mapsto\insetcove$.
Let $\setD_j(\matH)$ be the decoding set for the $j$th codeword $\encoder_0(j)$ in the eigenvalue decomposition coordinate such that
\begin{IEEEeqnarray}{rCl}
\bigcup\limits_{j=1}^{\NumCode} \setD_j(\matH) = \widetilde{\setS}_{\rxant,\rxant}\times \realset^{\rxant}_{ \geq }\label{eq:union-decoding-set}.
\end{IEEEeqnarray}
 Let $\error'_{\mathrm{avg}}$ denote the average probability of error over the auxiliary  channel.
 Then,
\begin{IEEEeqnarray}{rCl}
\IEEEeqnarraymulticol{3}{l}{
1-\error'}\notag\\
&\leq& 1-\error'_{\mathrm{avg}} \\
&=&\frac{1}{\NumCode}\Ex{\randmatH}{  \sum\limits_{j=1}^{\NumCode} \int\nolimits_{\setD_j(\randmatH)} \!\!q_{\randmatQ,\randmatD \given \randmatB =\matI_r+\herm{\randmatH}\encoder_0(j)\randmatH}(\matQ, \matD)d\matQ d\matD }\IEEEeqnarraynumspace\label{eq:converse-q-step1}\\
&\leq&\frac{\bl^{\rxant\bl}}{\Gamma_{\rxant}(\bl) \NumCode}  \Ex{\randmatH}{\sum\limits_{j=1}^{\NumCode} \int\nolimits_{\setD_j(\randmatH)}\prod\limits_{i=1}^{\rxant} g_i(d_i) d\matQ d\matD}  \label{eq:converse-q-step2}\\
&=& \frac{\bl^{\rxant\bl}}{\Gamma_{\rxant}(\bl) \NumCode}  \Ex{\randmatH}{\int\nolimits_{ \widetilde{\setS}_{\rxant,\rxant}\times \realset^{\rxant}_{\geq }  }\prod\limits_{i=1}^{\rxant} g_i(d_i) d\matQ d\matD}  \label{eq:converse-q-step3}\\
& \leq & \frac{\pi^{\rxant(\rxant-1)}\bl^{\rxant\bl}}{\Gamma_{\rxant}(\rxant) \Gamma_{\rxant}(\bl) \NumCode}  \Ex{\randmatH}{ \prod\limits_{i=1}^{\rxant}\int\nolimits_{\posrealset} g_i(x_i) d x_i}  \label{eq:converse-q-step4}
\end{IEEEeqnarray}
where~\eqref{eq:converse-q-step2} follows from~\eqref{eq:density-eigend-ub1} and~\eqref{eq:density-eigend-ub2}; \eqref{eq:converse-q-step3} follows from~\eqref{eq:union-decoding-set}; \eqref{eq:converse-q-step4} holds because the integrand does not depend on $\matQ$, because $\realset^{\rxant}_{\geq }\subset \realset^{\rxant}_{+}$
 and because the volume of $\widetilde{\setS}_{\rxant,\rxant}$ (with respect to the Lebesgue measure on $\widetilde{\setS}_{\rxant,\rxant}$) is $\pi^{\rxant(\rxant-1)}/\Gamma_{\rxant}(\rxant) $.
After algebraic manipulations, we obtain
\begin{IEEEeqnarray}{rCl}
\int\nolimits_{\posrealset} g_i(x_i) d x_i &=&
    \frac{b_0^{[\rxant-2i]^{+}+1} }{ \bl^{\bl+\rxant-2i+1} }\bigg[\Gamma(\bl+\rxant-2i+1, \bl+r-2i) \notag\\
    &&+ \left(\bl+\rxant-2i\right)^{\bl+\rxant -2i+1} e^{-(\bl+\rxant -2i)} \bigg].\IEEEeqnarraynumspace\label{eq:app-converse-q-integ}
\end{IEEEeqnarray}
Substituting~\eqref{eq:app-converse-q-integ} into~\eqref{eq:converse-q-step4} and using~\eqref{eq:upper-bd-eigenvalue-snr3}, we obtain
\begin{IEEEeqnarray}{rCl}
1-\error' 
&\leq & \frac{ c_{\csir}(\bl) }{\NumCode}.~\label{eq:app-ub-error2}
\end{IEEEeqnarray}
Note that the RHS of~\eqref{eq:app-ub-error2} is valid for every code.

\section{Proof of the Converse Part of Theorem~\ref{thm:zero-dispersion-nocsit}}
\label{app:proof-zero-dispersion-nocsit-ub}

In this appendix, we prove the converse asymptotic expansion for Theorem~\ref{thm:zero-dispersion-nocsit}. More
precisely, we show the following:

\begin{prop}\label{thm:zero-dispersion-nocsit-ub}
Let the pdf of the fading matrix $\randmatH$ satisfy the conditions in Theorem~\ref{thm:zero-dispersion-nocsit}.
Then
\begin{IEEEeqnarray}{rCL}
  \Rcsir^{\ast}(\bl,\error) \leq \Cnocsit + \bigO\mathopen{}\left(\frac{\log\bl}{n}\right).
  \label{eq:prop-converse-nocsit}
\end{IEEEeqnarray}
\end{prop}

\begin{IEEEproof}
Proceeding as in~\eqref{eq:inequality_beta-poly}--\eqref{eq:ub-rcsirt-original}, we obtain from Theorem~\ref{thm:converse-csir} that
\begin{IEEEeqnarray}{rCl}
\IEEEeqnarraymulticol{3}{l}{
(\bl-1)\Rcsir^\ast(\bl-1,\error ) } \notag\\
 \,\, &\le&  \bl \gamma - \log\mathopen{}\Big( \inf\limits_{\matQ \in \setU_{\txant}^{\mathrm{e}}}  \prob[S^{\csir}_\bl(\matQ) \leq \bl\gamma] -\error \Big) + \log c_{\csir}(\bl)\label{eq:basic-ineq-R-csir}\IEEEeqnarraynumspace
\end{IEEEeqnarray}
where $\gamma>0$ is arbitrary.
The third term on the RHS of~\eqref{eq:basic-ineq-R-csir} is upper-bounded by
\begin{IEEEeqnarray}{rCl}
\log c_{\csir}(\bl) &\le & \frac{\rxant^2}{2} \log\bl + \log\mathopen{}\left(\Ex{}{\left(1+\snr\fnorm{\randmatH}^2\right)^{\lfloor(\rxant+1)^2/4 \rfloor} }\right) \notag\\
&&+\,\bigO(1)\label{eq:bound-c-rx-n-asy-1} \\
&=& \frac{\rxant^2}{2} \log\bl + \bigO(1).\label{eq:bound-c-rx-n-asy-2}
\end{IEEEeqnarray}
Here, ~\eqref{eq:bound-c-rx-n-asy-1} follows from algebraic manipulations, and~\eqref{eq:bound-c-rx-n-asy-2} follows from the assumption~\eqref{eq:cond-bdd-pdf}, which ensures that the second term on the RHS of~\eqref{eq:bound-c-rx-n-asy-1} is finite.

To evaluate $\prob[S^{\csir}_\bl(\matQ) \leq \bl\gamma]$ on the RHS of~\eqref{eq:basic-ineq-R-csir}, we note that given $\randmatH = \matH$, the random variable $S^{\csir}_\bl(\matQ)$ is the sum of $\bl$ i.i.d. random variables.
Hence, using Theorem~\ref{thm:refine-be} (Appendix~\ref{sec:proof_of_eq:kramer_esseen_step}) and following similar steps as the ones reported in Appendix~\ref{sec:proof_of_eq:kramer_esseen_step}, we obtain
\begin{IEEEeqnarray}{rCl}
\prob[ S^{\csir}_\bl(\matQ) \leq \bl\gamma \given \randmatH = \matH]  &\geq& q_\bl(\functQ (\matH) ) + \bigO\mathopen{}\left(\frac{1}{\bl}\right) \IEEEeqnarraynumspace
 \label{eq:conv-csir-lb-prob-s-cond}
\end{IEEEeqnarray}
where the function $\functQ: \complexset^{\txant \times \rxant} \mapsto \realset$ is given by
\begin{equation}
\functQ (\matH) \define \frac{\gamma - \log\det\mathopen{}\left(\matI_{\rxant} + \herm{\matH} \matQ\matH \right)}{\sqrt{\tr\mathopen{}\big(\matI_{\rxant} - (\matI_{\rxant} + \herm{\matH} \matQ\matH)^{-2}\big)}}
\label{eq:def-f-argpn-matQ}
\end{equation}
the function $q_\bl(\cdot)$ was defined in~\eqref{eq:def-q-n-y}, and the $\bigO(1/\bl)$ term is uniform in $\matQ$, $\gamma$ and $\matH$.
Let
\begin{equation}
\UgammaQ\define \functQ(\randmatH).
\end{equation}
Averaging~\eqref{eq:conv-csir-lb-prob-s-cond} over $\randmatH$, we obtain
\begin{IEEEeqnarray}{rCl}
\IEEEeqnarraymulticol{3}{l}{
\prob[ S^{\csir}_\bl(\matQ) \leq \bl\gamma]}\notag\\
\,\,&\geq& \Ex{}{ Q(-\sqrt{\bl}  \UgammaQ )} \notag\\
&&-\, \Ex{}{\frac{[1-\bl U^2(\gamma, \matQ)]^{+} e^{-\bl U^2(\gamma, \matQ) /2}}{6\sqrt{\bl}}}  +  \bigO\mathopen{}\left(\frac{1}{\bl}\right). \IEEEeqnarraynumspace
 \label{eq:conv-csir-lb-prob-s-avg}
\end{IEEEeqnarray}

We proceed to lower-bound the first two terms on the RHS of~\eqref{eq:conv-csir-lb-prob-s-avg}.
%
%
To this end, we show in Lemma~\ref{lem:bound-pdf-der-pdf-csir} ahead that there exist $\delta_1\in(0,\Cnocsit)$ and $\delta>0$ such that $u\mapsto f_{\UgammaQ}(u)$, where $\pdfgeneral_{\UgammaQ}$ denotes the pdf of $\UgammaQ$, is continuously differentiable on $(-\delta,\delta)$, and that $f_{\UgammaQ}(u)$ and $f_{\UgammaQ}'(u)$ are uniformly bounded for every $\gamma\in~(\Cnocsit-\delta_1, \Cnocsit+\delta_1)$, every $\matQ\in \insetcove$, and every $u\in (-\delta,\delta)$.
We then apply Lemma~\ref{lem:expectation-phi} in Appendix~\ref{sec:proof_of_averaging_over_channel} with $A$ being a standard normal random variable and $B=U(\gamma, \matQ) $ to lower-bound the first term on the RHS of~\eqref{eq:conv-csir-lb-prob-s-avg} for every $\delta>0$ as
\begin{IEEEeqnarray}{rCl}
\IEEEeqnarraymulticol{3}{l}{
\Ex{}{ Q(-\sqrt{\bl}  \UgammaQ )}}\notag\\
 &\geq& \prob\mathopen{}\big[\log\det\mathopen{}\left(\matI_{\rxant} + \herm{\randmatH} \matQ\randmatH \right) \leq \gamma\big] - \frac{1}{\bl}\frac{2}{\delta^{2}}\quad \notag\\
    &&   -\,\frac{1}{\bl}\Big(\frac{1}{\delta}+\frac{1}{2}\Big) \sup\limits_{u\in(-\delta,\delta)}\!\! \max\mathopen{}\left\{ \pdfgeneral_{\UgammaQ} (u) ,  \big|\pdfgeneral'_{\UgammaQ} (u)\big|\right\} . \notag\\
    &&  \label{eq:bound-Q-nX-csir}
\end{IEEEeqnarray}
We upper-bound the second term on the RHS of~\eqref{eq:conv-csir-lb-prob-s-avg} for $\bl > \delta^{-2}$ as
\begin{IEEEeqnarray}{rCl}
\IEEEeqnarraymulticol{3}{l}{
\Ex{}{\frac{\big|1-\bl U^2(\gamma, \matQ)\big|^{+} e^{-\bl U^2(\gamma, \matQ) /2}}{6\sqrt{\bl}}}  }\notag\\
 &\leq & \frac{1}{6\sqrt{\bl}}   \! \sup\limits_{u\in(-\delta,\delta)}\!\pdfgeneral_{\UgammaQ} (u) \! \int\nolimits_{-1/\sqrt{\bl}}^{1/\sqrt{\bl}}  \underbrace{(1-\bl t^2)e^{-\bl t^2/2}}_{\leq 1} dt\label{eq:ub-ex-1-minus-nu-2-nocsi} \IEEEeqnarraynumspace \\
&\leq & \frac{1 }{3 \bl} \sup\limits_{u\in(-\delta,\delta)} \pdfgeneral_{\UgammaQ} (u) .\label{eq:ub-ex-1-minus-nu-3-nocsi}
\end{IEEEeqnarray}
The following lemma establishes that $f_{\UgammaQ}$ and $f'_{\UgammaQ}$ are indeed uniformly bounded.
\begin{lemma}
\label{lem:bound-pdf-der-pdf-csir}
Let $\randmatH$ have pdf $\pdfH$ satisfying Conditions \ref{item:th-zd-no-cond1} and~\ref{item:th-zd-no-cond2} in Theorem~\ref{thm:zero-dispersion-nocsit}.
Let $\functQ: \complexset^{\txant \times \rxant} \mapsto \realset$ be defined as in~\eqref{eq:def-f-argpn-matQ} and let $\UgammaQ $ with pdf $\pdfgeneral_{\UgammaQ}$ denote the random variable $\functQ (\randmatH)$. Then, there exist $\delta_1\in(0,\Cnocsit)$ and $\delta>0$ such that $u\mapsto f_{\UgammaQ}(u)$ is continuously differentiable on $(-\delta,\delta)$ and that
\begin{IEEEeqnarray}{rCl}
\sup\limits_{\gamma \in( \Cnocsit-\delta_1, \Cnocsit+\delta_1) }  \sup\limits_{\matQ \in \insetcove} \sup\limits_{u\in(-\delta,\delta)} \pdfgeneral_{\UgammaQ} (u) &<& \infty \label{eq:lemma-bound-pdf-1} \\
\sup\limits_{\gamma \in( \Cnocsit-\delta_1, \Cnocsit+\delta_1) } \sup\limits_{\matQ \in \insetcove}  \sup\limits_{u\in(-\delta,\delta)}  \big|\pdfgeneral'_{\UgammaQ} (u)\big| &<& \infty.\label{eq:lemma-bound-pdf-2}\IEEEeqnarraynumspace
\end{IEEEeqnarray}
\end{lemma}

\begin{IEEEproof}
See Appendix~\ref{app:proof-boundedness-pdf-deri}.
\end{IEEEproof}

Using~\eqref{eq:bound-Q-nX-csir},~\eqref{eq:ub-ex-1-minus-nu-3-nocsi}, and Lemma~\ref{lem:bound-pdf-der-pdf-csir} in~\eqref{eq:conv-csir-lb-prob-s-avg}, and then~\eqref{eq:conv-csir-lb-prob-s-avg} and~\eqref{eq:bound-c-rx-n-asy-2} in~\eqref{eq:basic-ineq-R-csir}, we obtain for every $\gamma \in( \Cnocsit-\delta_1, \Cnocsit+\delta_1) $ that
\begin{IEEEeqnarray}{rCl}
\IEEEeqnarraymulticol{3}{l}{
(\bl-1)\Rcsir^\ast(\bl-1,\error ) }\notag\\
\,\,&\leq&\bl \gamma - \log \mathopen{}\Big(  \inf\limits_{\matQ \in \insetcove} \prob[ \log\det\mathopen{}\left(\matI_{\rxant} + \herm{\matH} \matQ\matH \right) \leq  \gamma] -\error \quad\quad\notag\\
&& \hfill + \, \bigO\mathopen{}\left(1/\bl\right)  \Big)+  \bigO\mathopen{}\left(\log\bl\right) \quad\quad\\
&=&\bl\gamma -\log \mathopen{}\big( \cdistno(\gamma)-\error + \bigO\mathopen{}\left(1/\bl\right) \big)+ \bigO\mathopen{}\left(\log\bl\right) \IEEEeqnarraynumspace \label{eq:ub-R*-2}
\end{IEEEeqnarray}
where~\eqref{eq:ub-R*-2} follows from~\eqref{eq:P-out-alt-exp}.

We next set $\gamma$ so that
\vspace{-0.1cm}
\begin{IEEEeqnarray}{rCl}
 \cdistno(\gamma)-\error  + \bigO\mathopen{}\left(1/\bl\right) = 1/\bl.
 \label{eq:choose-gamma}
\end{IEEEeqnarray}
In words, we choose $\gamma$ so that the argument of the logarithm in~\eqref{eq:ub-R*-2} is equal to $1/\bl$.
Since the function $(\matQ,R)\mapsto F_{\matQ}(R)$ is continuous and $\insetcove$ is compact, by the maximum theorem~\cite[Sec.~VI.3]{berge63} the function $\cdistno(R) =\inf\nolimits_{\matQ\in\insetcove} F_{\matQ}(R)$ is continuous in $R$.
This guarantees that such a~$\gamma$ indeed exists.
We next show that, for sufficiently large $\bl$, this $\gamma$ satisfies
\begin{IEEEeqnarray}{rCl}
|\gamma - \Cnocsit| \le \bigO(1/\bl).
\label{eq:asy-exp-gamma}
\end{IEEEeqnarray}
This implies that, for sufficiently large $\bl$, $\gamma$ belongs to the interval $(\Cnocsit-\delta_1,\Cnocsit+\delta_1)$. We then obtain~\eqref{eq:prop-converse-nocsit} by combining~\eqref{eq:ub-R*-2} with~\eqref{eq:choose-gamma} and~\eqref{eq:asy-exp-gamma}, and dividing both sides of~\eqref{eq:ub-R*-2} by $\bl-1$.

To prove~\eqref{eq:asy-exp-gamma}, we note that by~\eqref{eq:condition-nocsit-subgradient} and the definition of $\liminf$, there exists a $\delta_2\in(0,\delta_1)$ such that
\begin{IEEEeqnarray}{rCl}
\inf\limits_{\gamma\in (\Cnocsit -\delta_2, \Cnocsit+\delta_2)}\frac{ \cdistno(\gamma) - \cdistno(\Cnocsit) }{\gamma - \Cnocsit} >0. \IEEEeqnarraynumspace
\label{eq:rephase-cond-pos-deriv}
\end{IEEEeqnarray}
Substituting~\eqref{eq:rephase-cond-pos-deriv} into~\eqref{eq:choose-gamma} and using that $\cdistno(\Cnocsit) =\error$, we obtain~\eqref{eq:asy-exp-gamma}.
This concludes the proof of Proposition~\ref{thm:zero-dispersion-nocsit-ub}.
\end{IEEEproof}

\subsection{Proof of Lemma~\ref{lem:bound-pdf-der-pdf-csir}}
\label{app:proof-boundedness-pdf-deri}
Throughout this section, we shall use $\constrm$ to indicate a finite constant that does not depend on any parameter of interest; its magnitude and sign may change at each occurrence.
The proof of this lemma is technical and makes use of concepts from Riemannian geometry.

%

Denote by $\{\setM_l\}$ the open subsets
\begin{IEEEeqnarray}{rCl}
\setM_l \define \{ \matH \in \complexset^{\txant\times\rxant} : \fnorm{\matH} < l \}
\end{IEEEeqnarray}
indexed by $l\in \amsbb{N}$.
We shall use the following \emph{flat Riemannian} metric~\cite[pp.~13 and~165]{jost11} on $\setM_l$
\begin{IEEEeqnarray}{rCl}
\langle\matH_1, \matH_2\rangle \define \Real\mathopen{}\big\{\tr\mathopen{}\left(\herm{\matH_1}\matH_2\right)\!\big\}.
\label{eq:Riemannian-metric}
\end{IEEEeqnarray}
Using this metric, we define the gradient $\gradient g$ of an arbitrary function $g : \setM_l \mapsto \realset $ as in~\eqref{eq:def-gradient-Crt}.
Note that the metric~\eqref{eq:Riemannian-metric} induces a norm on the tangent space of $\setM_l$, which can be identified with the Frobenius norm.

Our proof consists of two steps.
Let $f_{l} (u)$ denote the pdf of the random variable $\UgammaQ$ conditioned on $\randmatH \in \setM_l$.
We first show that there exist $l_0 \in \integerset$, $\delta>0$, and $\delta_1\in(0,\Cnocsit)$ such that $f_{l}(u)$ and $f_{l}'(u)$ are uniformly bounded for every $\gamma \in (\Cnocsit-\delta_1, \Cnocsit+\delta_1)$, every $\matQ\in\insetcove$, every $u\in[-\delta, \delta]$, and every $l\geq l_0$.
We then show that $u \mapsto f_{\UgammaQ}(u)$ is continuously differentiable on $(-\delta,\delta)$, and that for every $u\in(-\delta,\delta)$, the sequences $\{f_{l}(u)\}$ and $\{f'_l(u)\}$ converge uniformly to $f_{\UgammaQ}(u)$ and $f'_{\UgammaQ}(u)$, respectively, i.e.,
\begin{IEEEeqnarray}{rCl}
\lim\limits_{l\to\infty} \sup\limits_{u\in(-\delta,\delta)}\mathopen{}\left|f_l(u)- f_{\UgammaQ}(u)\right| &=& 0\label{eq:convergence-f_l}\\
\lim\limits_{l\to\infty} \sup\limits_{u\in(-\delta,\delta)} \mathopen{}\left|f'_l(u) - f_{\UgammaQ}'(u)\right| &=& 0\label{eq:convergence-f_l_d}
\end{IEEEeqnarray}
from which Lemma~\ref{lem:bound-pdf-der-pdf-csir} follows.


\subsubsection{Uniform Boundness of $\{\pdfgeneral_l\}$ and $\{\pdfgeneral'_l\}$}
\label{app:proof-uniform-bdd-pdfu}
To establish that $\{\pdfgeneral_l\}$ and $\{\pdfgeneral'_l\}$ are uniformly bounded, we shall need the following lemma.

\begin{lemma}
\label{lem:formula-Stokes}
Let $\setM$ be an oriented Riemannian manifold with Riemannian metric~\eqref{eq:Riemannian-metric} and let $\funct : \setM \mapsto \realset$ be a smooth function with $\mnorm{\gradient \funct } \neq 0$ on $\setM$. Let $P$ be a random variable on $\setM$ with smooth pdf~$\pdfgeneral$.
Then,
\begin{enumerate}[1)]
\item the pdf $\pdfgeneral_{\ast}$ of $\funct(P)$ at $u$ is
\begin{IEEEeqnarray}{rCl}
\label{eq:pdf-fu}
\pdfgeneral_{\ast}(u) = \int\nolimits_{\funct^{-1}(u) } \pdfgeneral \frac{\surform }{ \mnorm{ \gradient \funct } }
\end{IEEEeqnarray}
where $\funct^{-1}(u)$ denotes the preimage $\{x \in \setM : \funct(x) = u\}$ and  $\surform$ denotes the surface area form on $\funct^{-1}(u) $, chosen so that $\surform(\gradient \funct)>0$;

\item if the pdf $f$ is compactly supported, then the derivative of~$\pdfgeneral_{\ast}$~is
\begin{IEEEeqnarray}{rCl}
\pdfgeneral_{\ast}'(u) =  \int\nolimits_{\funct^{-1}(u)} \psi_1 \frac{\surform }{ \mnorm{ \gradient \funct } }
\label{eq:pdf-fu-der}
\end{IEEEeqnarray}
where $\psi_1$ is defined implicitly via
\begin{IEEEeqnarray}{rCl}
\psi_1 \volform  = d \mathopen{}\left(\pdfgeneral \frac{\surform}{ \mnorm{\gradient \funct} } \right)
\label{eq:def-psi-1-stokes}
 \end{IEEEeqnarray}
with $\volform$ denoting the volume form on $\setM$ and $d(\cdot)$ being exterior differentiation~\cite[p.~256]{munkres91-a}.
\end{enumerate}
\end{lemma}

\begin{IEEEproof}
To prove~\eqref{eq:pdf-fu}, we note that for arbitrary $a,b\in \realset$
\begin{IEEEeqnarray}{rCl}
\int\nolimits_{a}^{b} \pdfgeneral_{\ast}(u) du  &=& \int\nolimits_{\funct^{-1}((a,b)) }  \pdfgeneral  \volform \label{eq:proof-pdf-f-u-1}\\
&=& \int\nolimits_{a}^{b} \left(\int\nolimits_{\funct^{-1}(u) } \pdfgeneral \frac{ \surform }{ \mnorm{\gradient \funct}}  \right) du \label{eq:proof-pdf-f-u-2}
\end{IEEEeqnarray}
where~\eqref{eq:proof-pdf-f-u-2} follows from the smooth coarea formula~\cite[p.~160]{chavel06}.
This implies~\eqref{eq:pdf-fu}.

To prove~\eqref{eq:pdf-fu-der}, we shall use that for an arbitrary $\delta>0$,
\begin{IEEEeqnarray}{rCl}
\IEEEeqnarraymulticol{3}{l}{
\pdfgeneral_{\ast}(u + \delta) - \pdfgeneral_{\ast}(u) }\notag\\
\quad &=& \int\nolimits_{\funct^{-1}(u+\delta)} \pdfgeneral \frac{\surform }{ \mnorm{ \gradient \funct } } - \int\nolimits_{\funct^{-1}(u)} \pdfgeneral \frac{\surform }{ \mnorm{\gradient \funct } }\\
&=& \int\nolimits_{\funct^{-1}((u,u+\delta))  } d \mathopen{}\left(  \pdfgeneral \frac{\surform }{ \mnorm{ \gradient \funct } } \right) \label{eq:use-stokes}\\
&=&  \int\nolimits_{\funct^{-1}((u,u+\delta)) } \psi_1 \volform \label{eq:use-def-psi-1}
\end{IEEEeqnarray}
where in~\eqref{eq:use-stokes} we used Stoke's theorem~\cite[Th.~III.7.2]{{chavel06}}, that $\pdfgeneral$ is compactly supported, and that the restriction of the form~$\pdfgeneral \frac{\surform }{ \mnorm{ \gradient \funct } } $ to $\funct^{-1}((u,u+\delta))$ is also compactly supported;~\eqref{eq:use-def-psi-1} follows from the definition of~$\psi_1$ (see~\eqref{eq:def-psi-1-stokes}).
Equation~\eqref{eq:pdf-fu-der} follows then from similar steps as in~\eqref{eq:proof-pdf-f-u-1}--\eqref{eq:proof-pdf-f-u-2}.
\end{IEEEproof}

Using Lemma~\ref{lem:formula-Stokes}, we obtain
\begin{IEEEeqnarray}{rCl}
\pdfgeneral_{l}(u) = \int\nolimits_{\functQ^{-1}(u) \cap \setM_l} \frac{\pdfgeneral_{\randmatH}}{\prob[\randmatH \in \setM_l]} \frac{ \surform}{ \mnorm{ \gradient \functQ } }
\label{eq:formula-varphi}
\end{IEEEeqnarray}
and
\begin{IEEEeqnarray}{rCl}
\pdfgeneral'_l(u) = \int\nolimits_{\functQ^{-1}(u) \cap \setM_l}\frac{ \psi_1}{\prob[\randmatH \in \setM_l]}  \frac{ \surform}{\mnorm{\gradient \functQ} }
\label{eq:formula-varphi-d}
\end{IEEEeqnarray}
where $\psi_1$ satisfies
\begin{IEEEeqnarray}{rCl}
\psi_1 \volform  = d \mathopen{}\left(\pdfgeneral_{\randmatH} \frac{\surform}{\mnorm{ \gradient \functQ }} \right).
\label{eq:def-psi-1-stokes-real 1}
\end{IEEEeqnarray}
Since $\prob[\randmatH \in \setM_l]\to 1$ as $l\to\infty$, there exists a $l_0$ such that $\prob[\randmatH \in \setM_l] \geq  1/2$ for every $l\geq l_0$.

We next show that there exist $\delta>0$, $0<\delta_1<\Cnocsit$, such that for every $\gamma\in(\Cnocsit-\delta_1 , \Cnocsit+\delta_1)$, every $u\in(-\delta,\delta)$, every $\matQ \in \setU_{\txant}^{\mathrm{e}}$, every $\matH \in \functQ^{-1}(u) \cap \setM_l$, and every $l\geq l_0$
\begin{IEEEeqnarray}{rCl}
\pdfH(\matH) &\leq& \constrm \cdot \fnorm{\matH}^{-2\txant\rxant -3}\label{eq:bound-pdfH-1}\\
|\psi_1(\matH)| &\leq& \constrm \cdot \fnorm{\matH}^{-2\txant\rxant -3}
\label{eq:bound-psi-1}
\end{IEEEeqnarray}
and
\begin{IEEEeqnarray}{rCl}
A_l(u) \define \int\nolimits_{\funct_{\argpn, \matQ}^{-1}(u) \cap \setM_l } \frac{ \fnorm{\matH}^{-2\txant\rxant -3} \, \surform}{\fnorm{\gradient \functQ}}  \leq  \constrm.
\label{eq:bound-area-u-0}
\end{IEEEeqnarray}
The uniform boundedness of~$\{\pdfgeneral_l\}$ and $\{\pdfgeneral'_l\}$ follows then by using the bounds~\eqref{eq:bound-pdfH-1}--\eqref{eq:bound-area-u-0} in~\eqref{eq:formula-varphi} and~\eqref{eq:formula-varphi-d}.

\paragraph*{Proof of~\eqref{eq:bound-pdfH-1}}
Since $\pdfH(\matH)$ is continuous by assumption, it is uniformly bounded for every $\matH\in\setM_1$. Hence,~\eqref{eq:bound-pdfH-1} holds for every $\matH\in\setM_1$.
For $\matH \notin \setM_1$, we have by~\eqref{eq:cond-bdd-pdf}
\begin{IEEEeqnarray}{rCl}
\pdfH(\matH) \leq a \fnorm{\matH}^{-2\txant\rxant -\lfloor(1+\rxant)^2/2\rfloor -1} \leq a \fnorm{\matH}^{-2\txant\rxant -3}.\IEEEeqnarraynumspace
\end{IEEEeqnarray}
This proves~\eqref{eq:bound-pdfH-1}.

\paragraph*{Proof of~\eqref{eq:bound-psi-1}}

The surface area form $\surform$ on $\functQ^{-1}(u) \cap \setM_l$ is given by
\begin{IEEEeqnarray}{rCl}
\surform = \frac{\star d \functQ}{\mnorm{ \gradient \functQ}}
\label{eq:surface-form-dS}
\end{IEEEeqnarray}
where $\star$ denotes the Hodge star operator~\cite[p.~103]{jost11} induced by the metric~\eqref{eq:Riemannian-metric}.
Using~\eqref{eq:surface-form-dS} and the definition of the Hodge star operator, the RHS of \eqref{eq:def-psi-1-stokes-real 1} becomes
\begin{IEEEeqnarray}{rCl}
\IEEEeqnarraymulticol{3}{l}{ d\mathopen{}\Bigg(\frac{\pdfH }{\mnorm{ \gradient \functQ }^2}\Bigg) \wedge \star d \functQ + \frac{\pdfH }{\mnorm{ \gradient \functQ }^2} \wedge d \star d \funct_{\argpn,\matQ} }\notag\\
 \quad &=&\Bigg( \frac{\langle  \gradient \pdfH,\gradient\functQ \rangle}{ \mnorm{ \gradient \functQ  }^2 }
 -  \frac{\pdfH   \langle  \gradient \mnorm{\gradient \functQ }^2 , \gradient \functQ \rangle  }{ \mnorm{ \gradient \functQ }^4} \notag\\
 &&\quad - \,\frac{\pdfH \cdot \Delta \functQ }{\mnorm{\gradient  \functQ }^2} \Bigg) \volform\label{eq:bound-psi-1-init1-3}
\end{IEEEeqnarray}
where $\wedge$ denotes the wedge product~\cite[p.~237]{munkres91-a} and $\Delta$ denotes the Laplace operator~\cite[Eq.~(3.1.6)]{jost11}.\footnote{The Laplace operator used here and in~\cite[Eq.~(3.1.6)]{jost11} differs from the usual one on $\realset^{\bl}$, as defined in calculus, by a minus sign. See~\cite[Sec.~3.1]{jost11} for a more detailed discussion.}
%
From~\eqref{eq:bound-psi-1-init1-3} we get
\begin{IEEEeqnarray}{rCl}
|\psi_1| &=& \bigg|\frac{\langle  \gradient \pdfH,\gradient\functQ \rangle}{ \mnorm{ \gradient \functQ  }^2 }
 -  \frac{\pdfH   \langle  \gradient \mnorm{\gradient \functQ }^2 , \gradient \functQ \rangle  }{ \mnorm{ \gradient \functQ }^4}\notag\\
   &&\,\,\,-\, \frac{\pdfH \cdot \Delta \functQ }{\mnorm{\gradient  \functQ }^2} \bigg|\\
   &\leq & \frac{\fnorm{\gradient\pdfH}}{ \mnorm{ \gradient \functQ  } }  +  \frac{\pdfH   \fnorm{ \gradient \mnorm{\gradient \functQ }^2}}{ \mnorm{ \gradient \functQ }^3}
   +\frac{\pdfH \cdot |\Delta \functQ| }{\mnorm{\gradient  \functQ }^2}\notag\\
   &&\label{eq:bound-psi-1-start}
\end{IEEEeqnarray}
where the last step follows from the triangle inequality and the Cauchy-Schwarz inequality.

We proceed to lower-bound $\mnorm{\gradient  \functQ }$.
Using the definition of the gradient~\eqref{eq:def-gradient-Crt} together with the matrix identities~\cite[p.~29]{lutkepohl96}
\begin{IEEEeqnarray}{rCl l}
\det(\matI + \varepsilon \matA) &=& 1 + \varepsilon \tr(\matA) + \bigO(\varepsilon^2), \quad&\varepsilon \to 0 \label{eq:matrix-identity}\\
(\matI+ \varepsilon\matA)^{-1} &=& \matI -\varepsilon \matA +\bigO(\varepsilon^2), \quad &\varepsilon \to  0
\end{IEEEeqnarray}
for every bounded square matrix $\matA$, we obtain
\begin{IEEEeqnarray}{rCl}
\gradient\functQ (\matH) &=& -\frac{2 \matQ\matH \matPhi^{-3} }{\left(\tr\mathopen{}\big(\matI_{\rxant} - \matPhi^{-2}\big) \right)^{3/2}  }\notag\\
 && \times  \Big( \underbrace{\tr(\matI_{\rxant} -\matPhi^{-2})\matPhi^2 + (\gamma - \log\det \matPhi)\matI_{\rxant}}_{\define \matT}  \Big) \IEEEeqnarraynumspace
\label{eq:def-gradient-f}
\end{IEEEeqnarray}
where $\matPhi \define \matI_{\rxant} + \herm{\matH} \matQ\matH $.

Fix an arbitrary $\delta_1\in(0,\Cnocsit)$ and choose $\delta \in (0, (\Cnocsit-\delta_1)/\sqrt{\rxant})$.
%
%
%
We first bound~$\tr(\matI_{\rxant} - \matPhi^{-2})$ as
\begin{IEEEeqnarray}{rCl}
\rxant \geq \tr\mathopen{}\big(\matI_{\rxant} - \matPhi^{-2}\big) &\geq&  1 - (1+\lambda_{\max}(\herm{\matH}\matQ\matH))^{-2}. \label{eq:bound-on-trace-I-Phi-0}
\end{IEEEeqnarray}
It follows from the first inequality in~\eqref{eq:bound-on-trace-I-Phi-0} and from~\eqref{eq:def-f-argpn-matQ} that for every $u\in(-\delta, \delta)$
\begin{IEEEeqnarray}{rCl}
|\gamma - \log\det\matPhi| = |u| \sqrt{\tr(\matI_{\rxant} - \matPhi^{-2}) } \leq \delta \sqrt{\rxant}.
\label{eq:bound-argpn-logdet}
\end{IEEEeqnarray}
Using~\eqref{eq:bound-argpn-logdet} and that the determinant is given by the product of the eigenvalues, we obtain that, for every $ \gamma \in(\Cnocsit-\delta_1,  \Cnocsit-\delta_1)$ and every $u\in(-\delta, \delta)$,
\begin{IEEEeqnarray}{rCl}
\rxant \log( 1 + \lambda_{\max} (\herm{\matH}\matQ\matH )) &\geq& \log\det\matPhi \label{eq:relation-lambda-max-det}\\
 &\geq& \gamma - \sqrt{\rxant}\delta\\
 & \geq &\Cnocsit-\delta_1 - \sqrt{\rxant}\delta > 0\IEEEeqnarraynumspace
\end{IEEEeqnarray}
which implies that
\begin{IEEEeqnarray}{rCl}
\lambda_{\max} (\herm{\matH}\matQ\matH ) \geq e^{(\Cnocsit -\delta_1- \sqrt{\rxant}\delta)/\rxant} -1 >0.
\label{eq:bound-lambda-max}
\end{IEEEeqnarray}
Combing~\eqref{eq:bound-lambda-max} with the second inequality in~\eqref{eq:bound-on-trace-I-Phi-0}, we obtain
\begin{IEEEeqnarray}{rCl}
 \tr\mathopen{}\big(\matI_{\rxant} - \matPhi^{-2}\big)  \geq 1-    e^{-2(\Cnocsit -\delta_1- \sqrt{\rxant}\delta)/\rxant}.
\label{eq:bound-on-trace-I-Phi-1}
\end{IEEEeqnarray}
We use~\eqref{eq:bound-argpn-logdet} and~\eqref{eq:bound-on-trace-I-Phi-1} to lower-bound the smallest eigenvalue of the matrix $\matT$ defined in~\eqref{eq:def-gradient-f} as
\begin{IEEEeqnarray}{rCl}
\lambda_{\min}\mathopen{}\left(\matT \right) &=&\tr(\matI_{\rxant} -\matPhi^{-2}) \underbrace{\lambda_{\min}( \matPhi^2)}_{\geq 1} + (\gamma - \log\det \matPhi) \IEEEeqnarraynumspace \\
&\geq & \tr(\matI_{\rxant} -\matPhi^{-2}) - \delta \sqrt{\rxant}\\
&\geq &  1-    e^{-2(\Cnocsit -\delta_1- \sqrt{\rxant}\delta)/\rxant} - \delta \sqrt{\rxant} \label{eq:bound-eigenvalue-complex-mat-3}.
\end{IEEEeqnarray}
The RHS of~\eqref{eq:bound-eigenvalue-complex-mat-3} can be made positive if we choose $\delta$ sufficiently small, in which case $\matT$ is invertible. We can theorefore lower-bound $\mnorm{ \gradient \functQ }$ as 
\begin{IEEEeqnarray}{rCl}
\mnorm{ \gradient \funct_{\argpn, \matQ} } &=&  \frac{2}{\left(\tr\mathopen{}\big(\matI_{\rxant} - \matPhi^{-2}\big) \right)^{3/2}  } \fnorm{ \matQ\matH\matPhi^{-3} \matT }\\
&\geq & \frac{2}{\rxant^{3/2}}\fnorm{ \matQ\matH\matPhi^{-3}}\cdot \frac{1}{\fnorm{ \matT^{-1} }} \label{eq:bound-grad-f-norms-1}\\
&\geq & \frac{2}{\rxant^{3/2}}\fnorm{ \matQ\matH} \cdot \frac{1}{\fnorm{\matPhi^{3}}} \cdot \frac{1}{\fnorm{ \matT^{-1} }} \label{eq:bound-grad-f-norms-2}.
\end{IEEEeqnarray}
Here, we use the first inequality in~\eqref{eq:bound-on-trace-I-Phi-0} and the submultiplicativity of the Frobenius norm.
The term $\fnorm{ \matQ\matH}$ can be bounded as
\begin{IEEEeqnarray}{rCl}
\fnorm{ \matQ\matH} &\geq & \frac{\fnorm{\herm{\matH} \matQ\matH}}{\fnorm{\matH}}\\
&\geq & \frac{ \lambda_{\max}(\herm{\matH} \matQ \matH) }{ \fnorm{\matH}}\label{eq:bound-fnorm-QH-2-1} \\
&\geq & \frac{e^{(\Cnocsit - \delta_1 - \sqrt{\rxant}\delta )/\rxant} -1}{\fnorm{\matH}} \label{eq:bound-fnorm-QH-2}
\end{IEEEeqnarray}
where~\eqref{eq:bound-fnorm-QH-2} follows from~\eqref{eq:bound-lambda-max}.

The term~$\fnorm{\matPhi^{3}}$ in~\eqref{eq:bound-grad-f-norms-2} can be upper-bounded as
\begin{IEEEeqnarray}{rCl}
\fnorm{\matPhi^{3}} &\leq& \sqrt{\rxant}(1+ \lambda_{\max}(\herm{\matH}\matQ\matH))^3 \\
&\leq&  \sqrt{\rxant} (1+ \det\matPhi )^3 \label{eq:bound-fnorm-matPhi-3-2}\\
&\leq& \constrm. \label{eq:bound-fnorm-matPhi-3}
\end{IEEEeqnarray}
Here,~\eqref{eq:bound-fnorm-matPhi-3} follows from~\eqref{eq:bound-argpn-logdet} and because $\gamma \leq \Cnocsit+\delta$.
Finally, $\fnorm{ \matT^{-1} }$ in~\eqref{eq:bound-grad-f-norms-2} can be bounded as
\begin{IEEEeqnarray}{rCl}
\fnorm{ \matT^{-1} }  \leq \sqrt{\rxant} \lambda_{\max}(\matT^{-1}) = \frac{\sqrt{\rxant} }{\lambda_{\min}(\matT)}.
\label{eq:bound-fnorm-matT-1}
\end{IEEEeqnarray}
The RHS of~\eqref{eq:bound-fnorm-matT-1} is bounded because of~\eqref{eq:bound-eigenvalue-complex-mat-3}.
Substituting~\eqref{eq:bound-fnorm-QH-2},~\eqref{eq:bound-fnorm-matPhi-3} and~\eqref{eq:bound-fnorm-matT-1} into \eqref{eq:bound-grad-f-norms-2}, we conclude that
\begin{IEEEeqnarray}{rCl}
\| \gradient \functQ\|^{-1} \leq  \constrm\cdot \fnorm{\matH}.
\label{eq:bound-grad-f-norm-final}
\end{IEEEeqnarray}

Following similar steps as the ones reported in~\eqref{eq:matrix-identity}--\eqref{eq:bound-grad-f-norm-final}, we can show that
\begin{IEEEeqnarray}{rCl}
 \fnorm{ \gradient \mnorm{\gradient \functQ }^2}<\constrm  \cdot\mnorm{\gradient \functQ } \label{eq:bound-gradd-delta-1}
 \end{IEEEeqnarray}
 and
 \begin{IEEEeqnarray}{rCl}
  |\Delta \functQ|<\constrm. \label{eq:bound-gradd-delta}
\end{IEEEeqnarray}
Substituting~\eqref{eq:bound-grad-f-norm-final}--\eqref{eq:bound-gradd-delta} into~\eqref{eq:bound-psi-1-start} and using the bounds~\eqref{eq:cond-bdd-pdf} and~\eqref{eq:cond-bdd-pdf-deri}, we obtain~\eqref{eq:bound-psi-1}.

%

\paragraph*{Proof of~\eqref{eq:bound-area-u-0}}

We begin by observing that for every $\matH \in \functQ^{-1}(u) \cap \setM_l$, every $\gamma\in(\Cnocsit-\delta_1 , \Cnocsit+\delta_1)$, every $u\in(-\delta,\delta)$ and every $\matQ \in \setU_{\txant}^{\mathrm{e}}$
\begin{IEEEeqnarray}{rCl}
\fnorm{\matH}^2 &\geq& \frac{\tr(\herm{\matH}\matQ\matH)}{\tr(\matQ)} \label{eq:bound-fnorm-H-11}\\
&\geq& \frac{1}{\snr} \lambda_{\max}(\herm{\matH}\matQ\matH)\label{eq:bound-fnorm-H-12}\\
&\geq& \frac{1}{\snr} \left(e^{(\Cnocsit -\delta_1- \sqrt{\rxant}\delta)/\rxant} -1 \right) \define k_0.\label{eq:bound-fnorm-H-1}
\end{IEEEeqnarray}
Here,~\eqref{eq:bound-fnorm-H-11} follows from  Cauchy-Schwarz inequality;~\eqref{eq:bound-fnorm-H-12} follows because $\tr(\matQ)=\snr$ for every $\matQ\in\insetcove$;~\eqref{eq:bound-fnorm-H-1} follows from~\eqref{eq:bound-lambda-max}.
From~\eqref{eq:bound-fnorm-H-1} we conclude that
\begin{IEEEeqnarray}{rCl}
\Big(\functQ^{-1}((-\delta,\delta)) \cap \setM_l\Big)
\subset \setM' \define \{\matH \in \complexset^{\txant\times\rxant}: \fnorm{\matH} \geq \sqrt{k_0}\}. \nonumber\\
\label{eq:subset-funtQ-1}
\end{IEEEeqnarray}
To upper-bound $A_l(u)$, we note that
\begin{IEEEeqnarray}{rCl}
\int\nolimits_{-\delta}^\delta A_l(u) du &=& \int\nolimits_{\functQ^{-1}((-\delta,\delta))  \cap \setM_l }   \fnorm{\matH}^{-2\txant\rxant -3} \volform\label{eq:bound-area-l-u-1}\\
&\leq& \int\nolimits_{\setM'}   \fnorm{\matH}^{-2\txant\rxant -3}  \volform \label{eq:bound-area-l-u-2}\\
&=& \constrm \cdot \int\nolimits_{\sqrt{k_0}}^{\infty} x^{-4} dx\label{eq:bound-area-l-u-3}\\
& =&\constrm.
\label{eq:bound-area-l-u}
\end{IEEEeqnarray}
Here,~\eqref{eq:bound-area-l-u-1} follows from the smooth coarea formula~\cite[p.~160]{chavel06};~\eqref{eq:bound-area-l-u-2}  follows from~\eqref{eq:subset-funtQ-1};~\eqref{eq:bound-area-l-u-3} follows by writing the RHS of~\eqref{eq:bound-area-l-u-2} in polar coordinates and by using that, by~\eqref{eq:bound-lambda-max}, $k_0>0$.
By the mean value theorem, it follows from~\eqref{eq:bound-area-l-u} that there exists a $u_0\in(-\delta,\delta)$ satisfying
\begin{IEEEeqnarray}{rCl}
A_l(u_0) = \frac{\int\nolimits_{-\delta}^\delta A_l(u) du}{2\delta} \leq \constrm.
\label{eq:bound-A-l-u-0}
\end{IEEEeqnarray}
Next, for every $u\in(u_0, \delta)$ we have that
\begin{IEEEeqnarray}{rCl}
A_l(u) - A_l (u_0) &=&  \int\nolimits_{\functQ^{-1}(u) \cap \setM_l }\frac{  \fnorm{\matH}^{-2\txant\rxant -3}  }{\fnorm{\gradient \functQ }} \surform \notag\\
&& -\, \int\nolimits_{\functQ^{-1}(u_0) \cap \setM_l } \frac{ \fnorm{\matH}^{-2\txant\rxant -3}  }{\fnorm{\gradient \functQ }} \surform\label{eq:calc-deri-area-0}\\
 &=& \int\nolimits_{\functQ^{-1}((u_0, u)) \cap \setM_l}\!\! d \mathopen{}\left(  \frac{  \fnorm{\matH}^{-2\txant\rxant -3}  }{\fnorm{\gradient \functQ}} \surform \right) \label{eq:use-stokes-2nd} \IEEEeqnarraynumspace\\
 &=& \int\nolimits_{\functQ^{-1}((u_0, u)) \cap \setM_l   } \psi_2 \volform  \label{eq:calc-deri-area}
\end{IEEEeqnarray}
where $\psi_2$ is defined implicitly via
\begin{equation}
\psi_2 \volform =   d \mathopen{}\left(  \frac{ \fnorm{\matH}^{-2\txant\rxant -3} }{\fnorm{\gradient \functQ}} \surform \right) .
\label{eq:expression-psi-2}
\end{equation}
Here,~\eqref{eq:use-stokes-2nd} follows from Stokes' theorem. Following similar steps as the ones reported in~\eqref{eq:surface-form-dS}--\eqref{eq:bound-gradd-delta}, we obtain that
\begin{IEEEeqnarray}{rCl}
|\psi_2| \leq \constrm\cdot  \fnorm{\matH}^{-2\txant\rxant -1} .
\label{eq:bound-psi-2}
\end{IEEEeqnarray}
We can therefore upper-bound $A_l(u)$ as
\begin{IEEEeqnarray}{rCl}
A_l(u) &=&  A_l(u_0) +  \int\nolimits_{\functQ^{-1}((u_0, u)) \cap \setM_l   } \psi_2 \volform \label{eq:bound-A_l_u-1}\\
&\leq & \constrm + \int\nolimits_{\setM' }  \constrm\cdot  \fnorm{\matH}^{-2\txant\rxant -1}   \volform\label{eq:bound-A_l_u-2}\\
&\leq & \constrm +  \int\nolimits_{\sqrt{k_0}}^{\infty} \constrm\cdot x^{-2 } dx \label{eq:bound-A_l_u-2}\\
&=& \constrm\label{eq:bound-A_l_u-3}.
\end{IEEEeqnarray}
Here,~\eqref{eq:bound-A_l_u-1} follows from~\eqref{eq:calc-deri-area};~\eqref{eq:bound-A_l_u-2} follows from~\eqref{eq:bound-A-l-u-0},~\eqref{eq:bound-psi-2}, and~\eqref{eq:subset-funtQ-1}.
Note that the bound~\eqref{eq:bound-A_l_u-3} is uniform in $\gamma$, $\matQ$, $u$, and~$l$.
Following similar steps as the ones reported in~\eqref{eq:calc-deri-area-0}--\eqref{eq:bound-A_l_u-3}, we obtain the same result for $u\in(-\delta, u_0)$. This proves~\eqref{eq:bound-area-u-0}.

\subsubsection{Convergence of $f_l(u)$ and $f_l'(u)$}
In this section, we will prove~\eqref{eq:convergence-f_l} and~\eqref{eq:convergence-f_l_d}.
By Lemma~\ref{lem:formula-Stokes},
\begin{IEEEeqnarray}{rCl}
\pdfgeneral_{\UgammaQ}(u) = \int\nolimits_{\functQ^{-1}(u)}  \frac{\pdfgeneral_{\randmatH} \surform}{ \mnorm{ \gradient \functQ } }.
\label{eq:formula-varphi-pdf-u}
\end{IEEEeqnarray}
We have the following chain of inequalities
\begin{IEEEeqnarray}{rCl}
\IEEEeqnarraymulticol{3}{l}{
|\pdfgeneral_{l}(u) - \pdfgeneral_{\UgammaQ}(u) | }\notag\\
\quad&\leq& |\prob[ \randmatH \in \setM_l]\pdfgeneral_{l}(u)  -\pdfgeneral_{\UgammaQ}(u)|\notag\\
&& +\, |(1-\prob[ \randmatH \in \setM_l] )\pdfgeneral_{l}(u)|\label{eq:bound-diff-pdf-U-1}\\
&\leq& \int\nolimits_{\functQ^{-1}(u) \cap (\complexset^{\txant\times\rxant} \setminus \setM_l)}  \frac{\pdfgeneral_{\randmatH} \surform}{ \mnorm{ \gradient \functQ } } \notag\\
 &&+\, \constrm \cdot(1-\prob[ \randmatH \in \setM_l] )\label{eq:bound-diff-pdf-U-2}\\
&\leq & \constrm \cdot \int\nolimits_{\functQ^{-1}(u) \cap (\complexset^{\txant\times\rxant} \setminus \setM_l)}  \frac{ \fnorm{\matH}^{-2\txant\rxant-3} \surform}{ \mnorm{ \gradient \functQ } }\notag\\
&& + \,\constrm \cdot (1-\prob[ \randmatH \in \setM_l] )\label{eq:bound-diff-pdf-U-3}.
%
\end{IEEEeqnarray}
Here,~\eqref{eq:bound-diff-pdf-U-1} follows from the triangle inequality;~\eqref{eq:bound-diff-pdf-U-2} follows from~\eqref{eq:formula-varphi} and because $\{\pdfgeneral_{l}(u)\}$ is uniformly bounded;~\eqref{eq:bound-diff-pdf-U-3} follows from~\eqref{eq:bound-pdfH-1}. 
Following similar steps as the ones reported in~\eqref{eq:bound-area-l-u-1}--\eqref{eq:bound-A_l_u-3}, we upper-bound the first term on the RHS of~\eqref{eq:bound-diff-pdf-U-3} as
\begin{IEEEeqnarray}{rCl}
 \int\nolimits_{\functQ^{-1}(u) \cap (\complexset^{\txant\times\rxant} \setminus\setM_l)}  \frac{ \fnorm{\matH}^{-2\txant\rxant-3} \surform}{ \mnorm{ \gradient \functQ } }  \leq \frac{\constrm}{l}.
 \label{eq:ub-int-funcQ-u-b}
\end{IEEEeqnarray}
Substituting~\eqref{eq:ub-int-funcQ-u-b} into~\eqref{eq:bound-diff-pdf-U-3}, and using that $\prob[ \randmatH \in \setM_l] \to 1$ as $l\to\infty$, we obtain that
\begin{IEEEeqnarray}{rCl}
\lim\limits_{l\to\infty} \sup\limits_{u\in(-\delta,\delta)}\mathopen{}\left|f_l(u)- f_{\UgammaQ}(u)\right| &=& 0.
\label{eq:bound-diff-pdf-U-l}
\end{IEEEeqnarray}
This proves~\eqref{eq:convergence-f_l}.

To prove~\eqref{eq:convergence-f_l_d}, we proceed as follows. Let $C^1([-\delta,\delta])$ denote the set of continuously differentiable functions on the compact interval $[-\delta,\delta]$.
 The space $C^1([-\delta,\delta])$ is a Banach space (i.e., a complete normed vector space) when equipped with the~$C^1$ norm~\cite[p.~92]{hunter01}
 \begin{equation}
 \|f\|_{C^1([-\delta,\delta])} =\sup\limits_{x\in[-\delta,\delta]}(|f(x)| + |f'(x)|).
 \label{eq:def-C1-norm}
 \end{equation}
Following similar steps as in~\eqref{eq:calc-deri-area-0}--\eqref{eq:bound-A_l_u-3}, we obtain that $\{f'_l\}$ is continuous on $[-\delta,\delta]$, i.e., the restriction of~$\{f_l\}$ to  $[-\delta,\delta]$ belongs to $C^1([-\delta,\delta])$.
Moreover, following similar steps as in~\eqref{eq:bound-diff-pdf-U-1}--\eqref{eq:bound-diff-pdf-U-l}, we obtain that for all $m>l>0$
\begin{IEEEeqnarray}{rCl}
\lim\limits_{l\to\infty} \sup_{u\in[-\delta,\delta]} \mathopen{}\Big(|f_m(u)-f_{l}(u)|+ | f_{m}'(u) -f_{l}'(u)| \Big)=0. \IEEEeqnarraynumspace
\end{IEEEeqnarray}
This means that $\{f_l\}$ restricted to $[-\delta,\delta]$ is a Cauchy sequence, and, hence, converges in $C^1([-\delta,\delta])$ with respect to the $C^1$ norm~\eqref{eq:def-C1-norm}.
Moreover, by~\eqref{eq:bound-diff-pdf-U-l} the limit of $\{f_l\}$ is $f_{\UgammaQ}$.
Therefore, we have $f_{\UgammaQ} \in C^1([-\delta,\delta])$,  and $\{f_l'\}$ converges to $f'_{\UgammaQ}$ with respect to the sup-norm $\|\cdot\|_{\infty}$.
This proves~\eqref{eq:convergence-f_l_d}.

\section{Proof of the Achievability Part of Theorem~\ref{thm:zero-dispersion-nocsit}}
\label{app:proof-asy-ach-nocsit}

We prove the achievability asymptotic expansion for Theorem~\ref{thm:zero-dispersion-nocsit}. More
precisely, we prove the following:
\begin{prop}
\label{thm:proof-asy-ach-nocsit}
Assume that there exists a $\optcov \in \setU_{\txant}$ satisfying~\eqref{eq:C-epsilon-nocsit-comput}.
Let $F_{\matQ^*}(\cdot)$ be as in~\eqref{eq:def-F-matQ}.
Assume that the joint pdf of the nonzero eigenvalues of $\herm{\randmatH}\optcov\randmatH$ is continuously differentiable and that $F_{\matQ^*}(\cdot)$ is differentiable and strictly increasing at
$\Cnocsit$, i.e.,
\begin{IEEEeqnarray}{rCl}
F_{\matQ^*}'(\Cnocsit)  >0.
\label{eq:thm-dispersion-condition2}
\end{IEEEeqnarray}
Let $\txantop = \rank(\optcov)$. Then,
\begin{IEEEeqnarray}{rCl}
\label{eq:ach-asy-formula}
\Rnocsi^{\ast}(\bl,\error)\geq \Cnocsit - (1+\rxant\txantop) \frac{\log\bl}{\bl} +\bigO\mathopen{}\left(\frac{1}{n}\right).
\end{IEEEeqnarray}
\end{prop}

Note that the conditions on the distribution of the fading matrix~$\randmatH$ under which Proposition~\ref{thm:proof-asy-ach-nocsit} holds are less stringent than (and, because of Proposition~\ref{prop:continuous} on p.~\pageref{prop:continuous} and Lemma~\ref{lem:bound-pdf-der-pdf-csir} on p.~\pageref{lem:bound-pdf-der-pdf-csir}, implied by) the conditions under which Proposition~\ref{thm:zero-dispersion-nocsit-ub} (converse part of Theorem~\ref{thm:zero-dispersion-nocsit}) holds.

\begin{IEEEproof}
The proof follows closely the proof of the achievability part of Theorem~\ref{thm:asy-mimo-csirt}. Following similar steps as the ones reported in~\eqref{eq:app-prob-error2-inner-step0-t}--\eqref{eq:asy-analysis-error2-t}, we obtain
\begin{IEEEeqnarray}{rCl}
\prob\mathopen{}\left[\prod_{i=1}^{\rxant} B_i \leq \gamma_\bl\right]&\leq& \bl^{\rxant\txantop} \gamma_\bl^{\bl-\txantop-\rxant}.\label{eq:asy-analysis-error2-nocsit}
\end{IEEEeqnarray}
Setting $\tau =1/\bl$ and $\gamma_\bl = \exp(-\Cnocsit + \bigO(1/\bl))$ in Theorem~\ref{thm:actual_ach_bound_nocsit}, and using \eqref{eq:asy-analysis-error2-nocsit}, we obtain
\begin{IEEEeqnarray}{rCl}
\label{eq:ach_logM_MIMO_nocsit}
\frac{\log \NumCode}{\bl} &\geq&  \Cnocsit - (1+\rxant\txantop) \frac{\log\bl}{\bl} +\bigO\mathopen{}\left(\frac{1}{n}\right).
\end{IEEEeqnarray}

To conclude the proof, we show that there exists indeed a $\gamma_\bl = \exp(-\Cnocsit + \bigO(1/\bl))$ satisfying
\begin{IEEEeqnarray}{rCl}
  \prob\mathopen{}\big[\sin^2\{\matI_{\bl,\txantop}, \sqrt{\bl}\matI_{\bl,\txantop}\halfcov\randmatH +\randmatW\}\leq \gamma_\bl\big] \geq 1-\error+1/\bl \IEEEeqnarraynumspace
  \label{eq:ach-nocsi-condition-gamma-n}
 \end{IEEEeqnarray}
 where $\halfcov \in\complexset^{\txantop\times\txant}$ satisfies $\herm{\halfcov}\halfcov=\optcov$.
Hereafter, we restrict ourselves to $\gamma_{\bl} \in \big(e^{-\Cnocsit-\delta},e^{-\Cnocsit+\delta} \big)$ for some $\delta \in (0,\Cnocsit)$.
Let $\minantop \define \min\{\txantop,\rxant\}$.
Consider the SVD of $\halfcov\matH$
\begin{IEEEeqnarray}{rCl}
\label{eq:ach-nocsit-svd}
\halfcov\matH =\matL \underbrace{\left(
\begin{array}{cc}
\mathsf{\Sigma}_{\minantop}
& \mathbf{0}_{\minantop \times (\rxant-\minantop)}\\
\mathbf{0}_{ (\txantop-\minantop)\times \minantop} &\mathbf{0}_{(\txantop-\minantop)\times(\rxant-\minantop)}\\
\end{array}
\right)}_{\define \mathsf{\Sigma}} \herm{\matV} \IEEEeqnarraynumspace
\end{IEEEeqnarray}
where $\matL\in\complexset^{\txantop\times\txantop}$ and $\matV\in\complexset^{\rxant\times\rxant}$ are unitary matrices, $\mathsf{\Sigma}_{\minantop}=\diag\{\sqrt{\lambda_1},\ldots,\sqrt{\lambda_{\minantop}}\}$ with $\lambda_1,\ldots,\lambda_{\minantop}$ being the $\minantop$ largest eigenvalues of $\herm{\matH}\optcov\matH$, and $\mathbf{0}_{a,b}$ denotes the all zero matrix of size $a\times b$.
Conditioned on $\randmatH=\matH$, we have
\begin{IEEEeqnarray}{rCl}
\IEEEeqnarraymulticol{3}{l}{
\sin^2\{\matI_{\bl,\txantop}, \sqrt{\bl}\matI_{\bl,\txantop}\matU\matH +\randmatW\} }\notag\\
 \quad &=&  \sin^2\mathopen{}\big\{ \matI_{\bl,\txantop}\matL,  (\sqrt{\bl}\matI_{\bl,\txantop}\halfcov \matH +\randmatW) \matV\big\}\label{eq:ach-sin-equality}\\
&=& \sin^2\mathopen{}\big\{\widetilde{\matL} \matI_{\bl,\txantop}\matL,  \widetilde{\matL} (\sqrt{\bl}\matI_{\bl,\txantop}\halfcov \matH +\randmatW) \matV\big\}\label{eq:ach-sin-equality-2}\\
&=&\sin^2\mathopen{}\big\{\matI_{\bl,\txantop},  \sqrt{\bl}\matI_{\bl,\txantop}\mathbf{\Sigma} + \randmatW\big\}\label{eq:ach-sin-equality-3}
\end{IEEEeqnarray}
where
\begin{IEEEeqnarray}{rCl}
\widetilde{\matL} &\define& \left(
               \begin{array}{cc}
                 \herm{\matL} & \mathbf{0}_{(\bl-\txantop)\times\txantop} \\
                 \mathbf{0}_{\txantop\times(\bl-\txantop)} & \matI_{\bl-\txantop} \\
               \end{array}
             \right)
\end{IEEEeqnarray}
is unitary. Here,~\eqref{eq:ach-sin-equality} follows because $\spanm(\matA)=\spanm(\matA\matB)$ for every invertable matrix $\matB$; \eqref{eq:ach-sin-equality-2} follows because the principal angles between two subspaces are invariant under simultaneous rotation of the two subspaces; \eqref{eq:ach-sin-equality-3} follows because $\randmatW$ is isotropically distributed, which implies that $\widetilde{\matL}\randmatW\matV$ has the same distribution as $\randmatW$.

Let $\vece_j$ and $\randvecw_j$ be the $j$th column of $\matI_{\bl,\txantop}$ and $\randmatW$, respectively.
Then
\begin{IEEEeqnarray}{rCl}
\IEEEeqnarraymulticol{3}{l}{
 \prob\mathopen{}\left[\sin^2 \mathopen{}\big\{\matI_{\bl,\txantop},\sqrt{\bl}\matI_{\bl,\txantop}\mathbb{\Sigma} + \randmatW\big\}\leq \gamma_\bl\right]}\notag\\
\quad &\geq & \prob\mathopen{}\left[\prod\limits_{j=1}^{\minantop} \sin^2\mathopen{} \big\{\vece_j, \sqrt{\bl\Lambda_j}\vece_j + \randvecw_j \big\}\leq \gamma_\bl\right]\label{eq:ach-prob-sin-ub-mid1}\\
&=& \prob\mathopen{}\left[\prod\limits_{j=1}^{\minantop} \sin^2 \mathopen{}\big\{\vece_1, \sqrt{\bl\Lambda_j}\vece_1 + \randvecw_j \big\}\leq \gamma_\bl\right].\label{eq:ach-prob-sin-ub-mid2}
\end{IEEEeqnarray}
Here,~\eqref{eq:ach-prob-sin-ub-mid1} follows from Lemma~\ref{lem:angle-btw-subspaces} (Appendix~\ref{app:proof-angle-btw-subspaces}) and~\eqref{eq:ach-prob-sin-ub-mid2} follows by symmetry.
%
By repeating the same steps as in \eqref{eq:ach-dist-sin-theta-csit}--\eqref{eq:ach-csit-gamma-bl}, we obtain from~\eqref{eq:ach-prob-sin-ub-mid2} that there exists a $\gamma_\bl = \exp(-\Cnocsit + \bigO(1/\bl))$ that satisfies~\eqref{eq:ach-nocsi-condition-gamma-n}.
\end{IEEEproof}

\section{Proof of Theorem~\ref{thm:proof-asy-iso} (Dispersion of Codes with Isotropic Codewords)}
\label{app:proof-asy-iso}
Using Proposition~\ref{thm:proof-asy-ach-nocsit} with $\optcov$ replaced by $(\snr/\txant)\matI_{\txant}$, we obtain
\begin{IEEEeqnarray}{rCl}
\Rnoiso^{\ast}(\bl,\error) \geq \Ciso +\bigO\mathopen{}\left(\frac{\log\bl}{n}\right).
\label{eq:proof-asy-ach-iso}
\end{IEEEeqnarray}
Since~$\Rnoiso^{\ast}(\bl,\error) \leq \Rcriso^{\ast}(\bl,\error)$, the proof is completed by showing that
\begin{IEEEeqnarray}{rCl}
\label{eq:proof-asy-conv-iso}
\Rcriso^{\ast}(\bl,\error)\leq \Ciso +\bigO\mathopen{}\left(\frac{\log\bl}{n}\right).
\end{IEEEeqnarray}

To prove~\eqref{eq:proof-asy-conv-iso}, we evaluate the converse bound~\eqref{eq:thm-converse-rcsir-iso} in the large-\bl limit.
This evaluation follows closely the proof of~\eqref{eq:proof-asy-conv-csit} in Appendix~\ref{app:proof-asy-csit-conv}.
Let $\Lambda_1\geq\cdots\geq \Lambda_{\minant}$ be the ordered nonzero eigenvalues of $\herm{\randmatH}\randmatH$.
Following similar steps as in~\eqref{eq:inequality_beta-poly}--\eqref{eq:ub-rcsirt-original}, we obtain that for every $\gamma>0$
\begin{IEEEeqnarray}{rCl}
\IEEEeqnarraymulticol{3}{l}{
\Rcriso^{\ast}(\bl,\error) }\notag\\
\quad &\leq& \gamma - \frac{1}{\bl}\log \mathopen{}\Big(\prob[\Scsir_\bl((\snr/\txant)  \matI_{\txant}) \leq  \bl \gamma] -\error\Big) +\bigO\mathopen{}\left(\frac{1}{\bl}\right) \IEEEeqnarraynumspace
\label{eq:proof-asy-conv-iso-1}
\end{IEEEeqnarray}
with $\Scsir_\bl(\cdot)$ defined in~\eqref{eq:info_density_mimo_csir}.
To evaluate the second term on the RHS of~\eqref{eq:proof-asy-conv-iso-1}, we proceed as in~Appendix~\ref{sec:proof_of_eq:kramer_esseen_step} to obtain
\begin{IEEEeqnarray}{rCl}
 \prob[ \Scsir_\bl((\snr/\txant)  \matI_{\txant}) \leq  \bl \gamma \given \bm{\Lambda} =\veclambda] \geq q_\bl\mathopen{}\big( \altu_{\gamma}(\veclambda)\big) + \frac{\const_1}{\bl} \IEEEeqnarraynumspace
 \label{eq:lb-prob-s-iso}
\end{IEEEeqnarray}
for $\gamma$ in a certain neighborhood of~$\Ciso$.
Here,  the function $q_\bl(\cdot)$ is given in~\eqref{eq:def-q-n-y}; the function $\altu_{\gamma}(\cdot): \posrealset^{\minant} \mapsto \realset$ is defined as
\begin{equation}
\altu_{\gamma}(\bm{\lambda}) \define \frac{\gamma - \sum\nolimits_{j=1}^{\minant}\log(1+ \snr\lambda_j/\txant)}{\sqrt{\minant - \sum\nolimits_{j=1}^{\minant}(1+\snr\lambda_j/\txant)^{-2}}};
\label{eq:define-altu}
\end{equation}
$\bm{\Lambda}=[\Lambda_1,\ldots,\Lambda_{\minant}]$; and $\const_1$ is a finite constant independent of $\gamma$ and $\bm{\lambda}$.
 A lower bound on $\prob[ \Scsir_\bl((\snr/\txant)  \matI_{\txant}) \leq  \bl \gamma]$ follows then by averaging both sides of~\eqref{eq:lb-prob-s-iso} with respect to~$\bm{\Lambda}$
\begin{IEEEeqnarray}{rCl}
\prob[ \Scsir_\bl((\snr/\txant)  \matI_{\txant}) \leq  \bl \gamma ] \geq \Ex{}{q_\bl\mathopen{}\big( \altu_{\gamma} (\bm{\Lambda})\big)} + \frac{\const_1}{\bl}.
\label{eq:lb-prob-Sn-iso}
\end{IEEEeqnarray}
Proceeding as in~\eqref{eq:p-j-rt}--\eqref{eq:p-j-rt-final} and using the assumption that the joint pdf of $\Lambda_1,\ldots,\Lambda_\minant$ is continuously differentiable, we obtain that
for all $\gamma \in  (\Ciso-\delta,\Ciso+\delta)$
\begin{IEEEeqnarray}{rCl}
\Ex{}{q_\bl\mathopen{}\big( \altu_{\gamma}(\bm{\Lambda})\big)} \geq \prob\mathopen{}\left[ \sum\limits_{j=1}^{\minant}\log(1+\snr\Lambda_j/\txant)\leq\gamma\right]  + \frac{\const_2}{\bl} \IEEEeqnarraynumspace
\label{eq:evaluate-Ex-qn-iso}
\end{IEEEeqnarray}
for some $\delta>0$ and $\const_2>-\infty$.
Substituting~\eqref{eq:evaluate-Ex-qn-iso} into~\eqref{eq:lb-prob-Sn-iso}, we see that for every $\gamma \in (\Ciso-\delta,\Ciso+\delta)$
\begin{IEEEeqnarray}{rCl}
\IEEEeqnarraymulticol{3}{l}{
\prob[ \Scsir_\bl((\snr/\txant)  \matI_{\txant}) \leq  \bl \gamma]}\notag\\
 \quad &\geq& \prob\mathopen{}\left[ \sum\limits_{j=1}^{\minant}\log(1+\snr\Lambda_j/\txant)\leq\gamma\right] +\frac{\const_1+\const_2}{\bl}\\
&=& \cdistiso(\gamma) + \frac{\const_1+\const_2}{\bl}.
\end{IEEEeqnarray}
The proof of~\eqref{eq:proof-asy-conv-iso} is concluded by repeating the same steps as in~\eqref{eq:choose-gamma-csit}--\eqref{eq:asy-exp-gamma-csit}.
\end{appendices}

%

\begin{thebibliography}{10}
\providecommand{\url}[1]{#1}
\csname url@samestyle\endcsname
\providecommand{\newblock}{\relax}
\providecommand{\bibinfo}[2]{#2}
\providecommand{\BIBentrySTDinterwordspacing}{\spaceskip=0pt\relax}
\providecommand{\BIBentryALTinterwordstretchfactor}{4}
\providecommand{\BIBentryALTinterwordspacing}{\spaceskip=\fontdimen2\font plus
\BIBentryALTinterwordstretchfactor\fontdimen3\font minus
  \fontdimen4\font\relax}
\providecommand{\BIBforeignlanguage}[2]{{%
\expandafter\ifx\csname l@#1\endcsname\relax
\typeout{** WARNING: IEEEtran.bst: No hyphenation pattern has been}%
\typeout{** loaded for the language `#1'. Using the pattern for}%
\typeout{** the default language instead.}%
\else
\language=\csname l@#1\endcsname
\fi
#2}}
\providecommand{\BIBdecl}{\relax}
\BIBdecl

\bibitem{biglieri98-10a}
E.~Biglieri, J.~Proakis, and S.~Shamai~(Shitz), ``Fading channels:
  Information-theoretic and communications aspects,'' \emph{{IEEE} Trans. Inf.
  Theory}, vol.~44, no.~6, pp. 2619--2692, Oct. 1998.

\bibitem{tse05a}
D.~N.~C. Tse and P.~Viswanath, \emph{Fundamentals of Wireless
  Communication}.\hskip 1em plus 0.5em minus 0.4em\relax Cambridge, U.K.:
  Cambridge Univ. Press, 2005.

\bibitem{foschini98}
G.~J. Foschini and M.~J. Gans, ``On limits of wireless communications in a
  fading environment when using multiple antennas,'' \emph{Wirel. Personal
  Commun.}, vol.~6, pp. 311--335, 1998.

\bibitem{effros98-08}
M.~Effros and A.~Goldsmith, ``Capacity definitions and coding strategies for
  general channels with receiver side information,'' in \emph{Proc. IEEE Int.
  Symp. Inf. Theory (ISIT)}, Cambridge MA, Aug. 1998, p. p.~39.

\bibitem{han03a}
T.~S. Han, \emph{Information-Spectrum Methods in Information Theory}.\hskip 1em
  plus 0.5em minus 0.4em\relax Berlin, Germany: Springer-Verlag, 2003.

\bibitem{effros10-07}
M.~Effros, A.~Goldsmith, and Y.~Liang, ``Generalizing capacity: New definitions
  and capacity theorems for composite channels,'' \emph{{IEEE} Trans. Inf.
  Theory}, vol.~56, no.~7, pp. 3069--3087, Jul. 2010.

\bibitem{verdu94-07a}
S.~Verd\'{u} and T.~S. Han, ``A general formula for channel capacity,''
  \emph{{IEEE} Trans. Inf. Theory}, vol.~40, no.~4, pp. 1147--1157, Jul. 1994.

\bibitem{ozarow94}
L.~H. Ozarow, S.~S. (Shitz), and A.~D. Wyner, ``Information theoretic
  considerations for cellular mobile radio,'' \emph{{IEEE} Trans. Inf. Theory},
  vol.~43, no.~2, pp. 359--378, May 1994.

\bibitem{polyanskiy10-05}
Y.~Polyanskiy, H.~V. Poor, and S.~Verd{\'u}, ``Channel coding rate in the
  finite blocklength regime,'' \emph{{IEEE} Trans. Inf. Theory}, vol.~56,
  no.~5, pp. 2307--2359, May 2010.

\bibitem{polyanskiy11-08a}
Y.~Polyanskiy and S.~Verd\'{u}, ``Scalar coherent fading channel: dispersion
  analysis,'' in \emph{Proc. IEEE Int. Symp. Inf. Theory (ISIT)}, Saint
  Petersburg, Russia, Aug. 2011, pp. 2959--2963.

\bibitem{yang12-09}
W.~Yang, G.~Durisi, T.~Koch, and Y.~Polyanskiy, ``Diversity versus channel
  knowledge at finite block-length,'' in \emph{Proc. IEEE Inf. Theory Workshop
  (ITW)}, Lausanne, Switzerland, Sep. 2012, pp. 577--581.

\bibitem{hoydis13}
\BIBentryALTinterwordspacing
J.~Hoydis, R.~Couillet, and P.~Piantanida, ``The second-order coding rate of
  the {MIMO} {R}ayleigh block-fading channel,'' \emph{{IEEE} Trans. Inf.
  Theory}, 2013, submitted. [Online]. Available:
  \url{http://arxiv.org/abs/1303.3400}
\BIBentrySTDinterwordspacing

\bibitem{potter13-07}
C.~Potter, K.~Kosbar, and A.~Panagos, ``On achievable rates for {MIMO} systems
  with imperfect channel state information in the finite length regime,''
  \emph{{IEEE} Trans. Commun.}, vol.~61, no.~7, pp. 2772--2781, Jul. 2013.

\bibitem{yang14-07b}
W.~Yang, G.~Caire, G.~Durisi, and Y.~Polyanskiy, ``Finite blocklength channel
  coding rate under a long-term power constraint,'' in \emph{Proc. IEEE Int.
  Symp. Inf. Theory (ISIT)}, Honolulu, HI, USA, Jul. 2014.

\bibitem{caire99-05}
G.~Caire, G.~Taricco, and E.~Biglieri, ``Optimum power control over fading
  channels,'' \emph{{IEEE} Trans. Inf. Theory}, vol.~45, no.~5, pp. 1468--1489,
  May 1999.

\bibitem{petrov75}
V.~V. Petrov, \emph{Sums of Independent Random Variates}.\hskip 1em plus 0.5em
  minus 0.4em\relax Springer-Verlag, 1975, translated from the Russian by A. A.
  Brown.

\bibitem{telatar99-11a}
{\.I}.~E. Telatar, ``Capacity of multi-antenna {Gaussian} channels,''
  \emph{Eur. Trans. Telecommun.}, vol.~10, pp. 585--595, Nov. 1999.

\bibitem{tulino04a}
A.~M. Tulino and S.~Verd{\'u}, ``Random matrix theory and wireless
  communications,'' in \emph{Foundations and Trends in Communications and
  Information Theory}.\hskip 1em plus 0.5em minus 0.4em\relax Delft, The
  Netherlands: now Publishers, 2004, vol.~1, no.~1, pp. 1--182.

\bibitem{johnson95-2}
N.~Johnson, S.~Kotz, and N.~Balakrishnan, \emph{Continuous Univariate
  Distributions}, 2nd~ed.\hskip 1em plus 0.5em minus 0.4em\relax New York:
  Wiley, 1995, vol.~2.

\bibitem{abbe13-05}
E.~Abbe, S.-L. Huang, and {\.I}.~E. Telatar, ``Proof of the outage probability
  conjecture for {MISO} channels,'' \emph{{IEEE} Trans. Inf. Theory}, vol.~59,
  no.~5, pp. 2596--2602, May 2013.

\bibitem{shannon59}
C.~E. Shannon, ``Probability of error for optimal codes in a {G}aussian
  channel,'' \emph{Bell Syst. Tech.~J.}, vol.~38, no.~3, pp. 611--656, May
  1959.

\bibitem{barg02-09}
A.~Barg and D.~Y. Nogin, ``Bounds on packings of spheres in the {G}rassmann
  manifold,'' \emph{{IEEE} Trans. Inf. Theory}, vol.~48, no.~9, pp. 2450--2454,
  Sep. 2002.

\bibitem{polyanskiy10}
Y.~Polyanskiy, ``Channel coding: non-asymptotic fundamental limits,'' Ph.D.
  dissertation, Princeton University, 2010.

\bibitem{polyanskiy11-04}
Y.~Polyanskiy, H.~V. Poor, and S.~Verd{\'u}, ``Dispersion of the
  {Gilbert-Elliott} channel,'' \emph{{IEEE} Trans. Inf. Theory}, vol.~57,
  no.~4, pp. 1829--1848, Apr. 2011.

\bibitem{tomamichel13-05}
\BIBentryALTinterwordspacing
M.~Tomamichel and V.~Y.~F. Tan, ``$\epsilon$-capacities and second-order coding
  rates for channels with general state,'' \emph{{IEEE} Trans. Inf. Theory},
  May 2013, submitted. [Online]. Available:
  \url{http://arxiv.org/abs/1305.6789}
\BIBentrySTDinterwordspacing

\bibitem{yang13-07}
W.~Yang, G.~Durisi, T.~Koch, and Y.~Polyanskiy, ``Quasi-static {SIMO} fading
  channels at finite blocklength,'' in \emph{Proc. IEEE Int. Symp. Inf. Theory
  (ISIT 2013)}, Istanbul, Turkey, Jul. 2013.

\bibitem{molavianJazi13-10}
E.~Molavian{J}azi and J.~N. Laneman, ``On the second-order coding rate of block
  fading channels,'' in \emph{Proc. Allerton Conf. Commun., Contr., Comput.},
  Monticello, IL, USA, Oct. 2013, to appear.

\bibitem{james64}
A.~T. James, ``Distribution of matrix variates and latent roots derived from
  normal samples,'' \emph{Ann. Math. Statist.}, vol.~35, pp. 475--501, 1964.

\bibitem{munkres91-a}
J.~R. Munkres, \emph{Analysis on Manifolds}.\hskip 1em plus 0.5em minus
  0.4em\relax Redwood City, CA: Addison-Wesley, 1991.

\bibitem{tan13-11}
\BIBentryALTinterwordspacing
V.~Y.~F. Tan and M.~Tomamichel, ``The third-order term in the normal
  approximation for the {AWGN} channel,'' \emph{{IEEE} Trans. Inf. Theory},
  Dec. 2013, submitted. [Online]. Available:
  \url{http://arxiv.org/abs/1311.2337}
\BIBentrySTDinterwordspacing

\bibitem{3gpp-ts36212}
{3GPP TS 36.212}, ``Technical specification group radio access network; evolved
  universal terrestrial radio access ({E-UTRA}); multiplexing and channel
  coding (release 10),'' Dec. 2012.

\bibitem{robertson95}
P.~Robertson, E.~Villebrun, and P.~Hoeher, ``A comparison of optimal and
  sub-optimal {MAP} decoding algorithms operating in the log domain,'' in
  \emph{Proc. IEEE Int. Conf. Commun. (ICC)}, Seattle, USA, Jun. 1995, pp.
  1009--1013.

\bibitem{sesia11}
S.~Sesia, I.~Toufik, and M.~Baker, Eds., \emph{{LTE}--The {UMTS} Long Term
  Evolution: From Theory to Practice}, 2nd~ed.\hskip 1em plus 0.5em minus
  0.4em\relax UK: Wiley, 2011.

\bibitem{miao92}
J.~Miao and A.~Ben-Israel, ``On principal angles between subspaces in
  $\amsbb{R}^n$,'' \emph{Linear Algebra Appl.}, vol. 171, pp. 81--98, 1992.

\bibitem{afriat57}
S.~Afriat, ``Orthogonal and oblique projectors and the characteristics of pairs
  of vector spaces,'' \emph{Proc. Cambridge Phil. Soc.}, vol.~53, no.~4, pp.
  800--816, 1957.

\bibitem{absil06}
P.~Absil, P.~Koev, and A.~Edelman, ``On the largest principal angle between
  random subspaces,'' \emph{Linear Algebra Appl.}, vol. 414, no.~1, pp.
  288--294, Apr. 2006.

\bibitem{roh06}
J.~C. Roh and B.~D. Rao, ``Design and analysis of {MIMO} spatial multiplexing
  systems with quantized feedback,'' \emph{{IEEE} Trans. Signal Process.},
  vol.~54, no.~8, pp. 2874--2886, Aug. 2006.

\bibitem{neyman33a}
J.~Neyman and E.~S. Pearson, ``On the problem of the most efficient tests of
  statistical hypotheses,'' \emph{Philosophical Trans. Royal Soc. A}, vol. 231,
  pp. 289--337, 1933.

\bibitem{abramowitz72}
M.~Abramowitz and I.~A. Stegun, Eds., \emph{Handbook of Mathematical Functions
  with Formulas, Graphs, and Mathematical Tables}, 10th~ed.\hskip 1em plus
  0.5em minus 0.4em\relax New York: Dover: Government Printing Office, 1972.

\bibitem{rudin76}
W.~Rudin, \emph{Principles of Mathematical Analysis}, 3rd~ed.\hskip 1em plus
  0.5em minus 0.4em\relax Singapore: McGraw-Hill, 1976.

\bibitem{boyd04}
S.~Boyd and L.~Vandenberghe, \emph{Convex Optimization}.\hskip 1em plus 0.5em
  minus 0.4em\relax Cambridge, U.K.: Cambridge Univ. Press, 2004.

\bibitem{johnson95-1}
N.~Johnson, S.~Kotz, and N.~Balakrishnan, \emph{Continuous Univariate
  Distributions}, 2nd~ed.\hskip 1em plus 0.5em minus 0.4em\relax New York:
  Wiley, 1995, vol.~1.

\bibitem{qi12-02}
F.~Qi and Q.-M. Luo, ``Bounds for the ratio of two gamma functions---from
  {W}endel's and related inequalities to logarithmically completely monotonic
  functions,'' \emph{Banach {J}. {M}ath. {A}nal.}, vol.~6, no.~2, pp. 132--158,
  2012.

\bibitem{grimmett01}
G.~Grimmett and D.~Stirzaker, \emph{Probability and Random Processes},
  3rd~ed.\hskip 1em plus 0.5em minus 0.4em\relax New York, USA: Oxford
  University Press, 2001.

\bibitem{polyanskiy13}
Y.~Polyanskiy, ``Saddle point in the minimax converse for channel coding,''
  \emph{{IEEE} Trans. Inf. Theory}, vol.~59, no.~5, pp. 2576--2595, May 2013.

\bibitem{horn85a}
R.~A. Horn and C.~R. Johnson, \emph{Matrix Analysis}.\hskip 1em plus 0.5em
  minus 0.4em\relax Cambridge, U.K.: Cambridge Univ. Press, 1985.

\bibitem{edelman89-05a}
A.~Edelman, ``Eigenvalues and condition numbers of random matrices,'' Ph.D.
  dissertation, MIT, May 1989.

\bibitem{marshall79}
A.~W. Marshall and I.~Olkin, \emph{Inequalities: Theory of Majorization and its
  Application}.\hskip 1em plus 0.5em minus 0.4em\relax Orlando, FL: Academic
  Press, Inc., 1979.

\bibitem{berge63}
C.~Berge, \emph{Topological Spaces}.\hskip 1em plus 0.5em minus 0.4em\relax
  Edinburg, UK: Oliver and Boyd, 1963.

\bibitem{jost11}
J.~Jost, \emph{Riemannian Geometry and Geometric Analysis}, 6th~ed.\hskip 1em
  plus 0.5em minus 0.4em\relax Berlin, Germany: Springer, 2011.

\bibitem{chavel06}
I.~Chavel, \emph{Riemannian Geometry: A Modern Introduction}.\hskip 1em plus
  0.5em minus 0.4em\relax Cambridge, UK: Cambridge Univ. Press, 2006.

\bibitem{lutkepohl96}
H.~L\"{u}tkepohl, \emph{Handbook of Matrices}.\hskip 1em plus 0.5em minus
  0.4em\relax Chichester, England: John Wiley \& Sons, 1996.

\bibitem{hunter01}
J.~K. Hunter and B.~Nachtergaele, \emph{Applied Analysis}.\hskip 1em plus 0.5em
  minus 0.4em\relax Singapore: World Scientific Publishing Co., 2001.

\end{thebibliography}


\begin{IEEEbiographynophoto}
{Wei Yang}(S'09)
received the B.E. degree in communication engineering and M.E. degree in communication and information systems from the Beijing University of Posts and Telecommunications, Beijing, China, in 2008 and 2011, respectively. He is currently pursuing a Ph.D. degree in electrical engineering at Chalmers University of Technology, Gothenburg, Sweden. From July to August 2012, he was a visiting student at the Laboratory for Information and Decision Systems, Massachusetts Institute of Technology, Cambridge, MA.

Mr. Yang is the recipient of a Student Paper Award at the 2012 IEEE International Symposium on Information Theory (ISIT), Cambridge, MA, and the 2013 IEEE Sweden VT-COM-IT joint chapter best student conference paper award.
His research interests are in the areas of information and communication theory.
\end{IEEEbiographynophoto}

\begin{IEEEbiographynophoto}{Giuseppe Durisi}(S'02, M'06, SM'12) received the Laurea degree summa cum laude and the Doctor degree both from Politecnico di Torino, Italy, in 2001 and 2006, respectively. From 2002 to 2006, he was with Istituto Superiore Mario Boella, Torino, Italy. From 2006 to 2010 he was a postdoctoral researcher at ETH Zurich, Switzerland. Since 2010 he has been with Chalmers University of Technology, Gothenburg, Sweden, where is now associate professor. He held visiting researcher positions at IMST, Germany, University of Pisa, Italy, ETH Zurich, Switzerland, and Vienna University of Technology, Austria.

Dr. Durisi is a senior member of the IEEE. He is the recipient of the 2013 IEEE ComSoc Best Young Researcher Award for the Europe, Middle East, and Africa Region, and is co-author of a paper that won a "student paper award" at the 2012 International Symposium on Information Theory, and of a paper that won the 2013 IEEE Sweden VT-COM-IT joint chapter best student conference paper award. He served as TPC member in several IEEE conferences, and is currently publications editor of the IEEE Transactions on Information Theory. His research interests are in the areas of communication and information theory.
\end{IEEEbiographynophoto}

\begin{IEEEbiographynophoto}{Tobias Koch} (S'02, M'09) is a Visiting Professor with the Signal Theory and Communications Department of Universidad Carlos III de Madrid (UC3M), Spain. He received the M.Sc.\ degree in electrical engineering (with distinction) in 2004 and the Ph.D.\ degree in electrical engineering in 2009, both from ETH Zurich, Switzerland. From June 2010 until May 2012 he was a Marie Curie Intra-European Research Fellow with the University of Cambridge, UK. He was also a research intern at Bell Labs, Murray Hill, NJ in 2004 and at Universitat Pompeu Fabra (UPF), Spain, in 2007. He joined UC3M in 2012. His research interests include digital communication theory and information theory.

Dr.\ Koch is serving as Vice Chair of the Spain Chapter of the IEEE Information Theory Society in 2013--2014.
\end{IEEEbiographynophoto}

\begin{IEEEbiographynophoto}{Yury Polyanskiy}(S'08-M'10)
is an Assistant Professor of Electrical Engineering and Computer Science at MIT.
He received the M.S. degree in applied mathematics and physics from the Moscow Institute of Physics and Technology,
Moscow, Russia in 2005 and the Ph.D. degree in electrical engineering from Princeton
University, Princeton, NJ in 2010. In 2000-2005 he lead the development of the embedded software in the
Department of Surface Oilfield Equipment, Borets Company LLC (Moscow).
 Currently, his research focuses on basic questions in information theory, error-correcting codes, wireless communication and fault-tolerant circuits.
Over the years Dr. Polyanskiy won the 2013 NSF CAREER award, the
2011 IEEE Information Theory Society Paper Award and Best Student Paper Awards at the 2008 and 2010 IEEE International
Symposia on Information Theory (ISIT).
\end{IEEEbiographynophoto}
\end{document}